\documentclass[
 article,
 groupedaddress,
 showpacs,
 preprintnumbers,
 amsmath,
 amssymb,
 aps,
 prstab,
 floatfix,
]{revtex4-1}

\usepackage{graphicx}					% Graphics
\usepackage{verbatim}					% 
\usepackage{dcolumn}                    % Align table columns on decimal point
\usepackage[ruled,vlined]{algorithm2e}
\include{pythonlisting}
\usepackage{color}
\usepackage{xcolor}
\usepackage[colorlinks = true,
            linkcolor = blue,
            urlcolor  = blue,
            citecolor = blue,
            anchorcolor = blue]{hyperref}
\newcommand{\pd}{\partial}				    % Partial derivative
\newcommand{\dd}{\mathrm{d}}				% General derivative

\newcommand{\const}{\mathrm{const}}         % const
\newcommand{\K}{\mathcal{K}}				% Invariant of the map

\newcommand{\ds}{\displaystyle}				% Displaystyle
\newcommand{\cc}{\centering}			    % Centering

\begin{document}

%===============================================================================
\title{
Machine-assisted discovery of integrable symplectic mappings
% Integrable area-preserving plane maps with polygon invariants I.
% Piecewise linear integer force function
}
\author{T.~Zolkin}
\email{zolkin@fnal.gov}
\affiliation{Fermilab, PO Box 500, Batavia, IL 60510-5011}
\author{Y.~Kharkov}
\affiliation{Joint Center for Quantum Information and Computer Science,
NIST/University of Maryland, College Park, MD 20742}
\affiliation{Joint Quantum Institute,
NIST/University of Maryland, College Park, MD 20742}
\thanks{Currently at AWS Quantum Technologies}
\author{S.~Nagaitsev}
\email{nsergei@jlab.org}
\affiliation{Jefferson Lab, Newport News, VA 23606}
\affiliation{Old Dominion University, Norfolk, VA 23529}
\date{\today}
%===============================================================%

%===============================================================%
\begin{abstract}

We present a new automated method for finding integrable
symplectic maps of the plane.
These dynamical systems possess a hidden symmetry associated with
an existence of conserved quantities, i.e. integrals of motion.
The core idea of the algorithm is based on the knowledge that the
evolution of an integrable system in the phase space is restricted
to a lower-dimensional submanifold.
Limiting ourselves to polygon invariants of motion, we analyze the
shape of individual trajectories thus successfully distinguishing
integrable motion from chaotic cases.
For example, our method rediscovers some of the famous
McMillan-Suris integrable mappings and ultra-discrete Painlev\'e
equations.
In total, over 100 new integrable families are presented and
analyzed; some of them are isolated in the space of parameters,
and some of them are families with one parameter (or the ratio of
parameters) being continuous or discrete.
At the end of the paper, we suggest how newly discovered maps
are related to a general 2D symplectic map via an introduction
of discrete perturbation theory and propose a method on how to
construct smooth near-integrable dynamical systems based on
mappings with polygon invariants.

\end{abstract}
%===============================================================%

%\pacs{00.00.Aa ,
%      00.00.Aa ,
%      00.00.Aa ,
%      00.00.Aa }% PACS, the Physics and Astronomy Classification Scheme.
%\keywords{Suggested keywords}% Use showkeys class option if keyword
                             % display desired

%===============================================================%
\maketitle
\tableofcontents
%===============================================================%

\newpage
%===============================================================%
%===============================================================%
%===============================================================%
\section{Introduction}

%----------------------------------------------------------------
Integrable systems play a fundamental role in mathematics and
physics, particularly in dynamical system theory, due to
well-understood dynamics.
Integrability is generally defined as the existence of a sufficient
number of functionally independent conserved quantities (integrals
of motion) in involution, which are related to intricate hidden
symmetries. 
Hence, in many cases the construction of integrable systems often
requires some special fine-tuning. This makes them very difficult
to discover. 
Although integrable models occupy measure-zero in the space of
parameters, they play a crucial role in the theory of dynamical
systems.
In fact, they shed light on the properties of more generic
near-integrable cases via Kolmogorov-Arnold-Moser (KAM) theorem
and various perturbation theories. 

%----------------------------------------------------------------
While the concept of integrability is most often discussed in the
context of continuous-time systems, it also arises in discrete-time
maps~\cite{arnold1968ergodic}.
Mappings are of profound importance in the field of dynamical
system theory exhibiting a broad range of behaviours including
dynamical chaos, bifurcations, strange attractors, and
integrability~\cite{lichtenberg2013regular}.
On the one hand, they can be used as an approximation of a
continuous flow (since in many cases the discretized form can be
easier to study) or as a reduction to a lower dimensional manifold,
e.g., considering Poincar\'e cross-sections.
%----------------------------------------------------------------
A particularly valuable class of discrete dynamical systems are
symplectic maps, associated with Hamiltonian flows, e.g., a 2D
area-preserving map is equivalent to a Hamiltonian system with 1
degree of freedom and a potential energy that is periodic in time.
Symplectic maps arise as effective models in a wide range of
domains, such as celestial mechanics, plasma hydrodynamics, fluid
flows, and particle accelerators~\cite{Meiss_RevModPhys.64.795}.

%----------------------------------------------------------------
Although area-preserving maps of the plane have been studied for
decades, only a handful of integrable cases have been discovered.
The existence of integrals of motion restricts phase space dynamics
to a set of nested tori.
Any discovery of novel integrable maps is highly non-trivial and
historically has been accomplished by scientists making inspired
guesses and applying relevant analytical tools.
In many cases, even the proof of the integrability of a specific
mapping is quite challenging.

%----------------------------------------------------------------
An automated search for integrable systems remains a nontrivial
task: only a few successful examples exist to date where
machine-assisted searches resulted in new discoveries in nonlinear
dynamics.
For example, computer-assisted searches combined with high precision
simulations found new families of periodic orbits in a gravitational
three-body problem \cite{vsuvakov2013three,li2017more,li2018over}.
Recent works utilized machine-learning  methods to discover
integrals of motion of classical dynamical systems from phase space
trajectories, e.g. Refs.~\cite{bondesan2019learning, tegmark2021,
ha2021discovering, wetzel2020discovering}.
These examples employed a computer-assisted search to either
discover new families of trajectories or to rediscover previously
known invariants of motion in integrable systems. 
In contrast, our algorithm has found previously unknown integrable
symplectic maps, advancing the theory of integrable systems.
    
%----------------------------------------------------------------
In this paper, we propose and describe this new algorithm for
discovering area-preserving integrable maps of the plane. 
While scanning over a large space of potential candidate maps, our
algorithm traces out individual orbits and subsequently analyzes
shapes of invariant manifolds.
We restricted our search to the family of mappings which possess
integrals of motion with a polygon geometry, thus constraining our
search to piecewise linear transformations of the plane.
Piecewise linear mappings could be regarded as the simplest
possible dynamical systems with rich nonlinear dynamics, which
makes them an attractive model to study.
They could be viewed as fundamental building blocks of generic
integrable symplectic maps.
An arbitrary smooth 2D symplectic map could be approximated
as an $n$-piece piecewise linear map, but only in the uniform
sense, see Section~\ref{sec:Discrete}.
By applying a machine-assisted search, we found over a hundred new
integrable mappings; some of them are isolated in the space of
parameters, and some of them are families with one parameter (or
the ratio of parameters) being continuous ($\in \mathbb{R}^{(+)}$)
or discrete ($\in \mathbb{Z}^{(+)}$).

%----------------------------------------------------------------
The structure of this article is as follows.
In Section~\ref{sec:2Dmaps} we provide a historical overview of
integrability of symplectic mappings of the plane.
In Sections~\ref{sec:1Pmaps} and~\ref{sec:2Pmaps} we review linear
integrable mappings produced by 1- and 2-piecewise linear force
functions.
At the end of Section~\ref{sec:2Pmaps} we introduce nonlinear
integrable mappings with polygon invariants, and in
Section~\ref{sec:Search} we propose an algorithm for the
machine-assisted discovery of integrable systems with more
complicated force functions.
Sections~\ref{sec:3Pmaps} and \ref{sec:4Pmaps} presents the main
results of our search, and finally, in Section~\ref{sec:App} we
discuss some practical applications of mappings with polygon
invariants.
Several appendices contain supplementary materials, tables with
dynamical properties of mappings, and a pseudo-code for the search
algorithm.

\newpage
%===============================================================%
%===============================================================%
%===============================================================%
\section{\label{sec:2Dmaps}2D symplectic mappings and integrability:
                           historical overview}

%----------------------------------------------------------------
The most general form of an autonomous  2D mapping of the plane,
$\mathcal M: \mathbb{R}^2 \to \mathbb{R}^2$, can be written as
\[
\label{math:Gform}
\begin{array}{l}
     q' = F(q, p),      \\[0.2cm]
     p' = G(q, p),
\end{array}
\]
where $'$ denotes an application of the map, and $F$ and $G$ are
functions of both variables.
If the Jacobian of a 2D map, $J\equiv\pd_q F\,\pd_p G-\pd_p F\,\pd_q G$,
is equal to unity, then the map is {\it area-preserving} and thus
also {\it symplectic} (for higher dimensional mappings, the
volume-preservation is necessary but not sufficient for a map to be
symplectic).
A map is called {\it integrable} in a Liouville
sense~\cite{veselov1991integrable} if there exists a real-valued
continuous function $\K(p,q)$ called an {\it integral of motion}
(or {\it invariant}), for which the level sets $\K(p,q) = \const$
are curves (or sets of points) and
\[
\forall\,(q,p)\in\mathbb{R}^2: \qquad \K(p',q') = \K(p,q).
\]

%----------------------------------------------------------------
The first nonlinear symplectic integrable map of the plane was
discovered by E.~McMillan~\cite{mcmillan1971problem} when he studied
a very specific form of the map
\begin{equation}
\begin{array}{l}
     q' = p,            \\[0.2cm]
     p' =-q+f(p),
\end{array}
    \label{math:MTform}
\end{equation}
where $f(p)$ will be called the {\it force function}.
We will refer to this mapping as the {\it McMillan-H\'enon form}
or the MH form for short (please  do not confuse with {\it McMillan
integrable map} \cite{mcmillan1971problem}, which is a special case
when $f(p)$ is given by a ratio of two quadratic polynomials).

%----------------------------------------------------------------
At first, the choice of the MH form (\ref{math:MTform}) might look
very restrictive and unmotivated.
McMillan's original idea was based on its simplicity and clear
symmetries (which will be described in detail below), which still
allowed some degree of freedom in modifying the dynamics by
varying $f(p)$.
As it was demonstrated later by D.~Turaev~\cite{turaev2002polynomial},
almost every symplectic map of the plane (and even 2$n$D symplectic
map) can be approximated by iterations of MH maps.
In this article, we will restrict our consideration to this form;
however, our results can be easily generalized to different
representations of the map.

%----------------------------------------------------------------
Below we summarize several important results regarding the integrability
of mappings in a McMillan-H\'enon form:
\begin{enumerate}
    %----------------------------------------------------------------
    \item First, the family of integrable mappings of the plane in the
    MH form were discovered by McMillan~\cite{mcmillan1971problem} for
    a biquadratic invariant in the form
    \[
    (\mathrm{I}):\qquad    \mathcal{K}(p,q) =
        \mathrm{A}\,p^2q^2 + \mathrm{B}\,(p^2q + p\,q^2) +
        \Gamma\,(p^2 + q^2) + \Delta\,p\,q + \mathrm{E}\,(p+q)
    \]
    corresponding to the force function
    \[
    f(p) = \frac{\mathrm{B}\,p^2 + \Delta\,p + \mathrm{E}}
                {\mathrm{A}\,p^2 + \mathrm{B}\,p + \Gamma}
    \]
    and integrable for any values of parameters $\mathrm{A}$,
    $\mathrm{B}$, $\Gamma$, $\Delta$ and $\mathrm{E}$
    (Fig.~\ref{fig:IntMaps}a).
    
    %----------------------------------------------------------------
    \item Later, Y. Suris~\cite{suris1989integrable} studied the
    integrability of a difference equation in the form
    $
    x_{n+1} + x_{n-1} = f(x_n)
    $
    referred to as the {\it Suris form of the map} (sometimes referred
    to as the {\it standard type}).
    The Suris form of the mapping has a clear physical interpretation,
    since in the continuous-time limit it takes a form of a 1D Newton
    equation $\ddot x = f(x)-2\,x$, where the right hand side plays a
    role of a mechanical force. 
    Mapping in the Suris form is just another representation of the MH
    form by setting $q_n = x_n$ and $p_n = x_{n+1}$ (one can think of
    an analogy between the Lagrangian approach with a single differential
    equation of the second order, and the Hamiltonian approach with two
    differential equations of the first order,
    e.g. see~\cite{veselov1991integrable}).
    Suris proved that for analytic force functions $f(p)$ and
    $\mathcal{K}(p,q)$, the invariant of the integrable mapping can take
    only one of the forms:
    (I) biquadratic function of $p$ and $q$ (McMillan map),
    (II) biquadratic exponential or
    (III) trigonometric polynomial:
    \[
    \begin{array}{ll}
    (\mathrm{II}):    &\ds  \mathcal{K}(p,q) =
        \mathrm{A}\,e^{\alpha\,p}e^{\alpha\,q} +
        \mathrm{B}\,(e^{\alpha\,p} + e^{\alpha\,q}) +
        \Gamma\,(e^{\alpha\,p}e^{-\alpha\,q} + e^{-\alpha\,p}e^{\alpha\,q}) +
        \Delta\,(e^{-\alpha\,p} + e^{-\alpha\,q}) +
        \mathrm{E}\,e^{-\alpha\,p}e^{-\alpha\,q},               \\[0.25cm]
    (\mathrm{III}):   &\ds \mathcal{K}(p,q) =
        \Lambda_1\,(\cos[\omega\,p-\psi]+\cos[\omega\,q-\psi]) +
        \Lambda_2\,\cos[\omega\,(p+q)-\phi] +
        \Lambda_3\,\cos[\omega\,(p-q)],                         \\[0.25cm]
    \mathrm{and}&\ds
    f_\mathrm{II}(q) = \frac{1}{\alpha}\,\log\left[
    \frac{\Gamma\,e^{\alpha\,q} + \Delta + \mathrm{E}\,e^{-\alpha\,q}}
         {\mathrm{A}\,e^{\alpha\,q} + \mathrm{B} + \Gamma\,e^{-\alpha\,q}}
    \right],
    \qquad
    f_\mathrm{III}(q) = \frac{2}{\omega}\,\arctan\left[
    \frac{\Lambda_1\sin\Phi + \Lambda_2 \sin(\Psi-q) + \Lambda_3 \sin q}
         {\Lambda_1\cos\Phi + \Lambda_2 \cos(\Psi-q) + \Lambda_3 \cos q}
    \right].
    \end{array}
    \]
    Mappings (I -- III) are integrable for any choice of parameters, see
    some examples with invariant level sets in Fig.~\ref{fig:IntMaps} (a. -- c.).
    Results by Suris greatly reduce possible forms of $f(p)$ corresponding
    to integrable maps.

%----------------------------------------------------------------
\begin{figure}[t!]\centering
\includegraphics[width=\columnwidth]{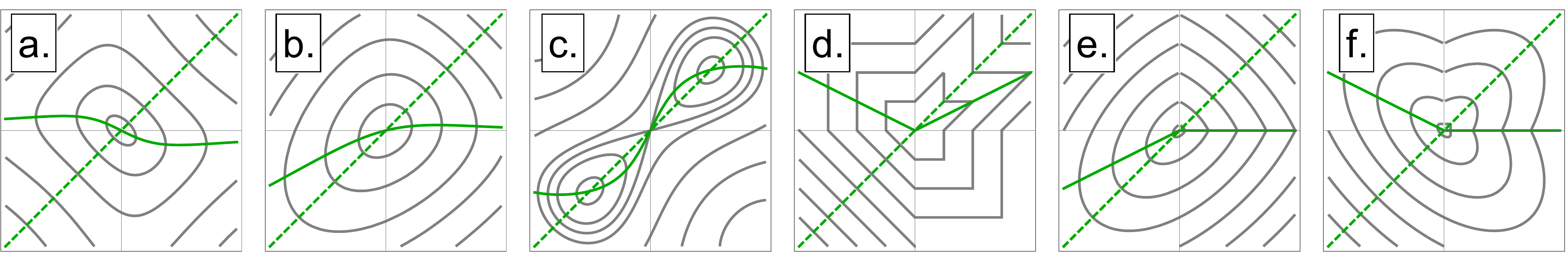}
\caption{\label{fig:IntMaps}
	Invariant level sets, $\K(p,q)=\const$, for integrable
        mappings in the MH form.
	McMillan-Suris mappings I -- III, respectively (a.) -- (c.),
	with
        $\mathrm{A},\Gamma,\Delta,\Lambda_{1,3} = -\Lambda_2 =1$ and $\mathrm{B},\mathrm{E},\psi,\phi=0$.
	Brown-Knuth map (d.) along with McMillan beheaded (e.) and
        two-headed (f.) ellipses.
	All mappings are shown for the range $q,p\in[-2,2]$.
        The dashed and solid dark green lines show the first and
        second symmetry lines.
    }
\end{figure}
%----------------------------------------------------------------

    %----------------------------------------------------------------
    \item In 1983 Morton Brown proposed an interesting problem in the
    section ``Advanced Problems'' of Amer. Math. Monthly
    \cite{brown1983}.
    
    {\it 6439. Let $\{x_n\}$ be a sequence of real numbers satisfying
    the relation $x_{n+1} = |x_n| - x_{n-1}$.
    Prove that $\{x_n\}$ is periodic with period 9.}
    
    One can notice that this map is in the Suris form (so it can be
    written in the MH form as well) with $f(p) = |p|$.
    
    In 1985 the solution by J. F. Slifker was published in
    \cite{brown1985} with a comment below:
    
    {\it Also solved by the proposer and sixty-one others.
    The problem turned out to be rather elementary.}
    
    Despite its ``simplicity,'' the map has proven to be interesting
    for dynamical system theory, combinatorics and topology.
    10 years later, in 1993, Brown extended his results to the formal
    publication \cite{brown1993}.
    He reconsidered the equation as the homeomorphism of the plane
    and provided a geometrical proof of periodicity including
    a description of the polygon invariant of motion
    (see~Fig.~\ref{fig:IntMaps}d):
    \[
        \K(p,q) =
        q +
        \big| q - |p| \big| +
        \Big| p - \big| q - |p| \big| \Big| +
        \Big| q - \big| p - |q| \big| \Big| +
        \bigg| p - |q| + \Big| q - \big| p - |q| \big| \Big| \bigg|.
    \]
    D. Knuth, who provided his own combinatorical proof of periodicity,
    wrote a letter to Brown:
    
    {\it When I saw advanced problem 6439, I couldn't believe that
    it was ``advanced'': a result like that has to be either false
    or elementary!
    
    But I soon found that it wasn't trivial.
    There is a simple proof, yet I can’t figure out how on earth
    anybody would discover such a remarkable result.
    Nor have I discovered any similar recurrence relations having
    the same property.
    So in a sense I have no idea how to solve the problem properly.
    Is there an “insightful” proof, or is the result simply true by
    chance?}
    
    While quite often this map is referred to as Knuth map, we would
    like to call it {\it Brown-Knuth map} to acknowledge all contributors.
    
    %----------------------------------------------------------------
    \item Finally, in his original article \cite{mcmillan1971problem},
    Edwin McMillan suggested another intricate integrable system with a
    piecewise quadratic invariant and a piecewise linear force function
    \[
    \K(p,q) = \left\{
        \begin{array}{l}
            p^2 + a\,p\,q + q^2,\qquad p,q \geq 0,          \\[0.2cm]
            p^2 - a\,p\,q + q^2,\qquad \mathrm{otherwise},
        \end{array}\right.
    \qquad\mathrm{and}\qquad
    f(p) = \left\{
        \begin{array}{ll}
            a\,p,   & q <    0,                             \\[0.2cm]
            0,      & q \geq 0,
        \end{array}\right.
    \]
    where $a\in \mathbb{R}$ is an arbitrary coefficient. 
    He referred to the corresponding invariant level sets as
    {\it beheaded} and {\it two-headed ellipses}
    (Fig.~\ref{fig:IntMaps}e).
    As we will see in Section~\ref{sec:ElemPII}, this map as well
    as Brown-Knuth map are examples of more general integrable
    systems with the force function $f(p)=a\,p + b\,|p|$ and invariants
    being collection of ellipses, hyperbolas and straight lines
    (segments).
    
    %----------------------------------------------------------------
    \item At the end, we should mention a special class of finite difference
    equations, which are integrable (in the sense that they admit a Lax
    pair representation), known as discrete Painlev\'e equations.
    Discrete Painlev\'e equations are closely related to integrable
    symplectic maps of the plane.
    We discuss the connection in Section~\ref{sec:Panleve}.
\end{enumerate}

\begin{figure}[bh!]\centering
\includegraphics[width=0.4\columnwidth]{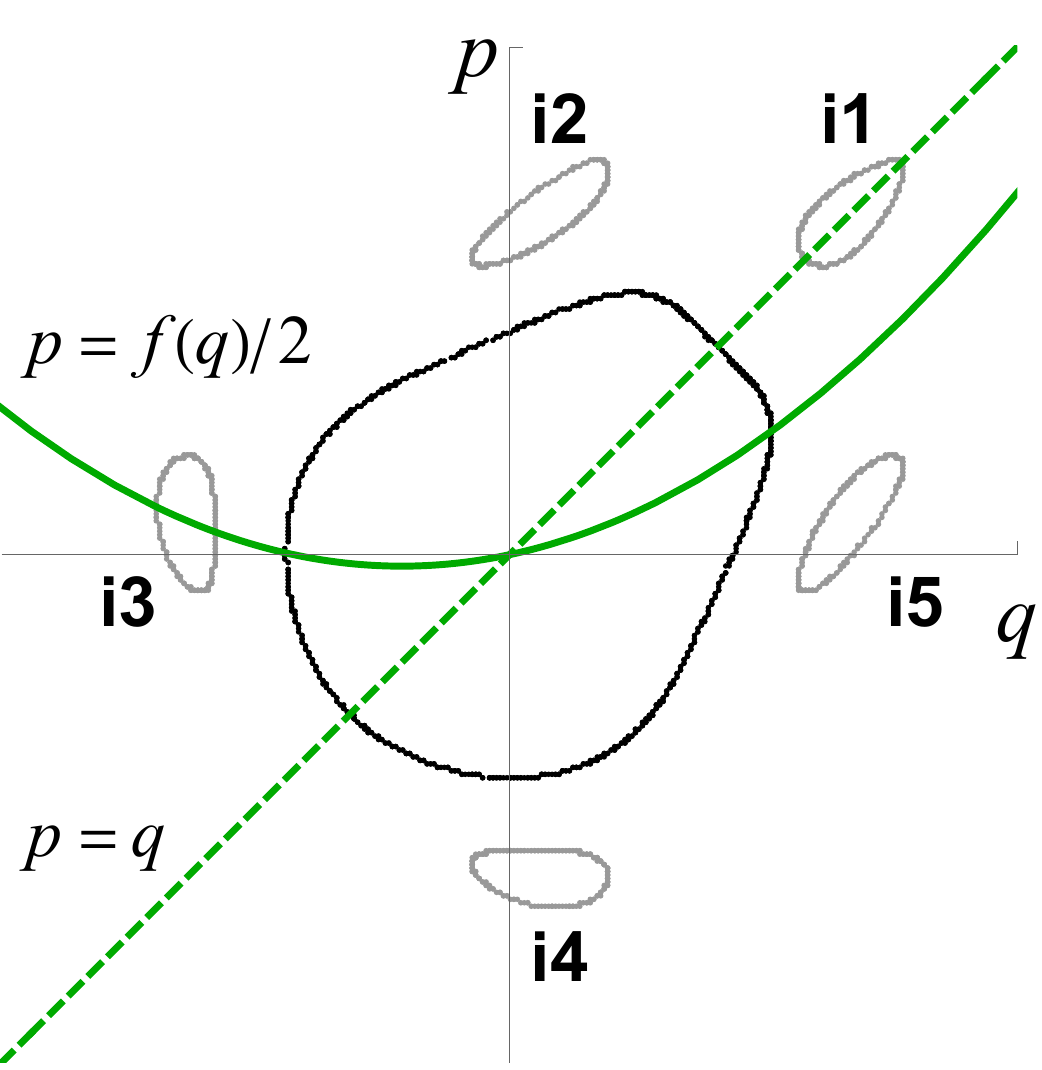}
\caption{\label{fig:McMSymm}
    Two constant level sets $\K(p,q)=\const$ for the chaotic
    H\'enon map, $f(p) = a\,p + b\,p^2$.
    The black curve shows a closed trajectory encompassing the
    origin and the gray curves show a set of islands (i.e.
    under iterations, points hop from island to island covering
    the closed gray curves labeled i1 -- i5).
    While the first (reflection) symmetry is quite obvious from
    the figure, we will focus on the second one.
    The upper branch of the blue curve, $\Phi(q)$, is vertically
    equidistant from the lower branch, $\Phi^{-1}(q)$.
    Island i3 satisfies the second symmetry in the same manner
    as does the blue curve.
    Islands i1 and i2 satisfy the symmetry in a sense that they
    are vertical reflections of islands i5 and i4, respectively
    (e.g. the lower part of i1 is equidistant from the upper
    part of i5 and the upper part of i1 is equidistant from the
    lower part of i5).
    The dashed and solid green lines are the first and second
    symmetry lines.
    }
\end{figure}

\newpage
%----------------------------------------------------------------
Several properties of MH mappings (\ref{math:MTform}) which will
help us to understand and interpret further results are:
\begin{itemize}
    %----------------------------------------------------------------
    \item A map in the MH form is symplectic and reversible for any
    continuous $f(p)$, with the inverse
          \[
            \begin{array}{ll}
                \mathcal{M}^{-1}:\quad   
                \begin{cases}
                q' =-p+f(q),\\
                p' = q.
                \end{cases}
            \end{array}
          \]
    \item For any map with $f(p)$ there is a {\it twin map} with the
    mirrored force function $-f(-p)$, so that their dynamics are
    identical up to a rotation of phase space by an angle of
    $\pi$, $(p,q)\rightarrow -(p,q)$.
    Thus, we will omit twin maps unless it is necessary.
    %----------------------------------------------------------------
    \item A MH map can be decomposed into two transformations
          $\mathcal{M} = \mathcal{F}\circ\mathrm{Rot}(\pi/2)$, where
          \[
            \begin{array}{ll}
                \mathrm{Rot}(\pi/2):\quad
                \begin{cases}
                q' = p,\\
                p' =-q,
                \end{cases}
            \end{array}
            \qquad\mathrm{and}\qquad
            \begin{array}{ll}
                \mathcal{F}:\quad 
                \begin{cases}
                q' = q,\\
                p' = p + f(q),                
                \end{cases}
            \end{array}
          \]
          with $\mathrm{Rot(\theta)}$ being a {\it rotation} (about
          the origin by an angle $\theta$):
          \[\mathrm{Rot}(\theta):\quad 
          \begin{bmatrix}
          q'\\
          p'
          \end{bmatrix} =
          \begin{bmatrix}
          \cos\theta & \sin\theta\\
         -\sin\theta & \cos\theta
          \end{bmatrix}
          \begin{bmatrix}
          q\\
          p
          \end{bmatrix}
          \]
          and $\mathcal{F}$ being a {\it thin lens} transformation.
          The name ``thin lens'' comes from optics and reflects
          that the transformation is localized so coordinates remain
          unchanged, $q'=q$, and momentum (or angle in optics) is
          changed according to location, $\Delta p \equiv p'-p=f(q)$.
          Both transformations are  area-preserving and symplectic.
          This decomposition is very important and explains the physical
          nature of the map representing a particle kicked periodically
          in time with the force $f(q)$.
          A general model of kicked oscillators includes famous
          examples of Chirikov
          map~\cite{chirikov1969research,chirikov1979universal},
          H\'enon map~\cite{henon1969numerical} and many others.
          In particular, this decomposition is often employed in
          accelerator physics;
          a model accelerator with one degree of freedom consisting
          of a linear optics insert and  a thin nonlinear lens (such
          as sextupole or McMillan integrable lens) can be rewritten
          in the MH form.
    %----------------------------------------------------------------
    \item A MH map can be decomposed into superposition of two
          transformations
          $\mathcal{M} = \mathcal{G}\circ\mathrm{Ref}(\pi/4)$ where
          \[
            \begin{array}{ll}
                \mathrm{Ref}(\pi/4):\quad
                \begin{cases}
                q' = p,\\
                p' = q,                    
                \end{cases}
            \end{array}
            \qquad\mathrm{and}\qquad
            \begin{array}{ll}
                \mathcal{G}:\quad 
                \begin{cases}
                q' = q,\\
                p' =-p + f(q),
                \end{cases}
            \end{array}
          \]
          with $\mathrm{Ref(\theta)}$ being a {\it reflection} about
          a line passing through the origin at an angle $\theta$
          with the $q$-axis
          \[\mathrm{Ref}(\theta):\quad
          \begin{bmatrix}
          q'\\
          p'
          \end{bmatrix} =
          \begin{bmatrix}
          \cos2\theta & \sin2\theta\\
          \sin2\theta &-\cos2\theta
          \end{bmatrix}
          \begin{bmatrix}
          q\\
          p
          \end{bmatrix}
          \]
          and $\mathcal{G}$ being a special {\it nonlinear vertical
          reflection}.
          Transformation $\mathcal{G}$ leaves the horizontal coordinate
          invariant, $q'=q$, while vertically 
          a point is reflected with
          respect to the line $p=f(q)/2$, i.e. equidistant condition
          \[p'-f(q)/2 =f(q/2)-p\] is satisfied.
          Both reflections are {\it anti-area-preserving}
          (the determinant of Jacobian is equal to $-1$) and are {\it
          involutions}, i.e. double application of the map is identical
          to transformation $\mathrm{Ref}^2(\theta)=\mathcal{G}^2=I_2$.
          In addition, each reflection has a line of fixed points:
          $p=q$ for $\mathrm{Ref}(\pi/4)$ (further referred to as the
          {\it first symmetry line}) and $p=f(q)/2$ for $\mathcal{G}$
          (further referred to as the {\it second symmetry line}), 
          see Fig.~\ref{fig:McMSymm}.
    %----------------------------------------------------------------      
    \item Fixed points of the map are given by the intersection of
          the two symmetry lines and always in I or
          III quadrants:
          \[
                p^* = q^* = f(q)/2.
          \]
    %----------------------------------------------------------------
    \item 2-cycles (if any) are given by the intersection of the
          second symmetry line $p=f(q)/2$ with its inverse
          $q=f(p)/2$ (so the intersections is always in II and IV
          quadrants).
          Additionally, if $f(p)$ is an odd function, 2-cycles are
          restricted to anti-diagonal $p=-q$.
    %----------------------------------------------------------------
    \item Finally, the last and most important property we would
          like to list here:
          any constant invariant level set of an integrable MH map
          (either a point, set of points, closed and open curves,
          or even collection of closed curves representing islands)
          is invariant under both reflections,
          $\mathrm{Ref}(\pi/4)$ and $\mathcal{G}$.
          If $f(p)$ is an odd function, invariant level sets possess
          additional symmetry $\mathcal{K}(p,q) = \mathcal{K}(-p,-q)$,
          meaning that the map coincides with its twin.
\end{itemize}

%----------------------------------------------------------------
Invariance under these transformations results in two geometrical
consequences, first noticed in relation to integrability by
E.~McMillan~\cite{mcmillan1971problem}.
First, invariant level sets are symmetric with respect to
transformation $\mathrm{Ref}(\pi/4)$:
\[
    \mathcal{K}(p,q) = \mathcal{K}(q,p).
\]
Second, nontrivial symmetry with respect to transformation
$\mathcal{G}$ is
\[
    \mathcal{K}(p,q) = \mathcal{K}(-p+f(q),q).
\]
In order to understand how this symmetry works, let's consider a
closed invariant curve $\K(p,q)=\const$ encompassing the fixed
point (blue curve in Fig.~\ref{fig:McMSymm}).
Solving implicit equation $\K(p,q)=\const$ for $p$ results in two
roots  $p=\Phi(q)$ or $p=\Phi^{-1}(q)$, which correspond to the
upper or lower branches in Fig.~\ref{fig:McMSymm}.
Next, using the mapping equation (MH form of map) the reader can
verify that the following important condition holds
\begin{equation}
\label{eq:McMillan_symmetry}
    f(q) = \Phi(q) + \Phi^{-1}(q).  
\end{equation}
The condition (\ref{eq:McMillan_symmetry}) implies that the symmetry line $p=f(q)/2$
``splits'' closed curve $\K(p,q)=\const$ into two halves:
upper $\Phi(q)$ and lower $\Phi^{-1}(q)$ branches with the
endpoints located at $\pd_q\Phi(q)=\pm\infty$.
For each value of $q$ where $\K(p,q)=\const$ is defined, the
upper and lower parts are equidistant from the second symmetry line
\[
    f(q)/2 - \Phi(q) = \Phi^{-1}(q) - f(q)/2.
\]

%----------------------------------------------------------------
McMillan wrote in regards to this condition:
{\it ``This is a remarkable result, of startling simplicity, which
fell out almost without effort on my part.
It leads to not just one, but to great families of functions $f(p)$
giving regions of stability.
It also has a simple geometrical interpretation.''}

%----------------------------------------------------------------
Here we would like to make two comments.
(i) The origin of these two symmetries is due to reversibility of
symplectic maps~\cite{lewis1961reversible};
any symplectic map with an inverse can be decomposed into
two involutions similar to a map in the MH form with the integral
of motion being invariant under both transformations.
(ii) Both symmetries also hold for survived
invariant tori in chaotic maps.
Fig.~\ref{fig:McMSymm} illustrates symmetry properties for
two classes of survived tori in H\'enon map: a closed curve and a
set of islands.

%----------------------------------------------------------------
%----------------------------------------------------------------

\newpage
%===============================================================%
%===============================================================%
%===============================================================%
\section{\label{sec:1Pmaps}1-piece maps}

%----------------------------------------------------------------
In this section, we will consider mappings in the MH form with $f$
being a simple linear function, $f(q)=k\,q$.
We will refer to such mappings as {\it 1-piece maps} or mappings
with {\it linear force function}, purposely avoiding the term
``linear map.''
We will reserve the term {\it linear map} (in contrast to {\it
nonlinear map}) in order to refer to independence of dynamics on
amplitude (i.e. rotation number is independent on amplitude for all
stable trajectories);
mappings in the MH form with nonlinear force function can possess
linear dynamics.
In order to avoid confusion, we will follow the terminology
established above.
As we will see, all 1-piece maps are integrable and linear.

%===============================================================%
%===============================================================%
\subsection{Polygon maps with integer coefficients}

\begin{figure}[ht!]\centering
\includegraphics[width=0.65\columnwidth]{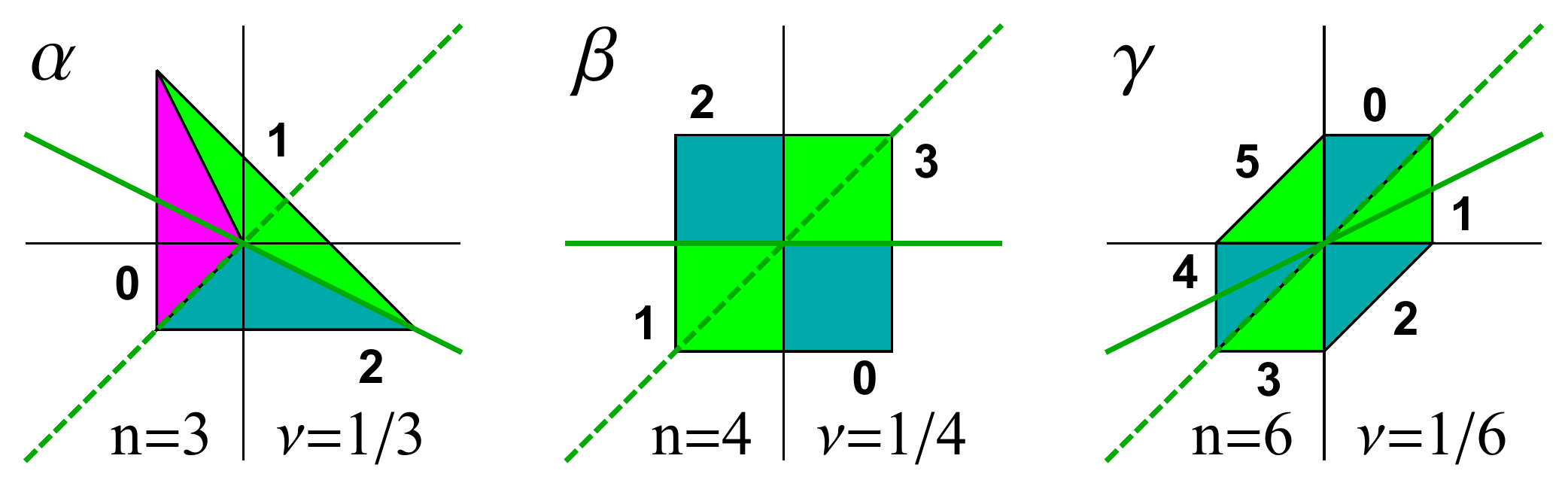}
\caption{\label{fig:PMaps1}
	Integer mappings with linear force function and polygon
	invariants $\alpha$, $\beta$ and $\gamma$ (we use the same
        naming convention as
        Ref.~\cite{cairns2014piecewise,cairns2016conewise}).
	Note that the invariant triangle was modified in order to
        satisfy both symmetries.
	Each figure shows one invariant polygon, with each colored
	region $i$ being mapped to region $i+1$.
	Here $n$ is the period and $\nu$ is the rotation number of
        the map, which represents the average increase in the angle
        per map iteration.
	The dashed and solid dark green lines show the first and
        second symmetry lines.
    }
\end{figure}

%----------------------------------------------------------------
According to {\it crystallographic restriction theorem}
\cite{bamberg2003crystallographic,kuzmanovich2002finite},
if $A$ is an integer $2 \times 2$ matrix and $A^n = I_2$ for some
natural $n \in \mathbb{N}$, where $I_2 = \mathrm{diag}\,(1,1)$
%\[
%I_2 =
%\begin{bmatrix}
%1 & 0 \\
%0 & 1 
%\end{bmatrix}
%\]
is an identity matrix, then the only possible solutions have
periods $n = 1,\, 2,\, 3,\, 4,\, 6$, which corresponds to 1-,2-,3-,4- and 6-fold
rotational symmetries.
Transformations with a period $n=1,2$ are simply $\pm I_2$, which
can be considered as special cases of rotation of the plane,
$\mathrm{Rot}(\theta)$, through the angles equal to $\theta=0$ or $\pi$
respectively.
Three other cases, $n = 3,4,6$, are given by mappings in the MH form
\[
\begin{array}{ll}
    \alpha:\quad 
    \begin{cases}
    q' = p,\\
    p' =-q - p
    \end{cases}
\end{array}\qquad\qquad
    \beta:\quad
    \begin{cases}
    q' = p,\\
    p' =-q
    \end{cases}\qquad\qquad
    \gamma:\quad
    \begin{cases}
    q' = p,\\
    p' =-q + p
    \end{cases}
\]
further referred to as $\alpha$, $\beta$ and $\gamma$ respectively,
and, with invariant polygons being concentric triangles, squares
and hexagons (see Fig.~\ref{fig:PMaps1}).

%----------------------------------------------------------------
%----------------------------------------------------------------

%===============================================================%
%===============================================================%
\subsection{Fundamental polygons of the first kind}

\begin{figure}[t!]\centering
\includegraphics[width=\columnwidth]{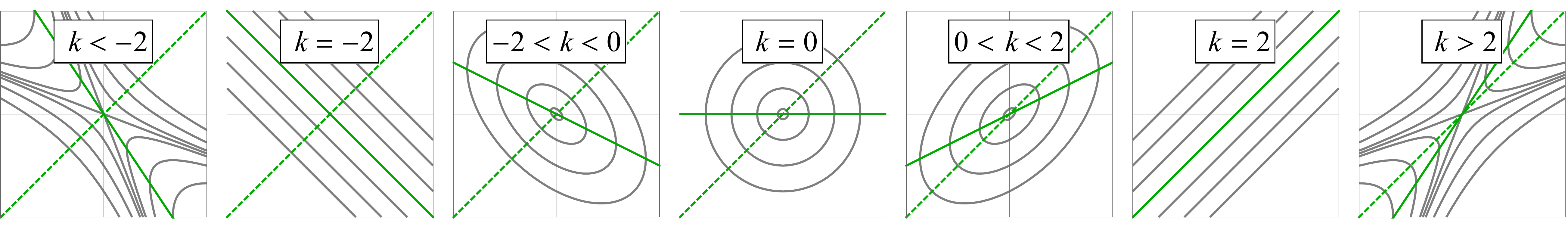}
\caption{\label{fig:LinearInvariant}
    Constant level sets of invariant for 1-piece map in the MH
    form, $\K(p,q)=p^2-k\,p\,q+q^2 = \const$ .
    For $|k|<2$ the level sets are ellipses, for $|k|>2$ the level
    sets are hyperbolas and for $|k|=2$ they are parallel lines.
    The dashed and solid green lines are the first and second
    symmetry lines.
    }
\end{figure}

%----------------------------------------------------------------
In order to understand the nature of invariant polygons from
crystallographic restriction theorem, we first will consider
a more general 1-piece map in the MH form
\begin{equation}
\label{math:1pL}
\begin{array}{l}
    q' = p,        \\[0.2cm]
    p' =-q+k\,p,
\end{array}
\end{equation}
where parameter $k\in\mathbb{R}$ is not restricted to integers.
This map is known to be integrable with quadratic invariant of
motion \cite{mcmillan1971problem}
\[
    \K(p,q) = p^2 - k\,p\,q + q^2.
\]
Mapping (\ref{math:1pL}) is unstable for $|k|>2$, that results
in level sets of the invariant being hyperbolas.
On the other hand, the mapping is stable for $|k|<2$ with level
sets corresponding to a family of concentric ellipses, see
Fig.~\ref{fig:LinearInvariant}.
Within the parameter range corresponding to stable maps, the
rotation number $\nu$ is independent of the amplitude and given by
$\nu = \arccos(k/2)/(2\pi)$.

%----------------------------------------------------------------
%----------------------------------------------------------------

%----------------------------------------------------------------
Thus, one can see that if $\arccos(k/2)$ is incommensurate with
$\pi$, the motion is quasi-periodic, and according to the Poincar\'e
recurrence theorem, iterations starting from the initial condition
$(q_0,p_0)$ will result in points densely covering the ellipse
$\K(p,q) = \K(p_0,q_0)$.
However, if $\arccos(k/2)$ is commensurate with $\pi$, motion is
strictly periodic and $\nu$ becomes a rational number.
As a result, point $(q_0,p_0)$ will return to its initial position
after a finite number of iterations and will not depict an ellipse.
In fact, linear mappings with rational $\nu$ are more than
integrable, they posses {\it superintegrability}, meaning that they
have more than one functionally independent invariant of motion
$\K(p,q)$.

\begin{figure}[h]\centering
\includegraphics[width=0.58\columnwidth]{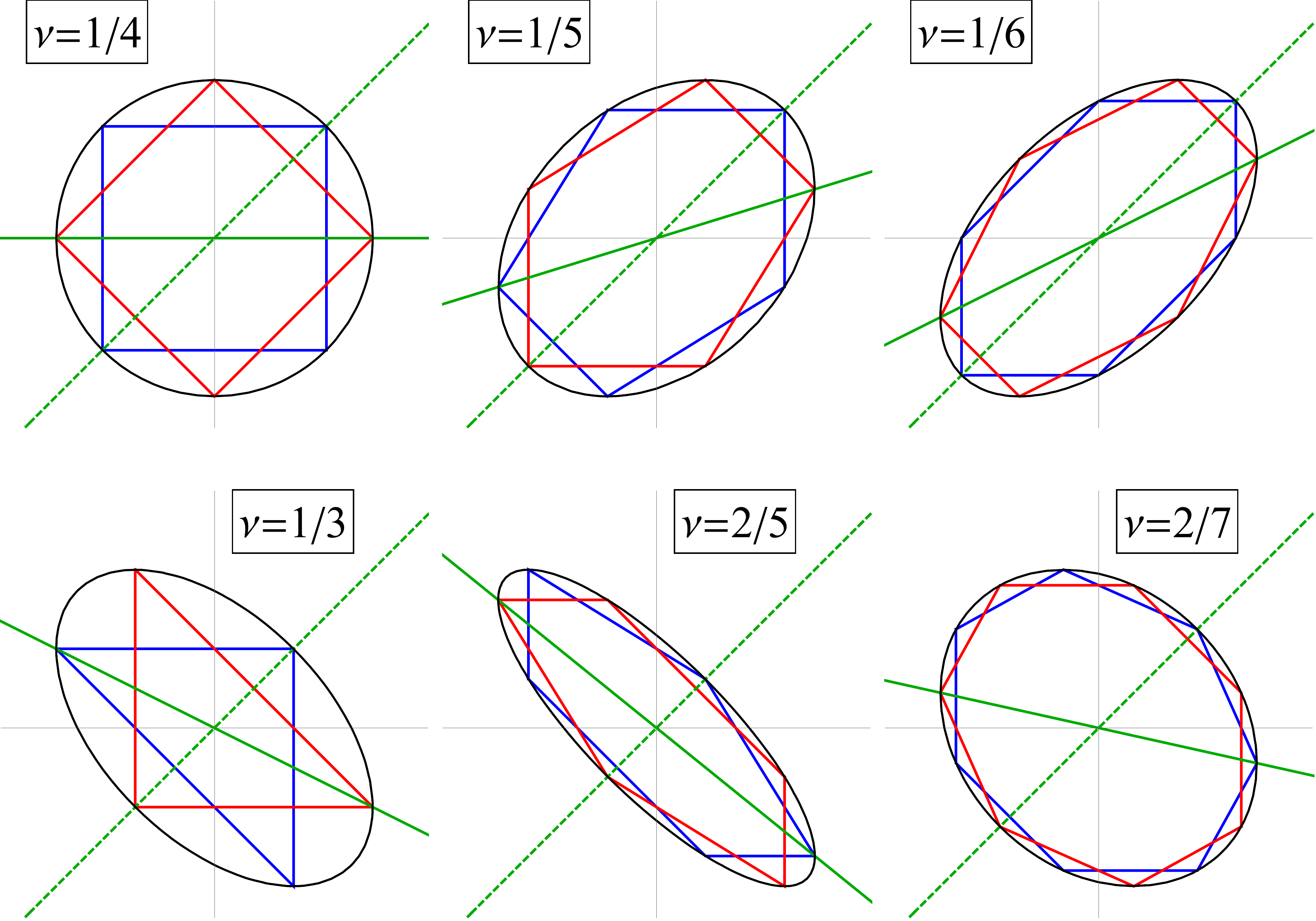}
\caption{\label{fig:ElementaryPolygons}
    Fundamental polygons of the first kind for different values
    of rotation number $\nu$.
    Each plot shows the invariant ellipse
    $\K(p,q)=p^2-k\,p\,q+q^2 = \const$ (black) and two invariant
    fundamental polygons inscribed into it (blue and red).
    The dashed and solid green lines correspond to the first and
    second symmetry lines.
    }
\end{figure}

%----------------------------------------------------------------
For example, for each map with rational $\nu\in \mathbb{Q}$, one
can construct two sets of polygon invariants by iterating an initial
point starting on the first, $p=q$, or second, $p=k\,q/2$, symmetry
lines and using iterations as vertices
(Fig.~\ref{fig:ElementaryPolygons}).
We will refer to such polygons as {\it first and second sets of
fundamental polygons of the first kind};
first or second sets refer to the position of the initial point on
first or second symmetry lines respectively, and ``the first kind''
refer to the fact that the force function consists of one piece only.
In the case of even period, two sets of polygons are actually
different while for odd period sets, polygons are identical up to
rotation by an angle $\pi$, $(q,p)\to(-q,-p)$.
In the former case, the first set of polygons has no vertices on the
second symmetry line (and two vertices on the first symmetry line)
while the second set has no vertices on the first symmetry line (and
two vertices on the second).
In the latter case, both sets of polygons have a single vertex on
each symmetry line.

%----------------------------------------------------------------
The only integer values of $k$ corresponding to stable mappings
($|k|<2$) are $k\in \{-1, 0, 1\}$.
For each integer value of parameter $k_1$, the value of $\arccos(k/2)$
(respectively $2\pi/3$, $\pi/2$ and $\pi/3$) is commensurate with
$\pi$, thus proving that maps $\alpha$, $\beta$ and $\gamma$ are the
only stable 1-piece maps in the MH form with integer coefficients.

%----------------------------------------------------------------
In addition, for $|k|=2$ invariant level sets
are straight lines.
We will denote these mappings as ``a'' for $k=2$ and ``b'' for
$k=-2$. 
While being unstable, both maps have polygonal chains (connected
series of line segments, in this case just line itself) as its
invariant.
In the following sections, we will see that unstable and stable
polygon maps can be combined together to form more complicated
systems.

%----------------------------------------------------------------
%----------------------------------------------------------------

\newpage
%===============================================================%
%===============================================================%
%===============================================================%
\section{\label{sec:2Pmaps}2-piece maps}

%===============================================================%
%===============================================================%
\subsection{Linear mappings}

%===============================================================%
\subsubsection{Polygon maps with integer coefficients}
%===============================================================%

%----------------------------------------------------------------
Another remarkably interesting result is given by CNR Theorem
(Cairns, Nikolayevsky and
Rossiter~\cite{cairns2014piecewise,cairns2016conewise}).
Suppose that $\mathcal{M}$ is a periodic continuous mapping of the
plane that is a linear transformation with integer coefficients in each half plane
$q \geq 0$ and $q < 0$.
Then $\mathcal M$ has a period
\[
n = 1,2,3,4,5,6,7,8,9\text{ or }12.
\]

%----------------------------------------------------------------
\begin{figure}[h!]\centering
\includegraphics[width=0.64\columnwidth]{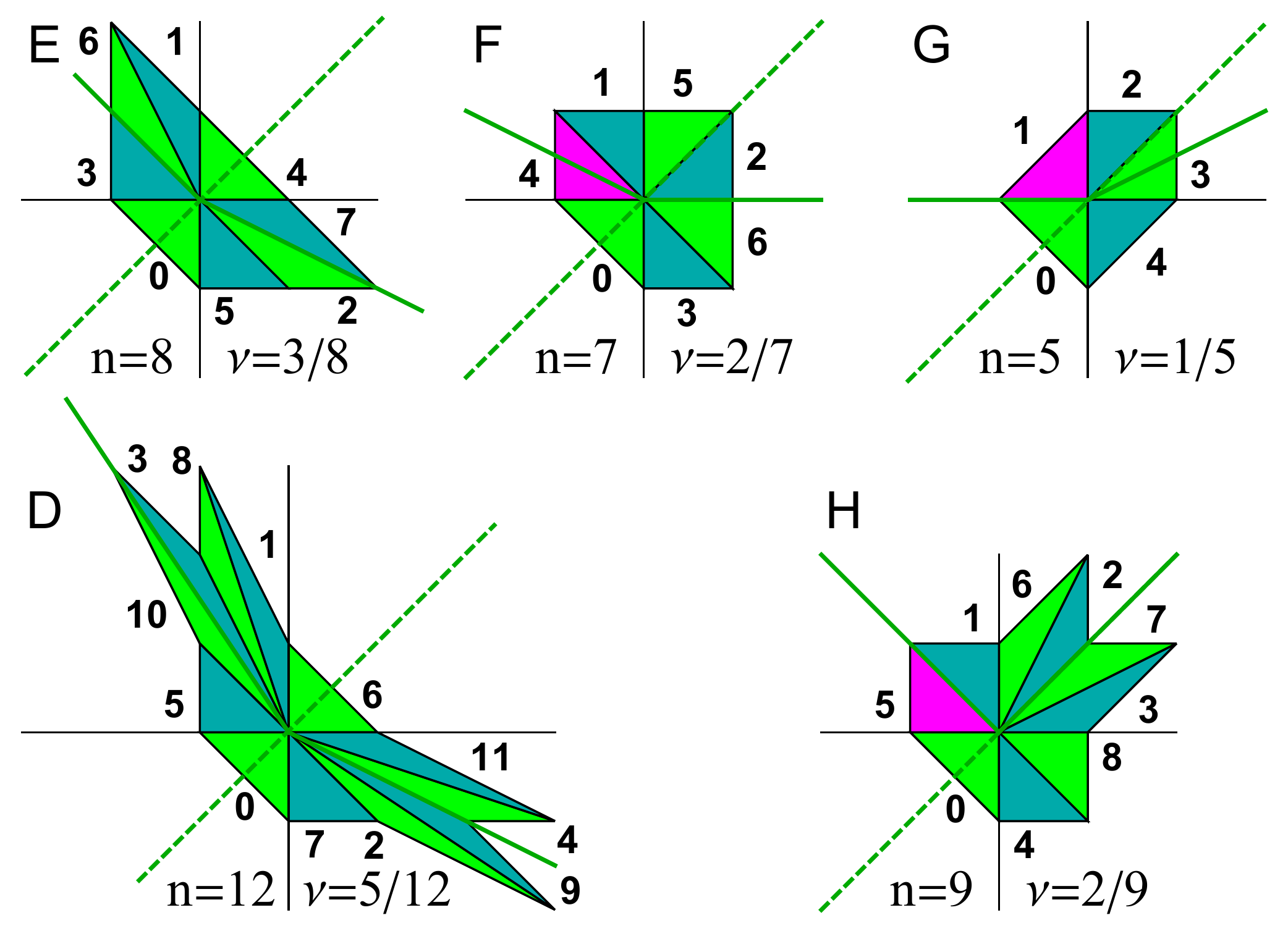}
\caption{\label{fig:PMaps2}
	Integer mappings with 2-piece force function and polygon
	invariants D -- H (named after~\cite{cairns2014piecewise}).
	Each figure shows one invariant polygon, with each colored
	region $i$ being mapped to the region $i+1$. Here
	$n$ and $\nu$ are the period and rotation number of the map.
	The solid and dashed dark green lines show the second and
        first symmetry lines.
    }
\end{figure}
%----------------------------------------------------------------

%----------------------------------------------------------------
All maps discovered by Cairns and others are in the MH form
with force function consisting of two linear segments
\begin{equation}
\label{math:2pForce}
    f(q) = %\frac{k_1+k_2}{2}\,q + \frac{k_2-k_1}{2}\,|q| =
    q \times
    \begin{cases}
        k_1, & q <   0,  \\
        k_2, & q\geq 0.
    \end{cases}
\end{equation}
In addition to previously described 1-piece maps (trivial cases
$k_1=k_2$) with $n=1,2,3,4,6$, now we have 5 more mappings
with $n=5,7,8,9$, Fig.~\ref{fig:PMaps2}.
These maps are stable, linear and integrable, with polygons being their 
invariant sets, and denoted with capital Latin letters D -- H.
Note how Cairns, Nikolayevsky and
Rossiter rediscovered Knuth map, H, from the first principles.

%===============================================================%
\subsubsection{\label{sec:ElemPII}Fundamental polygons of the second kind}
%===============================================================%

%----------------------------------------------------------------
Similar to the case of 1-piece mappings, in order to understand
the origin of polygon invariants we should consider a less restricted
map, corresponding to force function (\ref{math:2pForce}):
\begin{equation}
\label{math:2pL}
\begin{array}{l}
    \ds q' = p,        \\[0.2cm]
    \ds p' =-q+\frac{k_1+k_2}{2}\,p + \frac{k_2-k_1}{2}\,|p|,
\end{array}
\end{equation}
where parameters $k_{1,2}\in\mathbb{R}$.
This is a very nontrivial dynamical system considered in great details
in~\cite{beardon1995periodic,lagarias2005-I,lagarias2005-II,lagarias2005-III}.
Here, we will provide some results contributing to our understanding
of polygon maps.

\newpage
\begin{itemize}
    %----------------------------------------------------------------
    \item Map~(\ref{math:2pL}) is integrable.
    %----------------------------------------------------------------
    \item Map~(\ref{math:2pL}) is linear, i.e., there is no dependence
    of the rotation number on the amplitude.
    Since we are free to choose units of $p$ and $q$, using scaling
    transformation ($\epsilon > 0$)
    \[
    \begin{array}{l}
        \ds q' = \epsilon\,q,        \\[0.2cm]
        \ds p' = \epsilon\,p,
    \end{array}
    \]
    one can reconstruct dynamics on any ray of initial conditions
    (starting at the origin) from a single trajectory.
    %----------------------------------------------------------------
    \item The invariant of motion is a combination of segments of
    ellipses, straight lines or hyperbolas ``glued'' together and
    defined by
    $C_1 p^2 + C_2 p\,q + C_3 q^2 = \const$.
    In the plane of parameters $(k_1,k_2)$ there are {\it lines
    of constant topology}.
    Along these lines, the number of segments remains constant.
    %----------------------------------------------------------------
    \item Dynamics of map~(\ref{math:2pL}) can be either stable or
    unstable depending on the values of $k_{1,2}$, that creates a
    nontrivial fractal area on the stability diagram (see
    Ref.~\cite{beardon1995periodic} for more details).
    While we are not considering the fine structure of the fractal area,
    we would like to note that all pairs $(k_1,k_2)$ resulting in stable
    mappings are within the area defined by (Fig.~\ref{fig:2Pstab})
    \[
        p,q < 2 \qquad\land\qquad p\,q<4 \quad \mathrm{for} \quad p,q<0.
    \]
    Stable maps are either periodic when the rotation number $\nu$ is
    rational or quasi-periodic for irrational $\nu$.
  
%----------------------------------------------------------------
\begin{figure}[h!]\centering
\includegraphics[width=0.6\columnwidth]{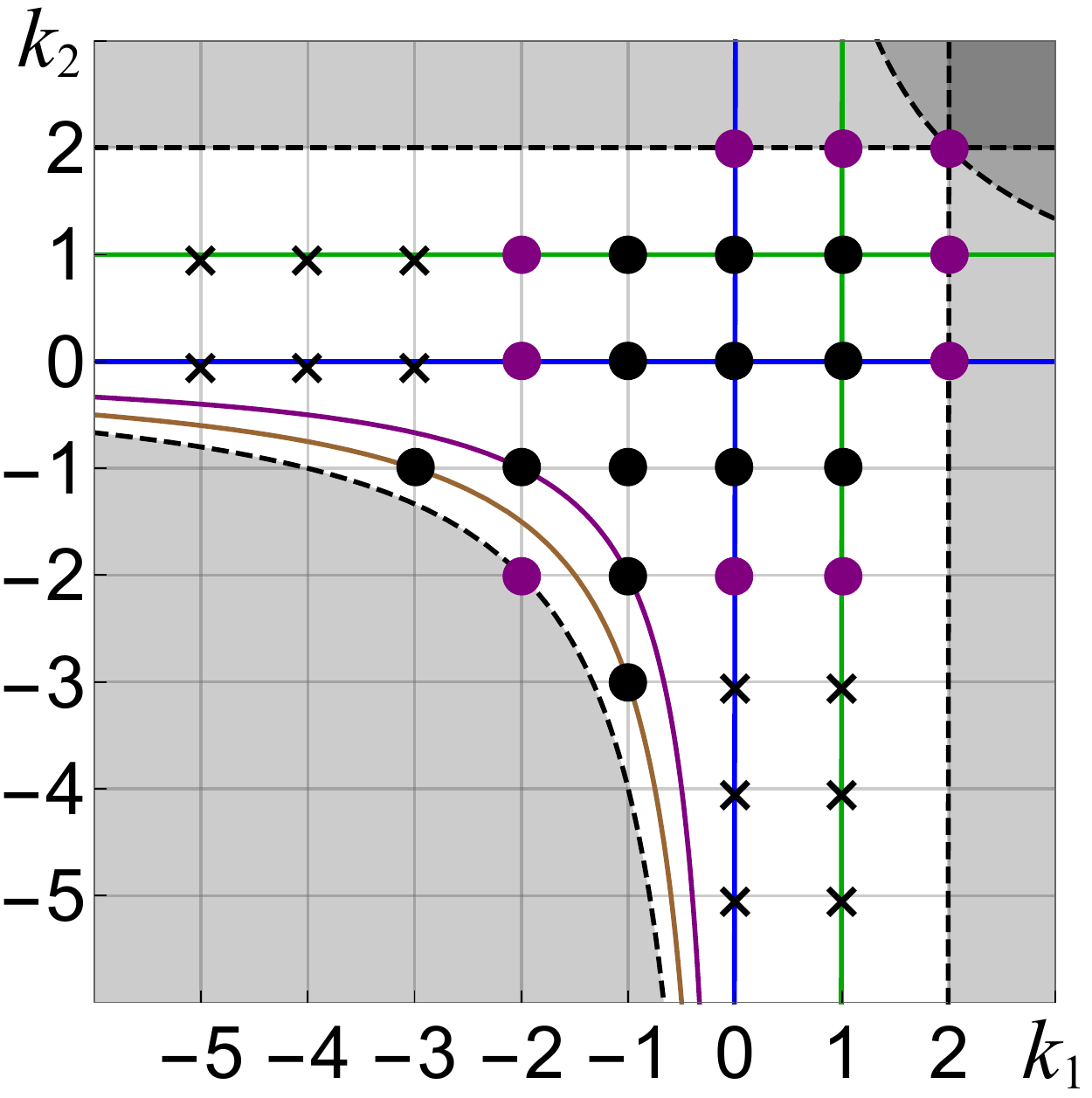}
\caption{\label{fig:2Pstab}
    The plane of parameters $(k_1,k_2)$ for 2-piece maps.
    The dashed lines show the area of global stability,
    $p,q<2\,\land\,p\,q<4$:
    the map can be stable (but not necessarily) only within the
    area of stability, shown in white.
    The solid curves show all lines of ``constant topology'' passing
    through integer nodes $(k_1,k_2)\in\mathbb{Z}^2$ within the stable
    zone: the red line corresponds to a single ellipse $k_1=k_2$, blue
    lines corresponds to two ellipses (McMillan beheaded and two-headed
    ellipse) $k_{1,2}=0$, green lines refer to three ellipses defined
    by $k_{1,2}=1$ and purple and brown hyperbolas are $k_1 k_2 = 2$
    and $k_1 k_2 = 3$ respectively.
    Integer nodes corresponding to stable mappings with polygon
    invariants are shown in black, unstable mappings with an invariant
    being a polygonal chain in purple, and the remaining integer nodes
    within the global region of stability are shown with crosses
    (unstable maps with invariant consisting of hyperbolas).
    }
\end{figure}
%----------------------------------------------------------------

    \newpage
    \item For example, the simplest case corresponds to the family
    of 1-piece maps, $k_1 = k_2$, (red line in Fig.~\ref{fig:2Pstab}).
    As we learned above, in this case the phase space consists of
    only one ellipse (or two sets of hyperbolas conjugate to each
    other) defined by
    $
    \K(p,q) = p^2 - k_1 p\,q + q^2
    $.
    As we move along the line $k_1 = k_2$, ellipses bifurcate into
    hyperbolas at $k_1=|2|$ and trajectories lose stability.
    
    Next line of constant topology is given by $k_2=0$ and line 
    $k_1=0$ corresponding to family of twin maps, see blue solid lines
    in Fig.~\ref{fig:2Pstab}.
    This is the aforementioned beheaded and two-headed ellipses
    discovered by McMillan with invariant defined by two ellipses
    glued together:
    \[
    \K(p,q) = \left\{
        \begin{array}{l}
            p^2 + k_1 p\,q + q^2,\qquad p,q \geq 0,          \\[0.2cm]
            p^2 - k_1 p\,q + q^2,\qquad \mathrm{otherwise}.
        \end{array}\right.
    \]
    Here is his explanation on how this map works:
    {\it In an earlier paragraph a promise was made that cases
    could be constructed with boundaries that more than double
    valued.
    I shall now fulfill the promise by describing a procedure
    by which this can be done.
    Take any known case with a center of symmetry at the origin,
    and erase the part of the boundary lying in the $+$, $+$
    quadrant.
    Replace $f(q)$ by $f(q) = 0$ to the right of the origin.
    Fill in the erased part of the boundary by reflecting the
    part in the $+$, $-$ quadrant about the $q$-axis.
    The resulting completed boundary is an invariant under the
    transformation with the original $f(q)$ to the left of the
    boundary, and $f(q)=0$ to the right.
    Starting with ellipses, we can get beheaded ellipses {\rm (first
    row of Fig.~\ref{fig:LinearInvariant2}, case $0<k_1<2$)}
    or double headed ellipses {\rm (first row of
    Fig.~\ref{fig:LinearInvariant2}, case $-2<k_1<0$)}, and here
    we see a four-valued function acting as an invariant boundary.
    }
    Note that in order to have only two ellipses being invariants
    of motion, the second ellipse have to be added to the $+$, $+$
    quadrant (or to $-$, $-$ quadrant for twin maps with $k_1=0$).
    Otherwise, an addition of an ellipse into the $+$, $-$ quadrant
    will result in its reflection in the quadrant $-$, $+$ (due
    to the first symmetry line $p=q$) making number of ellipses equal
    to three.

%----------------------------------------------------------------
\begin{figure}[b!]\centering
\includegraphics[width=\columnwidth]{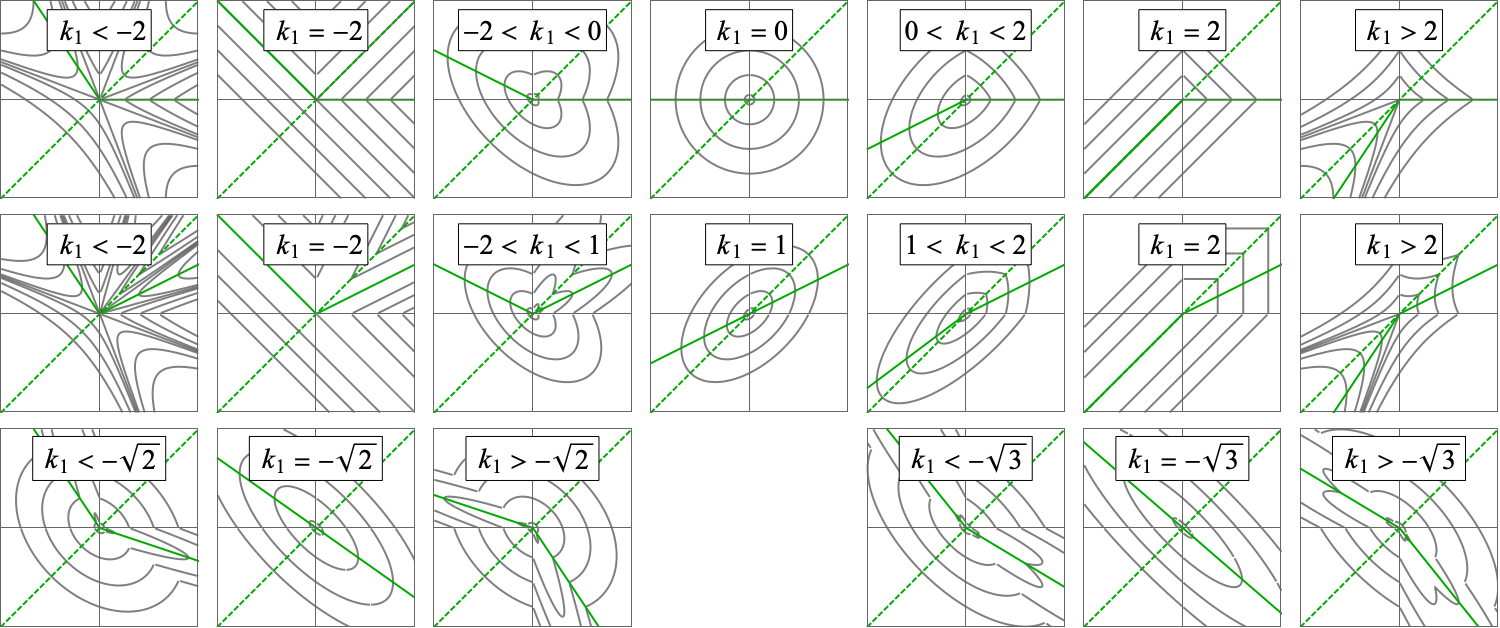}
\caption{\label{fig:LinearInvariant2}
    Invariant level sets along lines of constant topology for
    a 2-piece map in the MH form.
    The first and second rows illustrate the transformation of the
    invariant for cases $k_2=0$ and $k_2 = 1$
    (blue and green lines in Fig. \ref{fig:2Pstab}).
    The three left plots in the third row show bifurcation along
    hyperbola $k_1 k_2 = 2$ and the three right plots are for
    $k_1 k_2 = 3$
    (purple and brown hyperbolas in Fig. \ref{fig:2Pstab}).
    The solid and dashed green lines are the second and first
    symmetry lines.
    }
\end{figure}
%----------------------------------------------------------------

    In the case when the second segment of the force function has
    zero slope,  $k_2=0$, the map in Eq.~(\ref{math:2pL}) is stable
    for $|k_1|<2$ (see first row of Fig.~\ref{fig:LinearInvariant2}).
    For the slopes $k_1=\pm 2$ the invariant level sets become
    polygonal chains with 1- and 3-fold symmetries respectively.
    When stable, the rotation number is given by
    \[
        \nu = \frac{\arccos(k_1/2)}{\pi+2\,\arccos(k_1/2)}.
    \]
    As in the case $k_1=k_2$, when $\arccos(k/2)$ is commensurate
    with $\pi$, we have a rational rotation number $\nu\in\mathbb{Q}$
    resulting in a periodic map.
    The only integer values of $k_1$ satisfying this condition are
    $k_1\in\{-1,0,1\}$ (mappings F, $\beta$ and G respectively).
    
    In fact, this line of constant topology is just one of many
    defined by $k_2 = 2\,\cos(\pi/n)$ for $n \geq 2$.
    On each line the map is stable for $|k_1|<2$, it produces a
    polygonal chain as invariant for $k_1=\pm 2$ and when stable has
    a rotation number
    \[
        \nu = \frac{\arccos(k_1/2)}{\pi+n\,\arccos(k_1/2)}.
    \]
    Along each line of constant topology the invariant consists of
    $n$ ellipses (hyperbolas) with $n-1$ of them being in the first
    quarter.
    
    Among these lines there is another line that we should focus on:
    case $n=3$ with $k_2=1$ (and twin $k_1 = 1$).
    As we can see, in addition to line $k_2 = 0$, this is the only
    line with integer $k_1$.
    The invariant of motion is
    \[
    \K(p,q) = \left\{
        \begin{array}{l}
            p^2 - (2-k_1)(p\,q - q^2),\qquad p,q \geq 0\,\land\,p>q,          \\[0.2cm]
            (2-k_1)(p^2 - p\,q) + q^2,\qquad p,q \geq 0\,\land\,p<q,          \\[0.2cm]
            p^2 - k_1 p\,q + q^2,\qquad\qquad\quad\, \mathrm{otherwise},
        \end{array}\right.
    \]
    second row of Fig.~\ref{fig:LinearInvariant2}.
    From this line we have two more mappings with polygonal chain
    ($k_1=\pm2$) and three periodic maps with integer coefficients
    ($k_1=-1,0,1$, i.e maps $\alpha$ F and H).
    %----------------------------------------------------------------
    \item As one can notice, lines $k_1=k_2$ and $k_{1,2}=0,1$ are
    covering most of integer nodes $(k_1,k_2)\in\mathbb{Z}^2$ within
    the area of global stability (see Fig.~\ref{fig:2Pstab}).
    The remaining nodes belongs to two lines of constant topology
    from the family given by $k_1 k_2 = 4 \cos^2(\pi/(2n))$ with
    $k_{1,2}<0$ and $n\geq2$.
    Those lines are for $n=2$ and 3.
    Along those lines map is always stable and  periodic with the
    rotation number
    $
%       \nu = \frac{2\,n-1}{4\,n}.
        \nu = (2\,n-1)/(4\,n).
    $
    The invariant of motion is glued of $(2\,n-1)$ ellipse:
    the central ellipse, $(n-1)$ ellipses in II quadrant,
    and another $(n-1)$ in IV quadrant (placed symmetrically with
    respect to $p=q$).
    For example, for $n=2$ the invariant is
    \[
    \K(p,q) = \left\{
        \begin{array}{l}
            p^2 - (\frac{4}{k_1}-k_1)\,p\,q +(\frac{4}{k_1^2}-k_1)\,q^2,\qquad p,q \geq 0\,\land\,p>q,          \\[0.2cm]
            (\frac{4}{k_1^2}-k_1)\,p^2 - (\frac{4}{k_1}-k_1)\,p\,q + q^2,\qquad p,q \geq 0\,\land\,p<q,          \\[0.2cm]
            p^2 - k_1 p\,q + q^2,\qquad\qquad\qquad\qquad\quad\,\, \mathrm{otherwise}.
        \end{array}\right.
    \]
    Third row of Fig.~\ref{fig:LinearInvariant2} illustrates this
    case.
    For $k_{1,2} = -1$ from lines $n=2,3$ we have two remaining 
    periodic maps with integer coefficients, E and D.
\end{itemize}

\begin{figure}[b!]\centering
\includegraphics[width=0.63\columnwidth]{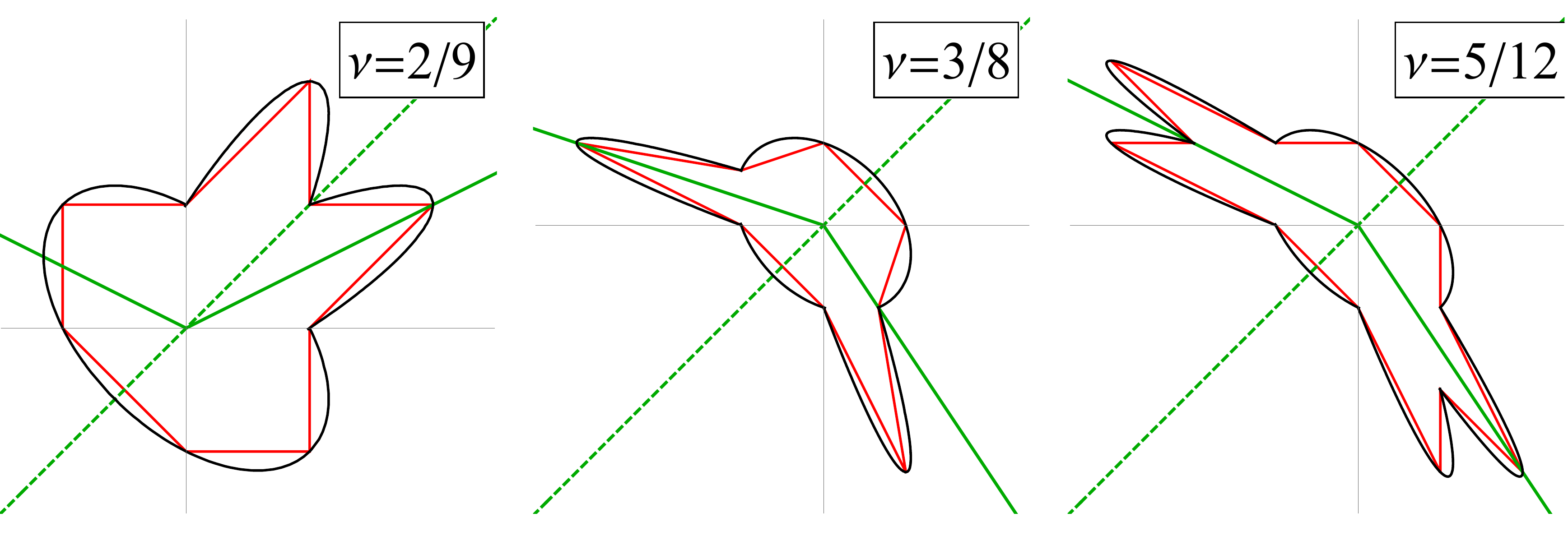}
\caption{\label{fig:ElementaryPolygons2}
    Shapes of fundamental polygons of the second kind corresponding
    to different families of maps represented in Fig.~\ref{fig:2Pstab}.
    Each plot shows invariant ellipses (black) and invariant polygons
    inscribed into it (red).
    From left to right:
    line $k_2=1$ for $k_1=-1$ (Brown-Knuth map, H),
    line $k_1 k_2=2$ for $k_1=-1$,
    line $k_1 k_2=3$ for $k_2=-3$ (map E).
    The dashed and solid  green lines correspond to the first and
    second symmetry lines.
    }
\end{figure}

%----------------------------------------------------------------
Thus, similarly to the case of the 1-piece map (degenerate line 
$k_1=k_2$), we can define {\it fundamental polygons of the second
kind}: for each line of constant topology, its stable maps with
a rational rotation number are degenerate and have more than one
invariant of motion including polygons.
In contrast to the case $k_1=k_2$, when we had two sets of polygons,
now we only have only a single set of fundamental polygons.
These polygons have two vertices on the second symmetry line for
maps with an even period or a single vertex on both symmetry lines
for odd periods.
Fig.~\ref{fig:ElementaryPolygons2} shows a few examples of this
degeneracy illustrating fundamental polygons of the second kind
inscribed in alternative invariant set of ellipses.
Note that the first example is for Brown-Knuth map showing an
alternative three-headed ellipse which is also an invariant of 
motion.

%----------------------------------------------------------------
%----------------------------------------------------------------

\newpage
%===============================================================%
\subsubsection{Ingredients of mappings with polygon invariant}
%===============================================================%

\begin{figure}[b!]\centering
\includegraphics[width=0.79\columnwidth]{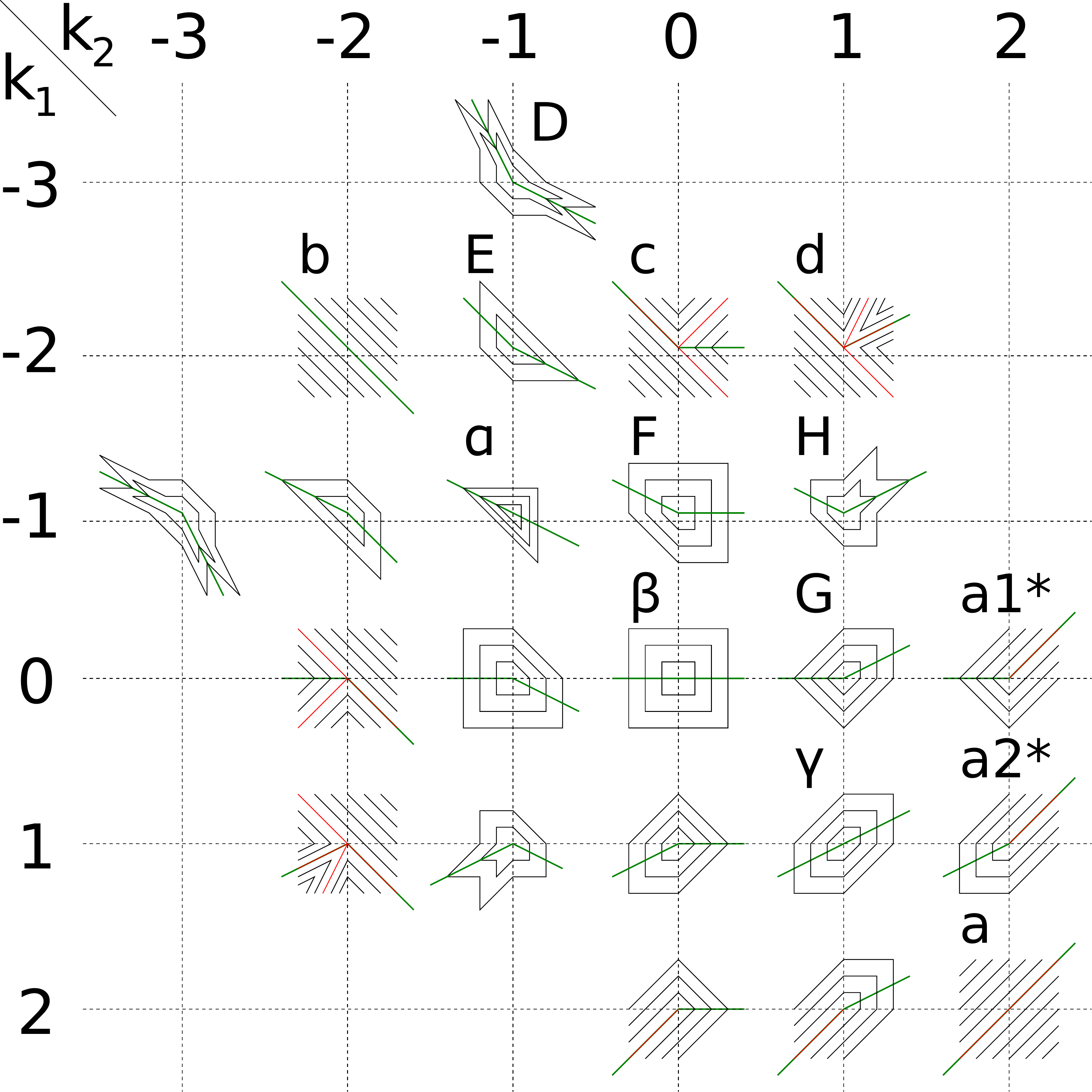}
\caption{\label{fig:PMaps2-L}
	Integer linear mappings with 2-piece force function and
        polygon invariants (stable and unstable maps).
	The black polygons (or polygonal chains for unstable mappings)
        show constant level sets of the invariant,
	red level sets are separatrix for unstable mappings with
        stable periodic dynamics, and the green line is the second
        symmetry line.
	}
\end{figure}

%----------------------------------------------------------------
Hereby, one can see that there is one condition and two mechanisms
determining  polygon shapes (including unstable mappings with
invariant level set being connected series of line segments ---
non-intersecting polygonal chains).
The condition is that the force function $f$ should be in a piecewise
linear form that follows from the second symmetry: the sum of
vertically opposed polygon sides is proportional to the value of the
force function.
Two mechanisms are:
\begin{itemize}
    \item {\bf Superintegrability in linear maps}. Rational rotation 
    number causes degeneracy when more than one invariant of motion
    exists, including polygons (e.g. fundamental polygons of the
    first and second kind).
    \item {\bf Loss of stability}. Ellipses are transitioned to
    hyperbolas via the straight lines in linear 2-piece maps, e.g.
    see cases $|k_1|=2$ for first two rows in
    Fig.~\ref{fig:LinearInvariant2}.
\end{itemize}
These mechanisms are respectively responsible for stable and
unstable polygons.
As we will see further, more complicated nonlinear mappings with
polygon invariants can be constructed, but at lower
($(q,p)\rightarrow 0$) or higher amplitudes
($(q,p)\rightarrow \infty$),
they reduce to one of the already described transformations;
we will observe that both mechanisms can be combined in one map
resulting in very nontrivial dynamics.
%----------------------------------------------------------------
Below, in Fig.~\ref{fig:PMaps2-L} we illustrate all integer linear
mappings with 2-piece force function and polygon invariants (stable
and unstable).

%----------------------------------------------------------------
%----------------------------------------------------------------

\newpage
%===============================================================%
%===============================================================%
\subsection{Nonlinear mappings}

%----------------------------------------------------------------
Here we begin our exploration of \textit{nonlinear} symplectic
mappings.
So far we considered only mappings with force function in the form
(\ref{math:2pForce}).
The simplest modification we can make is to add a constant {\it
shift parameter}, $d$:
\[
f(q) = \frac{k_1+k_2}{2}\,q + \frac{k_2-k_1}{2}\,|q| + d.
\]
The map becomes nonlinear and we can not reconstruct dynamics
along the ray of initial conditions using scaling transformation
$(q,p)\to\epsilon(q,p)$, $\epsilon > 0$.
Instead, now we have a choice of natural units $\epsilon = d$,
and one can show that without loss of generality for 2-piece maps,
$d$ can be restricted to $\pm 1$ or 0.

%----------------------------------------------------------------
Introduction of $d$ not only adds the dependence on amplitude, but
might result in a loss of integrability causing chaotic behavior.
A famous example is the Gingerbreadman~\cite{devaney1984piecewise} map
--- a chaotic two-dimensional map in the MH form with $f(q)=|q|+1$
(see first plot in Fig.~\ref{fig:ZooMaps}).
The remarkable property of this map is that some invariant tori
survived perturbation and after deformation remained polygons.
The chaotic dynamics is sectioned by concentric polygons making this
map the simplest model for area-preserving twist mappings with
zones of instability.
It shares many properties of the smooth mappings and has an advantage
of having an exact arithmetic on rational numbers.
Another mapping we would like to mention here is the {\it Rabbit}
map for $f(q) = |q| - 1$.
To the best of our knowledge this map has not been mentioned before, but we are certain
that Robert~Devaney, the discoverer of Gingerbreadman, knew about
it due to the close connection between the maps (second plot in
Fig.~\ref{fig:ZooMaps}).

%===============================================================%
\subsubsection{Zoo mappings}
%===============================================================%

\begin{figure}[b!]\centering
\includegraphics[width=\columnwidth]{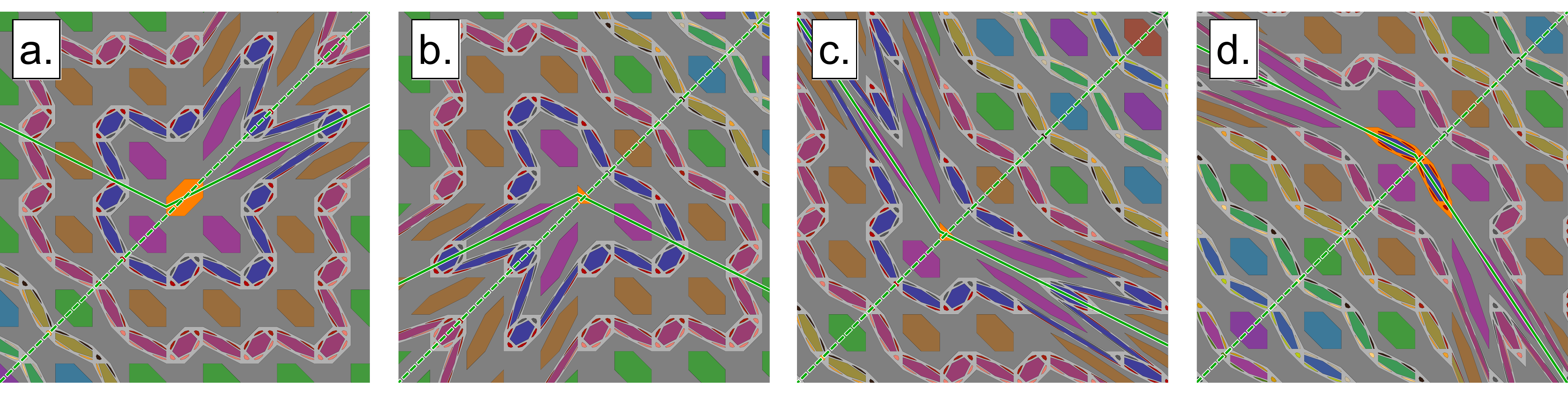}
\caption{\label{fig:ZooMaps}
	Zoo maps: Gingerbreadman (a.), Rabbit (b.), Octopus (c.)
        and Crab (d.) mappings.
	The light and dark gray areas represent alternating zones
        of instability which are sectioned by two families of
        concentric invariant polygons resembling animals.
	Within each zone of instability there are three different
	scenarios.
	First, orbits that result from initial conditions $(q_0,p_0)$
	within any gray zone such that both $q_0,p_0\in\mathbb{Q}$.
	Those are stable orbits with an exact periodic motion and
        rational rotation number resulting in an invariant set of
        points.
	The second scenario includes all other initial conditions
        within the gray zones with any or both $q_0,p_0$ being
        irrational, $q,p_0\in\mathbb{R}\setminus\mathbb{Q}$.
	Those are chaotic orbits never returning to initial
        conditions and densely covering the zone of instability.
	The third and last scenario is for any initial condition
	within the colored regions of phase space.
	Those are {\it chains of linear islands};
	the initial condition from these regions of phase space
        hops from island to island of the same group (each group is
        shown with its own color related to the relative orbit
        period).
	Such initial condition returns to itself and the motion is
        strictly periodic with the rotation number being
        {\it mode-locked} within the island.
	To view an ``animal'' you must rotate each image 135
        degrees clockwise.}
\end{figure}

%----------------------------------------------------------------
After reading the article of Cairns, Nikolayevsky and Rossiter
\cite{cairns2014piecewise}, our first idea, which predetermined
discoveries made in this article, was to produce objects similar
to Devaney map using newly discovered mappings D -- G.
First we tried mapping D, and amazingly it worked!
We introduce two new dynamical systems, the {\it Octopus} and the
{\it Crab mappings} for $f(q) = |q| - 2q \pm 1$ respectively
\cite{zolkinNOCE2017} (last two plots in Fig.~\ref{fig:ZooMaps}).
Informally, we grouped them together with Rabbit and Gingerbreadman,
and, called them {\it zoo maps}.
Mappings are again chaotic and ``sectioned'' by infinitely many
concentric invariant polygons.
Basic dynamics of these maps is described in the caption to
Figure~\ref{fig:ZooMaps} and dynamical properties of invariant
polygons are listed in Table~\ref{tab:Zoo} (Appendix~\ref{secAPP:Zoo}).

%----------------------------------------------------------------
%----------------------------------------------------------------

\newpage
%===============================================================%
\subsubsection{Nonlinear integrable maps with polygon invariants}
%===============================================================%

%----------------------------------------------------------------
\begin{table}[t!]
\centering
\begin{tabular}{p{3cm}p{3cm}p{3cm}p{4cm}p{4cm}}
%----------------------------------------------------------------
\hline\hline
Map       & $\nu_0$       & $\nu_1$             & $J_1$                 & $J_1'$                             \\\hline
          &               &                     &                       &                                    \\[-0.25cm]
b0        & $\frac{1}{2}$ & $\frac{1+3x}{2+8x}$ & $4\,\!x^2+2x$         & $\frac{3}{2}x^2+x$                 \\[0.25cm]
$\alpha$1 & $\frac{1}{3}$ & $\frac{1+3x}{3+8x}$ & $4\,\!x^2+3x+\frac{1}{2}$ & $\frac{3}{2}x^2+x+\frac{1}{6}$ \\[0.25cm]
$\alpha$2 & $\frac{1}{3}$ & $\frac{1+2x}{3+7x}$ & $\frac{7}{2}x^2+3x+\frac{1}{2}$ & $x^2+x+\frac{1}{6}$      \\[0.25cm]
$\beta$1  & $\frac{1}{4}$ & $\frac{1+2x}{4+7x}$ & $\frac{7}{2}x^2+4x+1$ & $x^2+x+\frac{1}{4}$                \\[0.25cm]
$\beta$2  & $\frac{1}{4}$ & $\frac{1+ x}{4+5x}$ & $\frac{5}{2}x^2+4x+1$ & $\frac{1}{2}x^2+x+\frac{1}{4}$     \\[0.25cm]
$\gamma$1 & $\frac{1}{6}$ & $\frac{1+ x}{6+5x}$ & $\frac{5}{2}x^2+6x+3$ & $\frac{1}{2}x^2+x+\frac{1}{2}$     \\[0.1cm]
\hline\hline
%-----------------------------------------------------------------
\end{tabular}
\caption{\label{tab:PMaps2-N}
Dynamical properties of integer nonlinear integrable mappings with
polygon invariants and 2-piece force function.
Here $\nu_0$ is the rotation number in the inner linear layer, 
$\nu_1$, $J_1$ and $J'_1$ are respectively rotation number, action
and partial action in the outer nonlinear layer,
$x$ is the linear amplitude in a nonlinear layer defined as horizontal or
vertical distance from separatrix to the trajectory under consideration.
}
\end{table}
%----------------------------------------------------------------

\begin{figure}[h!]\centering
\includegraphics[width=0.56\columnwidth]{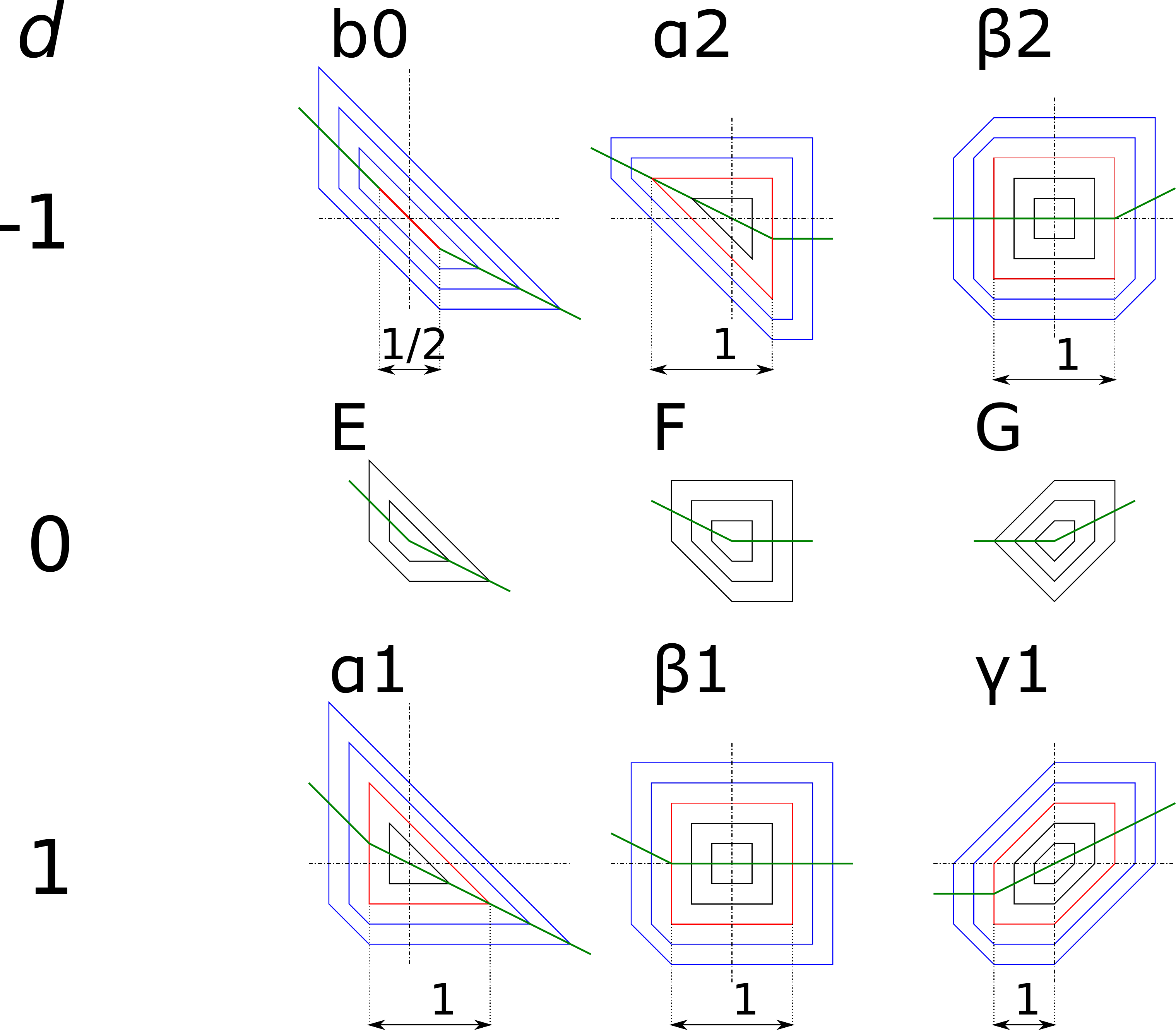}
\caption{\label{fig:PMaps2-N}
Integer nonlinear integrable maps with polygon invariants and
2-piece force function.
The top and bottom rows of the figures show level sets of the
invariant for $d = \mp 1$.
The middle row shows cases for $d = 0$, and, in each column
the transformations have the same slopes $k_{1,2}$ as for $d=0$
(see Fig.~\ref{fig:PMaps2-L});
note that the invariant level set of the outer nonlinear layer at
the large amplitude is asymptotic to the level set of the
corresponding map.
The black (blue) level sets are linear (nonlinear) trajectories.
The red level sets represent the separatrix isolating the inner
linear from the outer nonlinear layers.
The green line shows the second symmetry line, $p=f(q/2)$.
}
\end{figure}

%----------------------------------------------------------------
When $d=\pm 1$, not all mappings experience chaotic behavior like
Gingerbreadman or zoo maps.
Unexpectedly, transformations E, F and G produced nonlinear
integrable systems with polygon invariants if the vertical shift
parameter $d$ is added~\cite{zolkinNOCE2017}
(see Fig.~\ref{fig:PMaps2-N}).
Since $d \neq 0$, fixed point moves from the vertex to one of the
linear pieces (depending on a sign of $d$) of the force function.
As a result, the area of mode-locked motion around fixed point is
formed with invariant level sets being triangles, squares, hexagons
or line segment $p=-q$ with period 2 for map b0.
For larger amplitudes, the motion is still integrable, but rotation
number becomes amplitude dependent (trajectories shown in blue).
The inner linear and outer nonlinear layers are separated by
separatrix matching dynamics (shown in red).
Dynamical properties for all mappings are listed in
Table~\ref{tab:PMaps2-N} as functions of amplitude along $q$-axis
$x$, and, methods for calculation of rotation number as well as action-angle variables are provided in Appendix~\ref{secAPP:Act-Ang}.

%----------------------------------------------------------------
%----------------------------------------------------------------

%===============================================================%
%===============================================================%
%===============================================================%
\section{\label{sec:Search}Machine-assisted discovery of integrable
maps with polygon invariants}

\begin{figure}[t!]
\includegraphics[width=0.75\columnwidth]{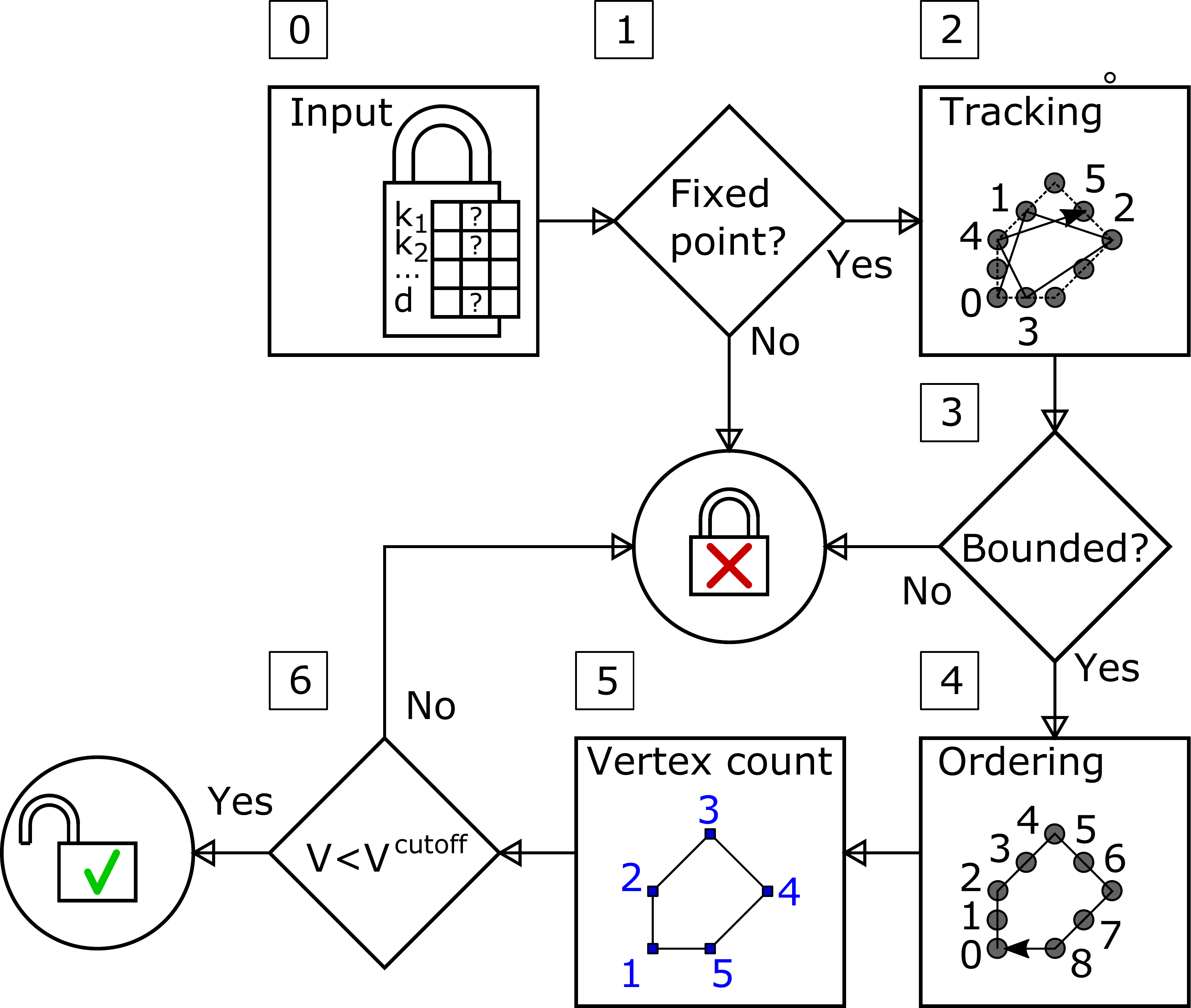}
%\vspace{-1cm}
\caption{\label{fig:drawing}
    Flowchart representing the machine-assisted discovery of
    integrable maps.
    The algorithm verifies if the map defined by the input (0.)
    is integrable with the invariants being concentric polygons.
    If the map has a fixed point (1.), the algorithm  performs
    tracking for different initial conditions (2.), and, if all
    orbits are bounded (3.) it proceeds to the construction of
    orbit polygons via ordering (4.).
    Finally, after counting the number of vertices $V$ (5.), if
    for each polygon $V < V^\mathrm{cutoff}$, the map is sent
    for analytic verification.
    The search process terminates when any conditional operation
    is false (NO).
}
\end{figure}

%----------------------------------------------------------------
After discovering 6 nonlinear polygon maps shown in Fig.~\ref{fig:PMaps2-N}
and performing few more numerical experiments, we realized that
the force function can be further generalized to have $n>2$ linear
segments, potentially producing new families of integrable systems.
In order to automate the search, we designed a relatively simple
algorithm that finds integrable mappings based on the analysis of
individual trajectories;
first, we verified some of its principles in~\cite{zolkinIPAC2019},
and then we significantly modified it and performed an extensive
search.
The detailed pseudo-code of the search algorithm is presented in
Appendix~\ref{secAPP:Alg} while its Python implementation is available
at
\href{https://github.com/yourball/polygon-maps}{github.com/yourball/polygon-maps}.
The algorithm is illustrated via flowchart in Fig.~\ref{fig:drawing}
and its blocks with processes/decisions are described below.

%----------------------------------------------------------------
\begin{enumerate}
%----------------------------------------------------------------
\item[0.]{\bf Input}
    We decided to focus on piecewise linear forces with integer
    coefficients,
    $f(q,\mathbf{k},\mathbf{l},d)$ with integer vectors
    $\mathbf{k},\mathbf{l}$ and $d$.
    While we performed a scan only for integer slopes $\mathbf{k}$,
    we were able to analytically generalize some of the results
    for more general values of the shift parameter $d\in\mathbb{R}$
    and lengths of segments $\mathbf{l}\in\mathbb{R}^+$.
%----------------------------------------------------------------
\item[1.]{\bf Fixed point}
    In order for a map to be {\it stable}, i.e. all phase space
    orbits are bounded, it should have at least one fixed point
    $\zeta^* = (q^*, p^*)$.
    Mappings that do not have a fixed point,
    $\forall\,q\in\mathbb{R}:\,q\neq f(q)/2$, are out of scope. 
%----------------------------------------------------------------
\item[2.]{\bf Tracking}
    If map is selected, we trace out its orbits
    $\vec{\zeta}=(\zeta_1,\ldots,\zeta_N)$ for different
    initial conditions $\zeta_0 = (q_0,\,p_0)$.
%----------------------------------------------------------------
\item[3.]{\bf Stability}
    Mappings with fixed point still can have unstable trajectories
    at larger amplitudes.
    If for some initial condition we detect an unbounded growth of
    the radial distance from the fixed point,
    $\exists\,i:\,\, |\zeta_i - \zeta^*| > r_\mathrm{max}$, the map
    is excluded from further analysis.
%----------------------------------------------------------------
\item[4.]{\bf Ordering}
    At the next stage the points of each individual trajectory
    are ordered according to their polar angle
    $\phi = \arctan{\left[(p-p^*)/(q-q^*)\right]}$, such that the
    angle $\phi$ is monotonically increasing.
    After arranging trajectory points according to $\phi$ we join
    consecutive points resulting in an {\it orbit polygon}.
%----------------------------------------------------------------
\item[5.]{\bf Vertex count}
    In order to count the number of vertices, $V$, for each orbit
    polygon we use scalar products of two vectors build on three
    consequent points 
    ($\mathbf{v}_1 = \zeta_i-\zeta_{i-1}$ and
    $\mathbf{v}_2 = \zeta_{i+1}-\zeta_i$),
    \[
        V = \sum_i \theta(\alpha_i-1),
        \qquad \qquad \alpha_i =
        \frac{\mathbf{v}_1\cdot\mathbf{v}_1}
             {|\mathbf{v}_1| |\mathbf{v}_2|},
    \]
    where $\theta(x) = \mathbf{1}_{x>0}$ is a Heaviside step function.
    In numeric experiments, when $\mathbf{v}_1\parallel\mathbf{v}_2$,
    the normalized scalar product might be slightly different from unity,
    so a small cutoff parameter should be introduced,
    $\epsilon < \left|( \mathbf{v_1} \cdot\mathbf{v_2})/
                      (|\mathbf{v_1}|    |\mathbf{v_2}|) - 1\right|$.
%----------------------------------------------------------------
\item[6.]{\bf Vertex cutoff}
    At the final stage we are left with stable integrable or chaotic
    maps which we would like to separate.
    The key insight is that we are looking for transformations with
    all invariant tori being polygons and assume that the phase space
    consists of concentric layers;
    in each layer the orbit polygons have the same finite number of
    vertexes which remains bounded when increasing the number of
    iterations $N\to\infty$.
    Conversely, for a chaotic trajectory the number of vertices
    linearly grows with the number of iterations,
    $V=\mathcal{O}(N)$.
    Such drastic difference in the growth of $V$ with the iteration
    number allows us to automatically separate between the two cases.
    In numeric experiment we check that for all initial conditions
    the number of vertices is bounded by some cutoff parameter,
    $V < V^\mathrm{cutoff} \leq N$.
\end{enumerate}

%----------------------------------------------------------------
%----------------------------------------------------------------

%----------------------------------------------------------------
In addition, for each mapping found by the search algorithm, we
manually verify the McMillan integrability condition, thus
rigorously proving their integrability.
Although the search algorithm described here is quite powerful,
there are some caveats to bear in mind:
\begin{itemize}
    %----------------------------------------------------------------
    \item Maps with strongly non-convex layers of polygon invariants
    can be missed by the algorithm. 
    In cases when the orbit polygon is not convex with respect to the
    fixed point $(q^*,\,p^*)$, the points of the trajectory can not be
    ordered according to the polar angle. 
    While we do not have a particular example, we should keep it in
    mind.
    %----------------------------------------------------------------
    \item Inside integrable nonlinear layers fibrated with polygons
    there are two types of orbits: those with rational and those with
    irrational rotation numbers.
    The latter are quasi-periodic with entire polygon being densely
    covered.
    On the other hand, the former are strictly periodic with orbit
    visiting only finite number of points, thus potentially showing
    a smaller number of vertexes.
    In Appendix~\ref{secAPP:Act-Ang} we show that all rotation
    numbers within a layer are in the form
    $\nu(q_0) = (\alpha + \beta\,q_0)/(\gamma + \delta\,q_0)$,
    where $\alpha,\,\beta,\,\gamma$ and $\delta$ are integer parameters.
    In order to avoid periodic orbits which complicate analysis, we can
    add a small irrational perturbation $\eta$ to all initial conditions,
    $q_0 \rightarrow q_0 + \eta$.
    In fact, from this formula one can see that in numeric experiments
    all observed orbits are periodic but this is not a problem as long
    as period is sufficiently large compare to the number of polygon
    sides, so points of orbit visit all sides of polygon covering them
    densely enough.
    %----------------------------------------------------------------
    \item Finally, while we've been only looking for maps with invariant
    layers being concentric polygons, while phase space can involve
    polygon islands.
    In the case of completely mode-locked islands ({\it linear islands})
    with all orbits visiting each island only once, our algorithm will
    successfully handle this case.
    But in the case of nonlinear islands (i.e. a trajectory visits
    sequentially each island under iterations and traces out invariant
    tori around centers of islands) our algorithm indeed will fail;
    as in the case of non-convex polygons, points of orbit can not be
    arranged by polar angle since now they are a multiple valued function.
\end{itemize}

%----------------------------------------------------------------
We performed the search for 3- and 4-piece integer force functions
using our algorithm across a wide range of the parameter values. 
The integer valued slopes of the segments of the force function are
selected from the range $-10 \leq k_i \leq 10$ and the shift parameter
$-50\leq d \leq 50$.
In the case of four-piece maps, the additional degree of freedom in
the space of parameters corresponds to the ratio of the lengths of
segments, $r\in [1/10,1/9,\ldots,1,2,\ldots,10]$. 
However, the search space can be further reduced when relying on our
prior knowledge about properties of piecewise linear mappings with
two segments.
A generic $n$-piece mapping at small and large amplitudes reduces
to a two-piece mapping.
As illustrated in Fig.~\ref{fig:PMaps2-L} two-piece mappings with
integer slope coefficients and polygon invariants are confined to
the parameter range $-3\leq k_{1,2}\leq 2$.
Therefore the values of the slopes of the outer segments in mappings
with polygon invariants are also restricted to $-3\leq k_{1,n}\leq 2$.
Similarly, in the vicinity of the fixed point $(q^*,\, p^*)$ at small
amplitudes $|q-q^*|\to 0$ and $|p-p^*|\to 0$, the mapping  behaves as
a single-piece or two-piece map, depending on whether the fixed point
belongs to the segment of the force function or coincides with one of
the vertexes separating adjacent segments.
Therefore, in order to guarantee that the mapping has polygon tori in
the vicinity of the fixed point, the range for the slope $k_i$ for
the corresponding segment is also confined to the same range.
%----------------------------------------------------------------
The values of the shift parameter $d$ and the length ratio $r$ are
less restrictive, because they are only responsible for the position
of the fixed point.
While we performed a scan only for integer $d$ and $r$ we were able
to analytically generalize some of these results for
$d,r \in \mathbb{R}$.

\newpage
%===============================================================%
%===============================================================%
%===============================================================%
\section{\label{sec:3Pmaps}3-piece maps}

\begin{figure}[h]
\includegraphics[width=\linewidth]{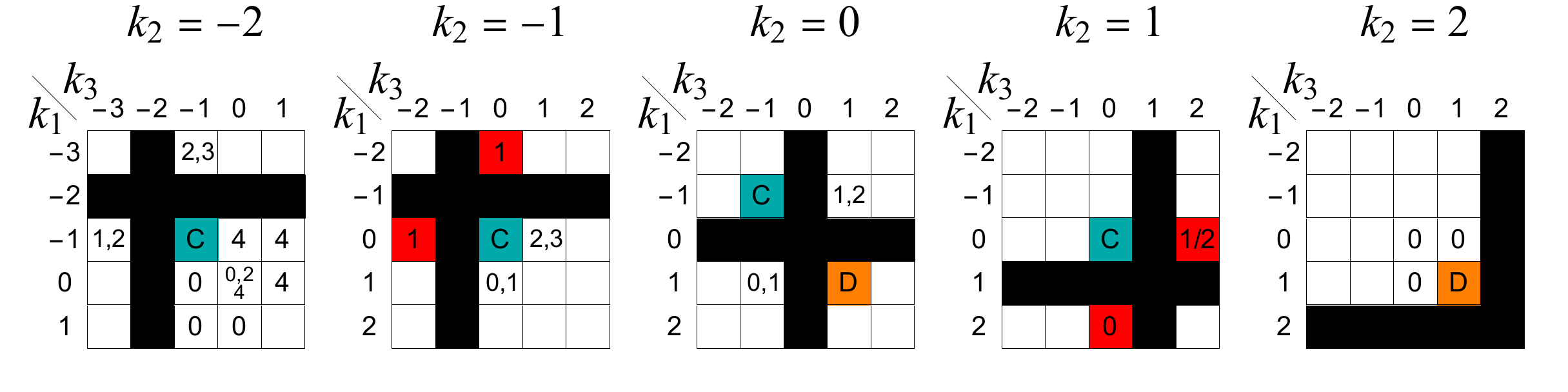}
%\vspace{-1cm}
\caption{\label{fig:3PTable}
    Integrable polygon mappings corresponding to a piecewise linear
    force function with three segments.
    Each table corresponds to its own value of $k_2$ and shows the
    value of the shift parameter $d$.
    Black cells represent degenerate cases when 3-piece force
    function becomes 2-piece ($k_1=k_2$ or $k_2=k_3$),
    white cells represent cases with isolated integer $d$ showing
    its values, orange cells represent mappings integrable for any
    integer $d\in\mathbb{Z}$ (``D''), cyan cells represent mappings
    integrable for continuous $d\in\mathbb{R}$ (``C'') and red cells
    represent unstable nonlinear polygon mappings where the numbers
    in the cell shows the values of $d$ corresponding to integrable
    maps.
}
\end{figure}

%----------------------------------------------------------------
In order to construct more complex piecewise linear symplectic
maps, one can add more segments to the force function $f$.
When the number of segments is equal to three, we have four
independent parameters: three slopes $k_{1,2,3}$ and a shift
parameter $d$.
By performing the rescaling of coordinates, one can see that without
loss of generality, the length of the middle segment can always be
assumed to be equal to one. 
In addition, unlike the case of 2-piece force function when $d=0,
\pm 1$, now the shift parameter belongs to reals.
Below, we will classify search results as isolated with respect to
$d$ (finite set of integer values), discrete (all integer values,
$d\in\mathbb{Z}$), and continuous (map is integrable for any
$d\in\mathbb{R}$).
Results of the machine-assisted search of new integrable maps are
briefly summarized in
Fig.~\ref{fig:3PTable} (the symmetry with respect to $k_1=k_3$ is
due to ``twin'' mappings).

%\vspace{-0.3cm}
%----------------------------------------------------------------
%----------------------------------------------------------------

\vspace{-0.55cm}
%===============================================================%
%===============================================================%
\subsection{Isolated {\it d}}

\begin{figure}[p!]\centering
\includegraphics[width=0.9\columnwidth]{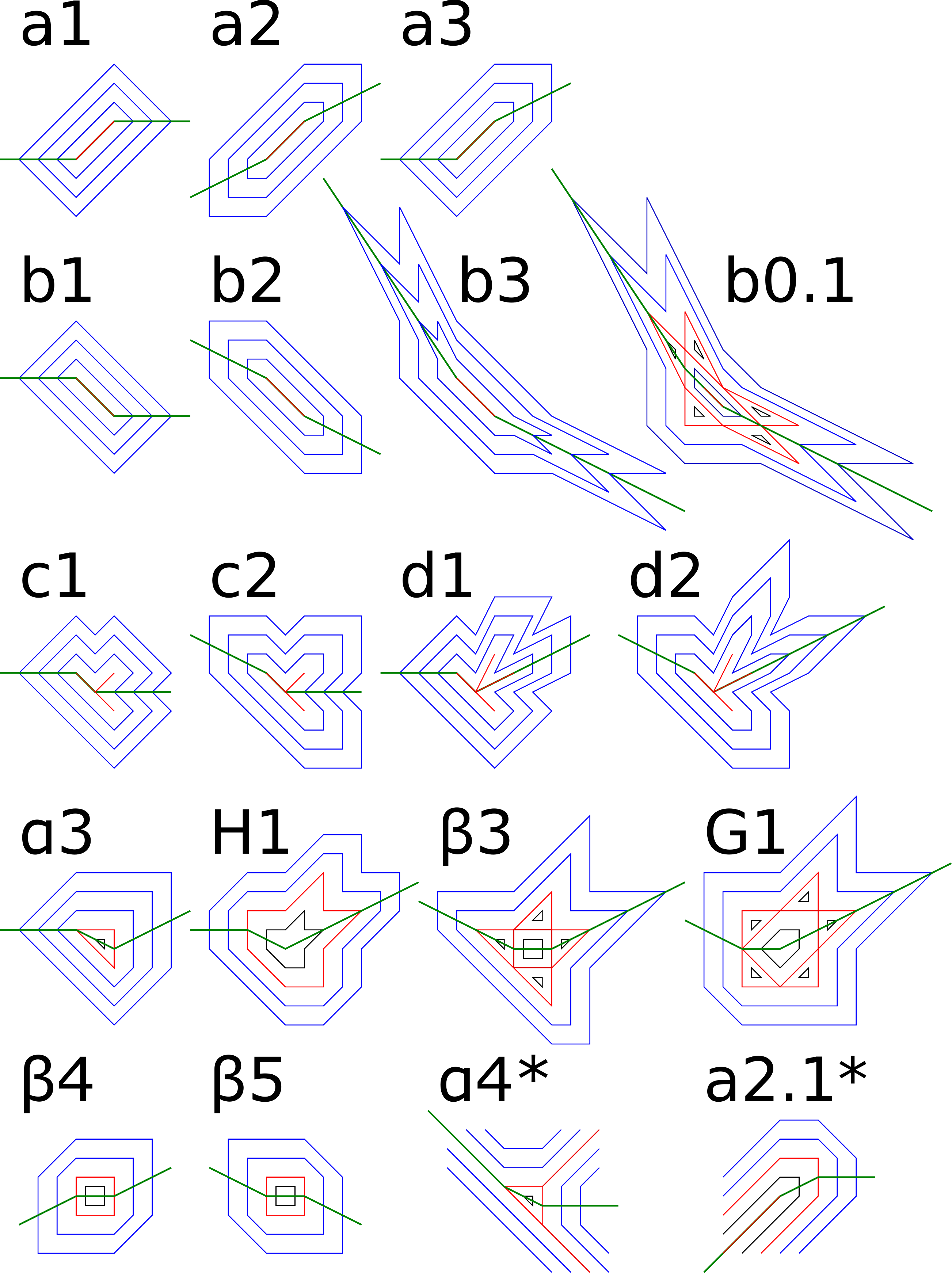}
\caption{\label{fig:PMaps3-I}
	Integrable polygon mappings with 3-piece force function
	and isolated $d$.
	Black and blue lines correspond to linear and nonlinear 
	invariant phase space trajectories, red lines are separatrices,
	and the thick green line is a second symmetry line, $p=f(q)/2$.
	Mappings b2 and $\beta$5 are also integrable for any $d\in\mathbb{R}$
	and
	mappings a2 and $\beta$4 are also integrable for any $d\in\mathbb{Z}$.
	Mappings $\alpha 4^*$ and a2.$1^*$ are unstable and were missed
	 by the search algorithm.
	}
\end{figure}

%\vspace{-0.3cm}
%----------------------------------------------------------------
Integer integrable polygon mappings with 3-piece force function
and isolated $d$ are shown in Fig.~\ref{fig:PMaps3-I} and rotation
numbers are provided in Table~\ref{tab:3p-I}.
As one can see, the addition of new pieces can stabilize previously
unstable 2 pieces maps a (a1,$2^*$), b, c and d (respectively
mappings a1--3, b1--3, c1--2 and d1--2).
Mappings $\alpha$3, $\beta$4--5 and H1 can be seen as previously
considered linear transformations ($\alpha$, $\beta$ and H) with
outer layer of nonlinear stable trajectories determined by the new
piece in the force function.

%----------------------------------------------------------------
Something interesting can be observed for mappings b0.1, $\beta$3
and G1.
For the map b0.1 we see that the inner and outer nonlinear layers
are separated by a chain of islands.
Unlike the ``usual'' chaotic chains observed in systems like
H\`enon~\cite{henon1969numerical} or
Chirikov~\cite{chirikov1969research,chirikov1979universal} mapping
(where under iterations phase-space point jumps from island to
island and trace out elliptic orbits around centers of islands),
these are {\it linear islands}.
They show true mode-locking similar to one observed in Arnold
tongue~\cite{arnol1961small} or even more similar to Gingerbread
man map~\cite{devaney1984piecewise};
when iterated, any initial condition within the chain returns
exactly to the same position after visiting each island only once.
Thus, the rotation number of the inner layer
\[
    \nu_0 = \frac{\frac{1}{2}+3x}{1+8x},\qquad
%    J_0 = 4x^2 + x,\qquad
%    J_0'= \frac{3}{2}x^2 + \frac{x}{2},\qquad
    x\in[0,\frac{1}{2}],
\]
and the one of the outer layer
\[
    \nu_1 = \frac{4+5y}{10+12y},\qquad
%    J_1 = 6y^2 + 10y + 4,\qquad
%    J_1'= \frac{5}{2}y^2 + 4y,\qquad
    y\in[0,\infty),
\]
where $x$ and $y$ represent the linear amplitudes, are matched on
the outer and inner separatrices, respectively $x=\frac{1}{2}$
and $y=0$.
They are equal to the rotation number inside the linear chain
consisting of 5 islands
\[
    \nu_0(1/2) = \nu_1(0) = \frac{2}{5}.
\]

%----------------------------------------------------------------
%----------------------------------------------------------------

\newpage
%----------------------------------------------------------------
%----------------------------------------------------------------
\begin{table}[th!]
\centering
\begin{tabular}{p{3cm}p{3cm}p{3cm}p{4cm}p{4cm}}
\hline\hline
Map      & $\nu_0$       & $\nu_1$               & $J_1$                 & $J_1'$                            \\\hline
         &               &                       &                       &                                   \\[-0.25cm]
a1       & $0$           & $\frac{x}{2+4x}$      & $2x^2+2x$             & $\frac{1}{2}x^2$                  \\[0.25cm]
a2       & $0$           & $\frac{x}{2+6x}$      & $3x^2+2x$             & $\frac{1}{2}x^2$                  \\[0.25cm]
a3       & $0$           & $\frac{x}{2+5x}$      & $\frac{5}{2}x^2+2x$   & $\frac{1}{2}x^2$                  \\[0.25cm]
b1       & $\frac{1}{2}$ & $\frac{1+x}{2+4x}$    & $2x^2+2x$             & $\frac{1}{2}x^2+x$                \\[0.25cm]
b2       & $\frac{1}{2}$ & $\frac{1+2x}{2+6x}$   & $3x^2+2x$             & $x^2+x$                           \\[0.25cm]
b3       & $\frac{1}{2}$ & $\frac{1+5x}{2+12x}$  & $6x^2+2x$             & $\frac{5}{2}x^2+x$                \\[0.25cm]
c1       & $\frac{1}{3}$ & $\frac{2+x}{6+4x}$    & $2x^2+6x$             & $\frac{1}{2}x^2+2x$               \\[0.25cm]
c2       & $\frac{1}{3}$ & $\frac{2+2x}{6+7x}$   & $\frac{7}{2}x^2+6x$   & $x^2+2x$                          \\[0.25cm]
d1       & $\frac{1}{4}$ & $\frac{2+x}{8+5x}$    & $\frac{5}{2}x^2+8x$   & $\frac{1}{2}x^2+2x$               \\[0.25cm]
d2       & $\frac{1}{4}$ & $\frac{2+2x}{8+9x}$   & $\frac{9}{2}x^2+8x$   & $x^2+2x$                          \\[0.25cm]
$\alpha$3& $\frac{1}{3}$ & $\frac{1+x}{3+5x}$    & $\frac{5}{2}x^2+3x+\frac{1}{2}$
                                                                                & $\frac{1}{2}x^2+x+\frac{1}{6}$                                                                   \\[0.25cm]
$\beta$3 & $\frac{1}{4}$ & $\frac{2+2x}{8+9x}$   & $\frac{9}{2}x^2+8x+3$ & $x^2+2x+\frac{3}{4}$              \\[0.25cm]
$\beta$4 & $\frac{1}{4}$ & $\frac{1+x}{4+6x}$    & $3x^2+4x+1$           & $\frac{1}{2}x^2+x+\frac{1}{4}$    \\[0.25cm]
$\beta$5 & $\frac{1}{4}$ & $\frac{1+2x}{4+6x}$   & $3x^2+4x+1$           & $x^2+x+\frac{1}{4}$               \\[0.25cm]
H1       & $\frac{2}{9}$ & $\frac{2+x}{9+5x}$    & $\frac{5}{2}x^2+9x+\frac{9}{2}$
                                                                                & $\frac{1}{2}x^2+2x+1$      \\[0.25cm]
G1       & $\frac{1}{5}$ & $\frac{2+2x}{10+9x}$  & $\frac{9}{2}x^2+10x+5$& $x^2+2x+1$                        \\[0.1cm]
\hline\hline
\end{tabular}
\vspace{-0.15cm}
\caption{Rotation numbers of the inner linear, $\nu_0$, and outer
        nonlinear layers, $\nu_1$, for integer integrable polygon
        mappings with 3-piece force function and isolated $d$.
        Last two columns show action, $J$, and partial action
        (Hamiltonian), $J'$, in nonlinear layer.
        Parameter $x\in[0,\infty)$ is horizontal (vertical) distance
        from the separatrix to considered trajectory. 
        }
\label{tab:3p-I}
\end{table}
%----------------------------------------------------------------
%----------------------------------------------------------------

%----------------------------------------------------------------
Note that polygons inside the chain of islands satisfy both
symmetries as a level set of the entire group and are invariant
under transformation;
for each island, there is another one which is its vertical
reflection with respect to the second symmetry line, and one
reflected with respect to the main diagonal (unless the island is
its own reflection with respect to the symmetry line).
Due to the fact that dynamics inside the island chain is degenerate
(all rotation numbers are the same), invariant contours can not be
simply traced out in the numerical experiment by looking at the
mapping iterations, but can be reconstructed from the symmetries in
order to match the separatrices.

%----------------------------------------------------------------
Mappings $\beta$3 and G1 don't have an inner nonlinear layer, and
the chain of islands is attached to a linear layer, so the entire
combined  phase space is mode-locked.
As expected, the number of islands is equal to the period inside 
linear zone (4 for $\beta$3 and 5 for G1) and rotation numbers are
$1/4$ and $1/5$ respectively.

%----------------------------------------------------------------
Finally, two more mappings, $\alpha 4^*$ and $a2.1^*$, are shown
at the end of the last row.
These cases were not found by an algorithm, since the outer layer
is unstable, and were constructed by authors using symmetries.

%----------------------------------------------------------------
%----------------------------------------------------------------
\begin{table}[bh!]
\centering
\begin{tabular}{p{3cm}p{3cm}p{4cm}p{3cm}p{4cm}}
\hline\hline
Map                 & $\nu_0$
    & $\nu_1,\,J_1,\,J_1'$                  & $\nu_\mathrm{isl}$, $\#_C$              & $\nu_2,\,J_2,\,J_1'$                          \\\hline
    &               &                       &                           &                                               \\[-0.25cm]
%a2                  & ---
%    & $0$                                   & ---                       & $\frac{z}{2+6z},\,z\in[0,\infty)$             \\[0.25cm]
a2-$k,\,(k\geq1)$   & $\frac{1}{6},\,x\in[0,k]$
    & $\frac{k}{6k+y},\,y\in[0,1]$          & $\frac{k}{6k+1},\,k$      &  $\frac{2k+z}{2(6k+1)+6z},\,z\in[0,\infty)$   \\[0.25cm]
                    &
    & $\frac{1}{2}y^2+6ky$                  &                           & $3z^2+2(6k+1)z + J_1(1)$                      \\[0.25cm]
                    &
    & $ky$                                  &                           & $\frac{1}{2}z^2+2kz + J_1'(1)$                \\[0.25cm]
$\beta$4-1          & ---
    & $\frac{1}{5},\,y\in[0,1]$             & $\frac{1}{5},\,1$         & $\frac{2+z}{10+6z},\,z\in[0,\infty)$          \\[0.25cm]
$\beta$4-$k,\,(k>1)$& $\frac{1}{6},\,x\in[0,k-1]$
    & $\frac{k-1+y}{6(k-1)+5y},\,y\in[0,1]$ & $\frac{k}{6k-1},\,k$      & $\frac{2k+z}{2(6k-1)+6z},\,z\in[0,\infty)$    \\[0.25cm]
                    &
    & $\frac{5}{2}y^2+6(k-1)y$              &                           & $3z^2+2(6k-1)z + J_1(1)$                      \\[0.25cm]
                    &
    & $\frac{1}{2}y^2+(k-1)y$               &                           & $\frac{1}{2}z^2+2kz + J_1'(1)$                \\[0.1cm]
\hline\hline
\end{tabular}
\vspace{-0.2cm}
\caption{Rotation number, $\nu_i$, action, $J_i$, and partial
        action, $J_i'$, for the $i$-th layer of the integer integrable
        polygon mappings with 3-piece force function and discrete
        $d$.
        $\nu_\mathrm{isl}$ and $\#_C$ are the rotation number in mode-locked area
        and number of chains in group of linear islands, respectively.
        }
\label{tab:3p-D}
\end{table}
%----------------------------------------------------------------
%----------------------------------------------------------------

%----------------------------------------------------------------
\begin{figure}[th!]\centering
\includegraphics[width=0.8\columnwidth]{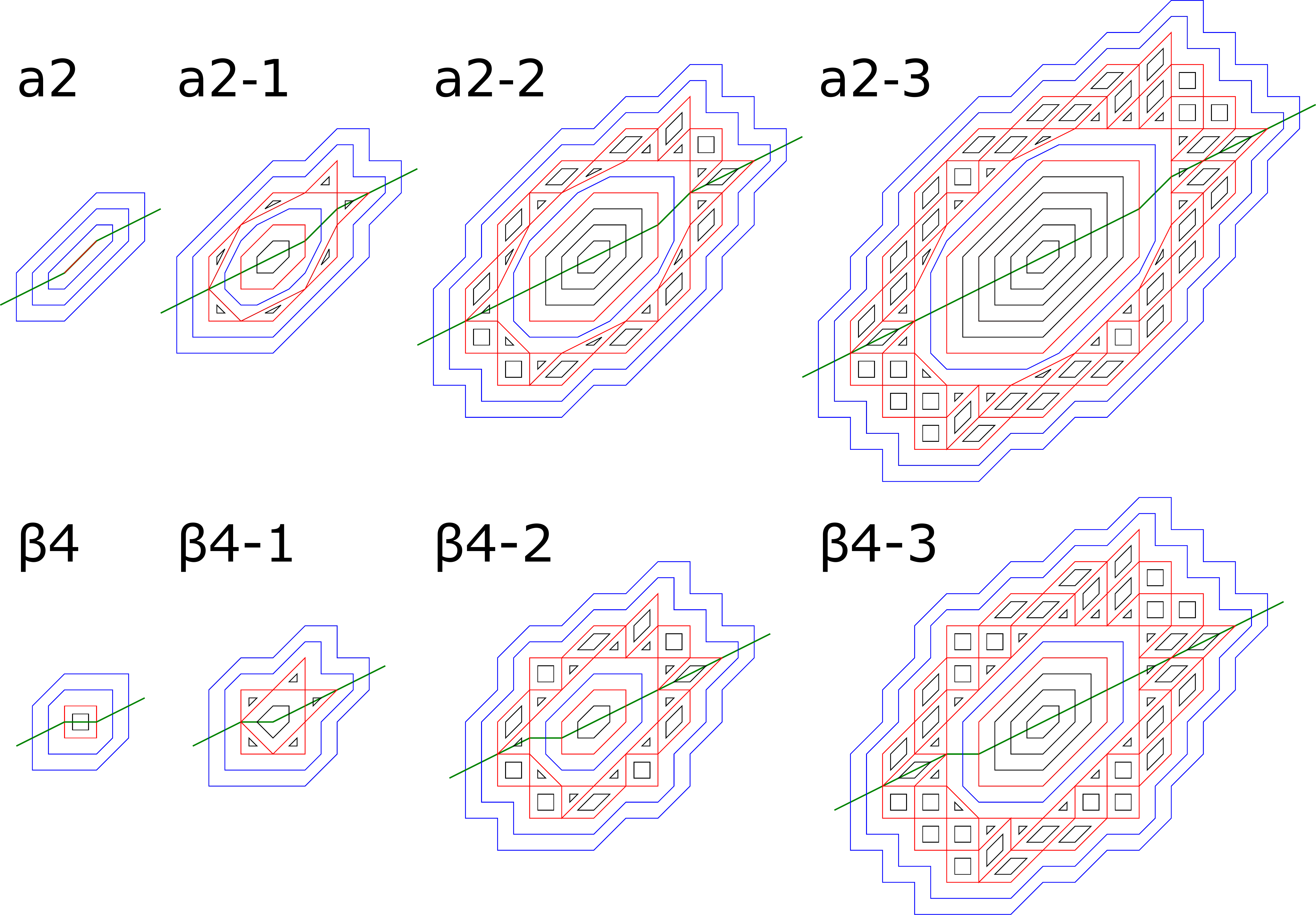}
\caption{\label{fig:PMaps3-D}
	Integer integrable polygon mappings with 3-piece force function
	and discrete $d\in\mathbb{Z}$.
	Each row shows first 4 members of a family.
	}
\end{figure}
%----------------------------------------------------------------

\vspace{-1cm}
%===============================================================%
%===============================================================%
\subsection{\label{sec:3P-D}Discrete {\it d}, (D)}

\begin{figure}[bh!]\centering
\includegraphics[width=0.85\columnwidth]{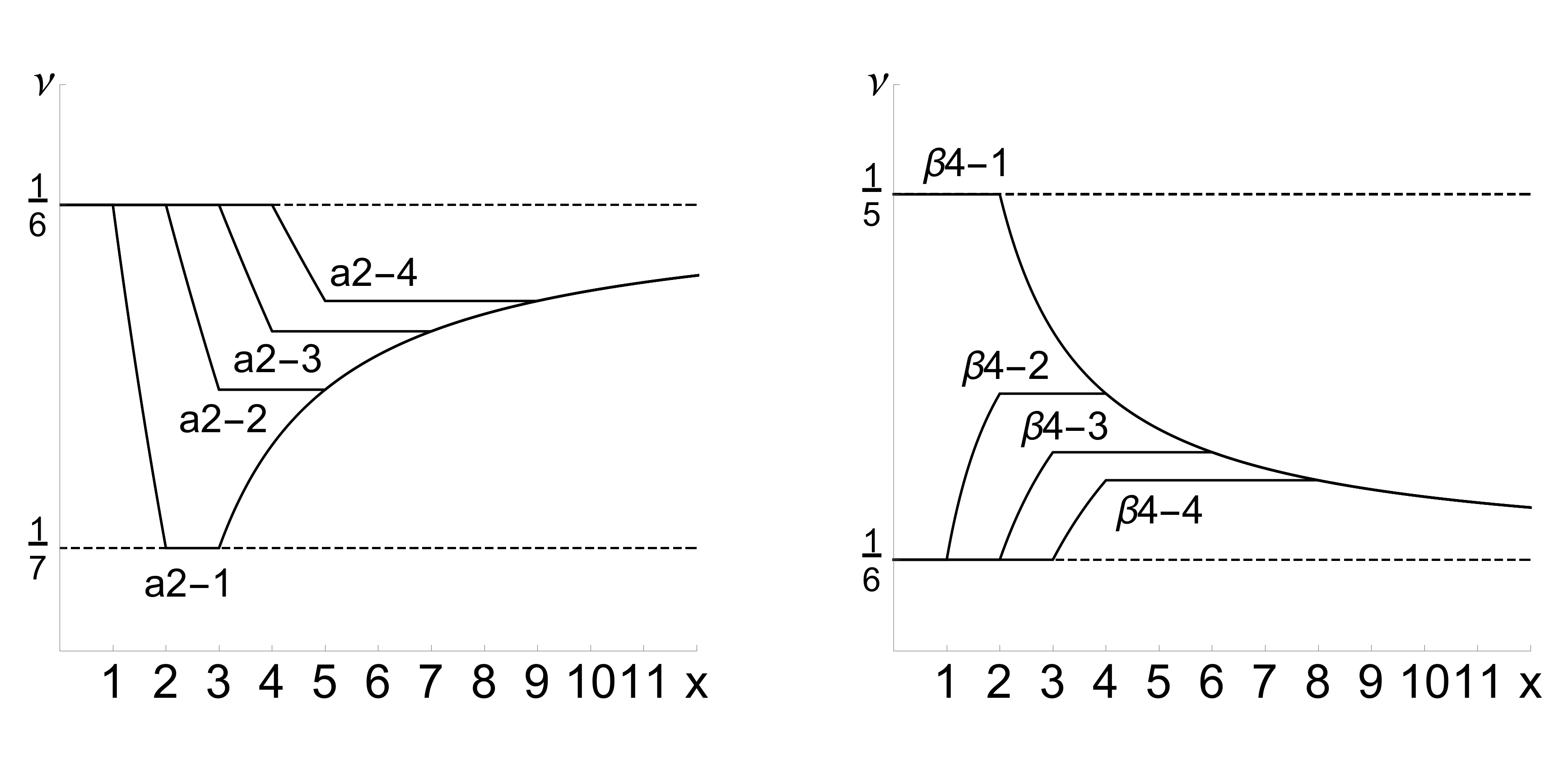}
\caption{\label{fig:PMaps3-DNu}
	Rotation number $\nu$ as a function of amplitude $x$
	(horizontal or vertical distance measured from fixed
	point to invariant polygon) for families of mappings
	a2-$k$ and $\beta4$-$k$ with $k=1,\ldots,4$.
	}
\end{figure}

%----------------------------------------------------------------
Mappings a2 and $\beta$4 are integrable for any $d\in\mathbb{Z}$,
forming two families of transformations with discrete parameter.
Invariant level sets are shown in Fig.~\ref{fig:PMaps3-D}.
The rotation number is provided in Table~\ref{tab:3p-D} and
Fig.~\ref{fig:PMaps3-DNu} shows it as a function of amplitude.
Inner and outer layers become separated with chains of linear
islands, with increment (or decrement) of $d$ from cases a2 and
$\beta$4, and the number of chains increases linearly with $d$.
On these {\it groups of linear islands}, the rotation number is
again mode-locked.

%----------------------------------------------------------------
%----------------------------------------------------------------

%===============================================================%
%===============================================================%
\subsection{Continuous {\it d}, (C)}

\begin{figure}[ph!]\centering
\includegraphics[width=0.75\columnwidth]{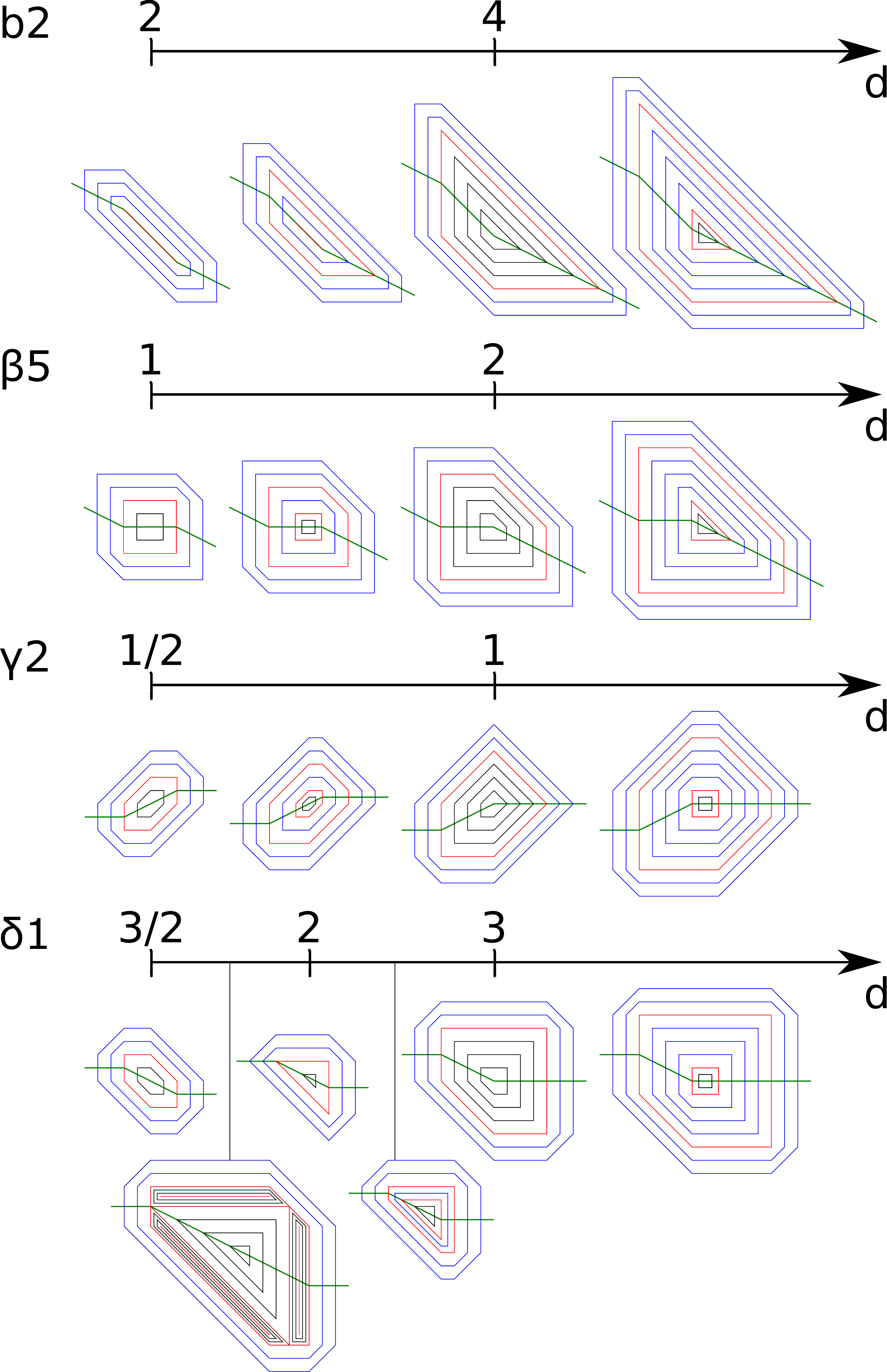}
\caption{\label{fig:PMaps3-C}
    Bifurcation diagram for integer integrable polygon mappings
    with 3-piece force function and continuous $d\in\mathbb{R}$.
    Each row shows changes of phase space with shift parameter $d$.
    Phase space plots at exact values of $d$ correspond to
    bifurcations in a system and plots in between them show
    prototype of map for some value of $d$ in the corresponding
    interval of parameter.
	}
\end{figure}

%----------------------------------------------------------------
In this subsection we present the last class of integer mappings
with 3-piece force function which is integrable for $d\in\mathbb{R}$.
Previously mentioned mappings b2 and $\beta$5 as well as two more,
$\gamma2$ and $\delta$1, are integrable for any real value of
the shift parameter.
As $d$ is continuously changed, the fixed point is moving along
the second symmetry line passing each segment and vertex of the
force function.
Bifurcation diagrams for mappings' phase space is shown in
Fig.~\ref{fig:PMaps3-C} and rotation numbers in different layers
are given in Table~\ref{tab:3p-C}.

%----------------------------------------------------------------
Something interesting can be observed for the mapping $\delta 1$.
When $d>3$ map has three layers: inner linear area with period 4
and two outer nonlinear layers.
As $d$ approaches 3 from above, linear area shrinks to zero and
first nonlinear layer ``mode-locks'' to linear area with rotation
number being $2/7$;
it reflects the bifurcation in the system since fixed point moved
to the second vertex of the force function.
Decreasing value of $d$ further form 3 to 2, one see how separatrix
split forming three layers again with inner layer being period 3.
At $d=2$ inner nonlinear layer disappears again in another
bifurcation.
Finally, in region of $d$ in between 2 and $3/2$ inner linear
area transforms from triangle to hexagon in a very nontrivial
way involving appearance of the chain of linear islands
(see Fig.~\ref{fig:PMaps3-C}).
The surprising fact is that rotation number in nonlinear layer
for $d\in[3/2,3]$ does not depend on the value of $d$.

%----------------------------------------------------------------
%----------------------------------------------------------------
\begin{table}[bh!]
\centering
\begin{tabular}{p{1.5cm}p{1.5cm}p{3.5cm}p{3.5cm}p{3.5cm}p{3.5cm}}
\hline\hline
b2      &           & $d=2$                 & $2<d<4$                       & $d=4$                 & $d>4$                     \\\hline
        &           &                       &                               &                       &                           \\[-0.25cm]
        & $\nu_0$   & $\frac{1}{2}$         & $\frac{1}{2}$                 & $\frac{3}{8}$         & $\frac{1}{3}$             \\[0.25cm]
        & $\nu_1$   & ---                   & $\frac{4-d+6x}{2(4-d)+16x}$   & ---                   & $\frac{d-4+3x}{3(d-4)+8x}$\\[0.25cm]
        &           &                       & $x\in[0,d/2-1]$               &                       & $x\in[0,1]$               \\[0.25cm]
        & $\nu_2$   & $\frac{1+2y}{2+6y}$   & $\frac{d-1+2y}{3d-4+6y}$      & $\frac{3+2y}{8+6y}$   & $\frac{d-1+2y}{3d-4+6y}$  \\[0.25cm]
        &           & $y\in[0,\infty)$      & $y\in[0,\infty)$              & $y\in[0,\infty)$      & $y\in[0,\infty)$          \\[0.4cm]
$\beta$5&           & $d=1$                 & $1<d<2$                       & $d=2$                 & $d>2$                     \\\hline
        &           &                       &                               &                       &                           \\[-0.25cm]
        & $\nu_0$   & $\frac{1}{4}$         & $\frac{1}{4}$                 & $\frac{2}{7}$         & $\frac{1}{3}$             \\[0.25cm]
        & $\nu_1$   & ---                   & $\frac{2-d+2x}{4(2-d)+7x}$    & ---                   & $\frac{d-2+2x}{3(d-2)+7x}$\\[0.25cm]
        &           &                       & $x\in[0,d-1]$                 &                       & $x\in[0,1]$               \\[0.25cm]
        & $\nu_2$   & $\frac{1+2y}{4+6y}$   & $\frac{d+2y}{3d+1+6y}$        & $\frac{2+2y}{7+6y}$   & $\frac{d+2y}{3d+1+6y}$    \\[0.25cm]
        &           & $y\in[0,\infty)$      & $y\in[0,\infty)$              & $y\in[0,\infty)$      & $y\in[0,\infty)$          \\[0.4cm]
$\gamma$2&          & $d=1/2$               & $1/2<d<1$                     & $d=1$                 & $d>1$                     \\\hline
        &           &                       &                               &                       &                           \\[-0.25cm]
        & $\nu_0$   & $\frac{1}{6}$         & $\frac{1}{6}$                 & $\frac{1}{5}$         & $\frac{1}{4}$             \\[0.25cm]
        & $\nu_1$   & ---                   & $\frac{1-d+x}{6(1-d)+5x}$     & ---                   & $\frac{d-1+x}{4(d-1)+5x}$ \\[0.25cm]
        &           &                       & $x\in[0,2d-1]$                &                       & $x\in[0,1]$               \\[0.25cm]
        & $\nu_2$   & $\frac{1+2y}{6+8y}$   & $\frac{d+y}{4d+1+4y}$         & $\frac{1+y}{5+4y}$    & $\frac{d+y}{4d+1+4y}$     \\[0.25cm]
        &           & $y\in[0,\infty)$      & $y\in[0,\infty)$              & $y\in[0,\infty)$      & $y\in[0,\infty)$          \\[0.4cm]
$\delta$1&          & $3/2\leq d \leq 2$    & $2<d<3$                       & $d=3$                 & $d>3$                     \\\hline
        &           &                       &                               &                       &                           \\[-0.25cm]
        & $\nu_0$   & $\frac{1}{3}$         & $\frac{1}{3}$                 & $\frac{2}{7}$         & $\frac{1}{4}$             \\[0.25cm]
        & $\nu_1$   & ---                   & $\frac{3-d+2x}{3(3-d)+7x}$    & ---                   & $\frac{d-3+2x}{4(d-3)+7x}$\\[0.25cm]
        &           &                       & $x\in[0,d-2]$                 &                       & $x\in[0,1]$               \\[0.25cm]
        & $\nu_2$   &$\frac{1+y}{3+4y}$     & $\frac{d-1+y}{4d-5+4y}$       & $\frac{2+y}{7+4y}$    & $\frac{d-1+y}{4d-5+4y}$   \\[0.25cm]
        &           & $y\in[0,\infty)$      & $y\in[0,\infty)$              & $y\in[0,\infty)$      & $y\in[0,\infty)$          \\[0.1cm]
\hline\hline
\end{tabular}
\caption{Rotation number, $\nu_i$, for the $i$-th layer of the integer
        integrable polygon mappings with 3-piece force function and
        continuous $d$.
        Different columns corresponds to different values (or
        intervals) of the shift parameter $d$.
        For each nonlinear layer, the rotation number is provided
        along with the interval of amplitude $x,y$.
        Amplitude in each layer is measured as a horizontal or
        vertical distance from the inner separatrix to the invariant
        curve under consideration.}
\label{tab:3p-C}
\end{table}
%----------------------------------------------------------------
%----------------------------------------------------------------

%----------------------------------------------------------------
%----------------------------------------------------------------

%===============================================================%
%===============================================================%
%===============================================================%
\section{\label{sec:4Pmaps}4-piece maps}

%----------------------------------------------------------------
In this section we will consider force function with four segments.
Adding an extra segment results in total of 6 independent parameters
defining the force function: 4 slopes $k_{1,2,3,4}$, shift parameter
$d$ and ratio $r = l_2/l_3$.
The scale transformation allows us to normalize one of two finite
pieces ($l_2$ or $l_3$) to unity so only one of them is independent;
alternatively we will use the ratio $r$ when it appears to be more
convenient.
In next three subsections we will present mappings isolated with
respect to $d$ (finite set of integer values) and isolated with
respect to $r$ as well, discrete $r$ (all positive integer values,
$r\in\mathbb{N}^+$) or continuous $r$ (map is integrable for any
$r\in\mathbb{R}^+$).
Results of a search are summarized in Tables~\ref{tab:4PTable1} and
\ref{tab:4PTable2}.
In addition, more cases which were harder to classify are presented
in the last Subsection~\ref{sec:4P-Tree}.

%----------------------------------------------------------------
%----------------------------------------------------------------
\begin{table}[h]
\begin{center}
\begin{tabular}{m{2.5cm}|m{1.33cm}m{1.33cm}m{1.33cm}||m{1.33cm}m{1.33cm}m{1.33cm}|m{2.5cm}}
\hline\hline
                    &         & \cc $r$ &         &         & \cc $r$ &         &                       \\
$[k_1,k_2,k_3,k_4]$ & \cc 1:1 & \cc 1:2 & \cc 1:3 & \cc 1:3 & \cc 1:2 & \cc 1:1 & $[k_1,k_2,k_3,k_4]$   \\\hline
$[-1,-2,-1,-2]$     &         & \cc 4,7 &         &         &         &         & $[-2,-1,-2,-1]$       \\[0.15cm]
$[0,-2,0,-1]$       & \cc 2   & \cc 0   & \cc 2   &         &         & \cc 4   & $[-1,0,-2,0]$         \\[0.15cm]
$[0,-1,1,-1]$       & \cc 2   &         &         & \cc 4   &         & \cc 2   & $[-1,1,-1,0]$         \\[0.15cm]
$[1,-2,0,-1]$       &         & \cc 0   &         &         &         &         & $[-1,0,-2,1]$         \\[0.15cm]
$[1,-1,0,-1]$       & \cc 1   &         & \cc 1   &         &         & \cc 4   & $[-1,0,-1,1]$         \\[0.15cm]
$[1,0,1,-1]$        & \cc 1,2 &         &         & \cc 4   &         & \cc 2,1 & $[-1,1,0,1]$          \\[0.15cm]
$[1,0,2,0]$         &         &         &         &         & \cc 0   &         & $[0,2,0,1]$           \\[0.15cm]
$[1,2,0,1]$         &         & \cc 0   &         &         &         &         & $[1,0,2,1]$           \\
\hline\hline
\end{tabular}
\end{center}
\caption{
        Integrable polygon mappings with 4-piece force function
        and isolated integer $d$ and $r$.
        Such mappings were found only for $r=1,2,3$ and corresponding
        values of parameter $d$ (in units of $l_2$) are presented for
        each quadruplet of integer slopes.
        }
\label{tab:4PTable1}
\end{table}   
%----------------------------------------------------------------
%----------------------------------------------------------------

%----------------------------------------------------------------
%----------------------------------------------------------------
\begin{table}[h]
\begin{center}
\begin{tabular}{m{2.5cm}|m{2.1cm}m{2.1cm}||m{2.1cm}m{2.1cm}|m{2.5cm}}
\hline\hline
$[k_1,k_2,k_3,k_4]$ & $d$ & $r$ & $r$ & $d$ & $[k_1,k_2,k_3,k_4]$   \\\hline
$[0,-1,-2,-1]$      & 0         & 1              & $\mathbb{N}^+$ & $4+3/r$ & $[-1,-2,-1,0]$    \\[0.15cm]
                    & 1         & $\mathbb{R}^+$ & $\mathbb{R}^+$ & $4+2/r$ &                   \\[0.15cm]
$[0,-2,-1,0]$       & 0         & $\mathbb{R}^+$ & $\mathbb{R}^+$ & $3+4/r$ & $[0,-1,-2,0]$     \\[0.15cm]
                    & $2+1/r$   & $\mathbb{R}^+$ & $\mathbb{R}^+$ & $2+2/r$ &                   \\[0.15cm]
                    & $4+2/r$   & $\mathbb{R}^+$ & $\mathbb{R}^+$ & 1       &                   \\[0.15cm]
$[0,-1,1,0]$        & 2         & $\mathbb{R}^+$ & $\mathbb{R}^+$ & $1+1/r$ & $[0,1,-1,0]$      \\[0.15cm]
$[0,1,2,0]$         & 1         & $\mathbb{R}^+$ & $\mathbb{R}^+$ & 0       & $[0,2,1,0]$       \\[0.15cm]
$[1,-1,-2,-1]$      & 0         & 1              & $\mathbb{N}^+$ & $4+3/r$ & $[-1,-2,-1,1]$    \\[0.15cm]
$[1,-1,0,-1]$       & 0         & 1              & $\mathbb{N}^+$ & $2+3/r$ & $[-1,0,-1,1]$     \\[0.15cm]
$[1,0,-2,-1]$       &           &                & $2,4,6,\ldots$ & $4+3/(2r)$ & $[-1,-2,0,1]$  \\[0.15cm]
$[1,-2,-1,0]$       & 0         & $\mathbb{R}^+$ & $\mathbb{R}^+$ & $3+4/r$ & $[0,-1,-2,1]$     \\[0.15cm]
$[1,0,-1,0]$        & 0         & $\mathbb{R}^+$ & $\mathbb{R}^+$ & $3+2/r$ & $[0,-1,0,1]$      \\[0.15cm]
                    & 1         & $\mathbb{R}^+$ & $\mathbb{R}^+$ & $3+1/r$ &                   \\[0.15cm]
                    & $1+1/r$   & $\mathbb{R}^+$ & $\mathbb{R}^+$ & $2+1/r$ &                   \\[0.15cm]
$[1,2,1,0]$         & 0         & $\mathbb{R}^+$ & $\mathbb{R}^+$ & 1       & $[0,1,2,1]$       \\[0.15cm]
                    & $1/r$     & $\mathbb{N}^+$ & 1              & 0       &                   \\
\hline\hline
\end{tabular}
\end{center}
\caption{
        Integrable polygon mappings with 4-piece force function
        and isolated integer $d$ and discrete of continuous $r$.
        Values of parameters $d$ (in units of $l_2$) and $r$ are
        presented for each quadruplet of integer slopes.
        }
\label{tab:4PTable2}
\end{table}   
%----------------------------------------------------------------
%----------------------------------------------------------------

\newpage
%===============================================================%
%===============================================================%
\subsection{Isolated {\it d} and {\it r}}

\begin{figure}[b!]\centering
\includegraphics[width=\columnwidth]{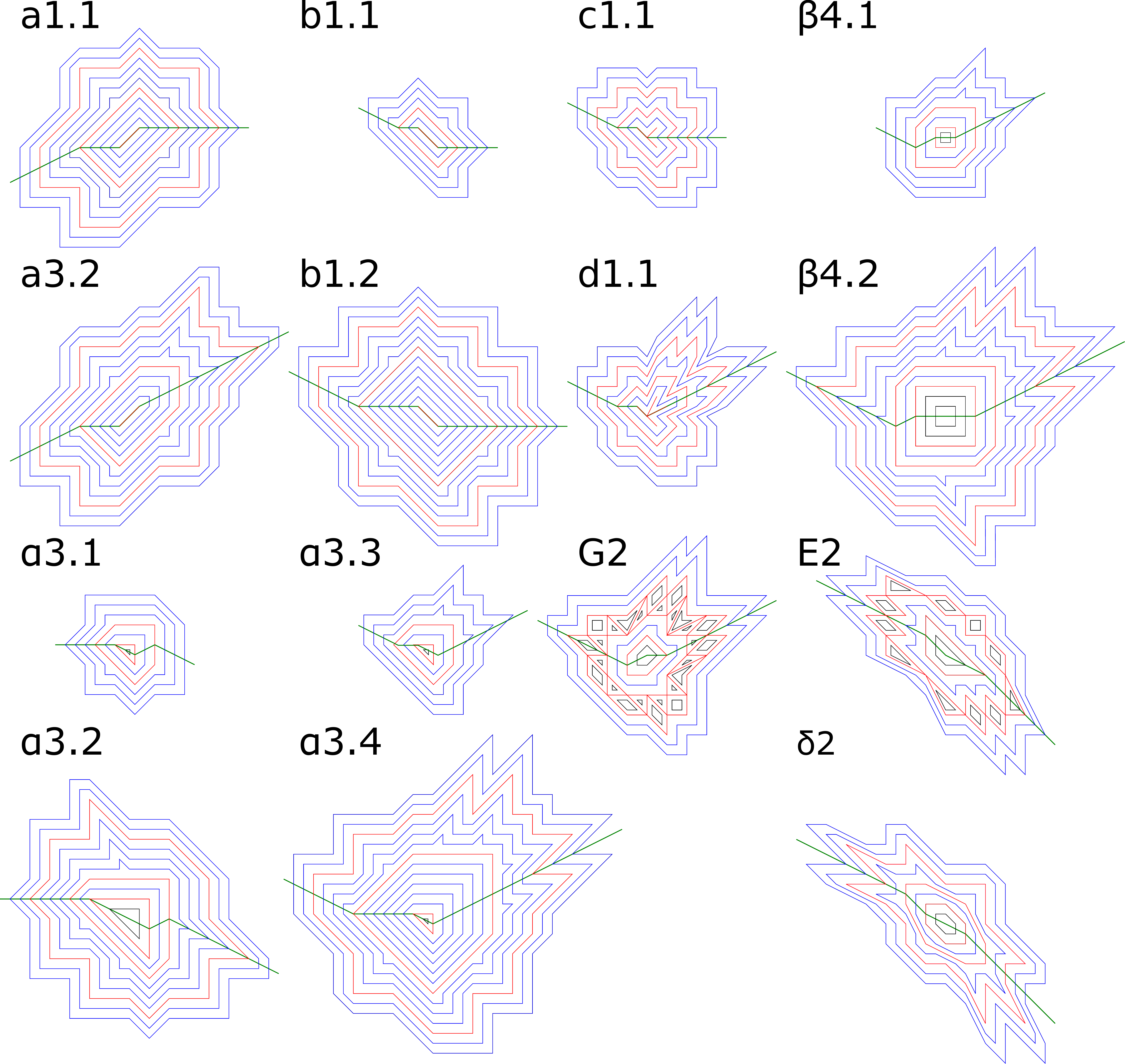}
\caption{\label{fig:PMaps4-I}
	Integer integrable polygon mappings with 4-piece force function
	and isolated values of shift parameter, $d$, and ratio of finite
	pieces, $r$.
	}
\end{figure}

%----------------------------------------------------------------
4-piece mappings with isolated integer values of $d$ and $r$ are
shown in Fig.~\ref{fig:PMaps4-I} and action angle variables are
provided in Tables~\ref{tab:PMaps4-I1} and \ref{tab:PMaps4-I2};
these and all further tables with dynamical properties of 4-piece
maps are moved to Appendix~\ref{secAPP:Nu} for the reader's
convenience.

%----------------------------------------------------------------
Naively we expected that layers of phase space change along with
the force function and separatrices of motion are associated with
vertices of $f(q)$, for example look at mappings b1.1,
$\alpha$3.1, $\alpha$3.3 or $\beta$4.1.
But results of our search are showing that dynamics can be more
complicated.
As one can see for mappings a1.1, a3.2, c1.1 and few others,
an additional layer can appear on outer pieces of force function.
This is due to the fact that polygons in one of them have sides
decreasing with the amplitude;
as a result, at some critical amplitude length of the side becomes
equal to zero and invariant level sets transition to new layer.
We would like to name this phenomena as
{\it long-range catastrophe}.

%----------------------------------------------------------------
%----------------------------------------------------------------

\newpage
%===============================================================%
%===============================================================%
\subsection{Isolated {\it d}, discrete {\it r}}

\begin{figure}[b!]\centering
\includegraphics[width=0.95\columnwidth]{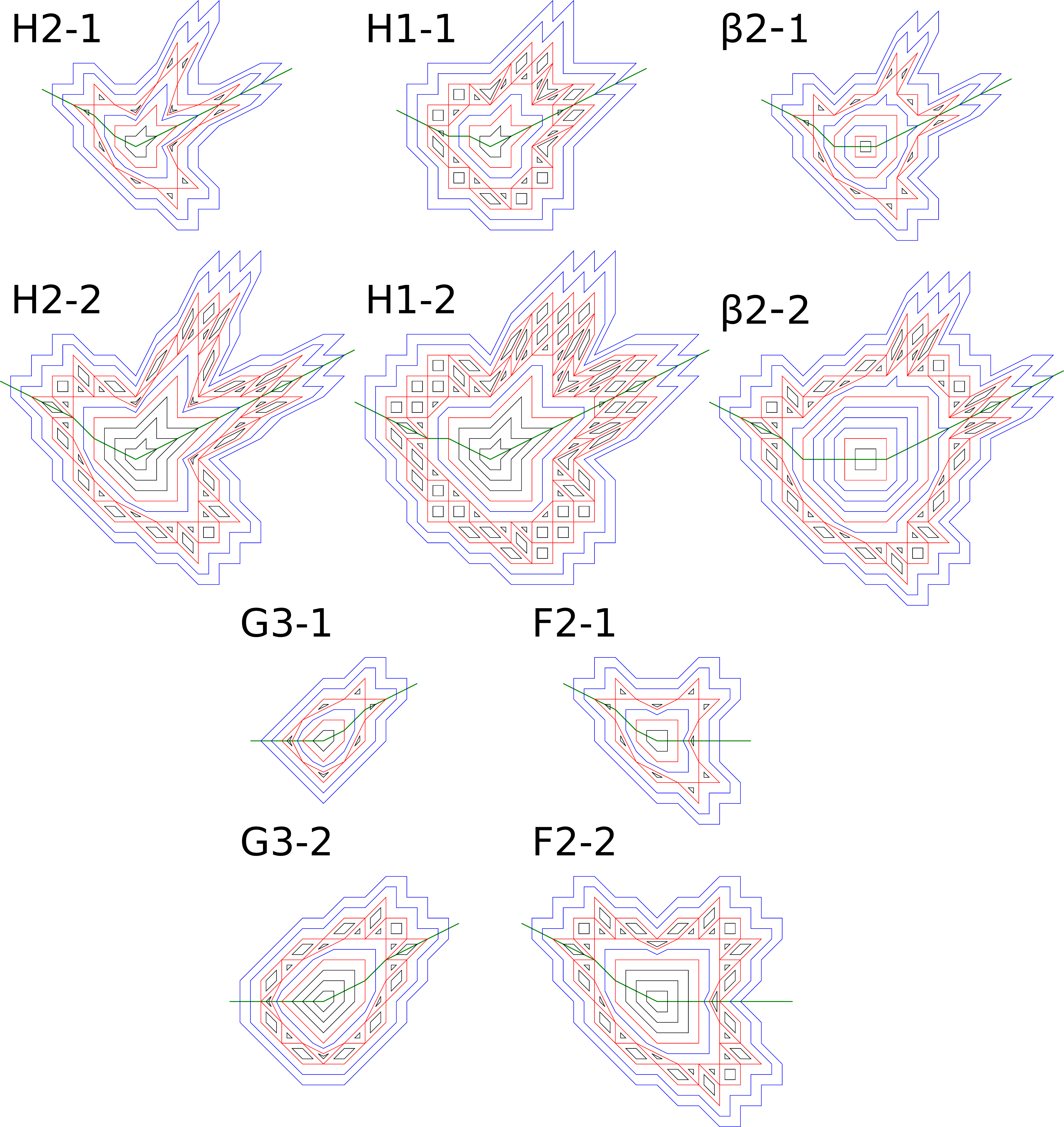}
\caption{\label{fig:PMaps4-D}
	Integer integrable polygon mappings with 4-piece force function
	and isolated value of the shift parameter, $d$, and integer
	ratio of finite pieces $r\in\mathbb{N}^+$ (for the map $\beta$2-1
	$r\in2\mathbb{N}^+=2,4,6,\ldots$).
	For each family of mappings two first prototypes are presented.
	}
\end{figure}

%----------------------------------------------------------------
In this subsection we present systems integrable with any positive
integer ratio of $r = l_1/l_2\in\mathbb{N}^+$ but isolated integer
values of the shift parameter $d$.
Five different families of mappings with 2 prototypes from each
are shown in Fig.~\ref{fig:PMaps4-D}.
These maps are similar to 3-piece force function mappings with
discrete $d$ (D) (Subsection~\ref{sec:3P-D}): they all have
chains of linear islands separating outer layers with nonlinear
motion, and, the number of chains within the linear island is
linearly growing with $r$.
Rotation number $\nu_i$ for different layers are listed in
Tables~\ref{tab:PMaps4-D1} and \ref{tab:PMaps4-D2}.

%----------------------------------------------------------------
%----------------------------------------------------------------

\newpage
%===============================================================%
%===============================================================%
\subsection{Isolated {\it d}, continuous {\it r}}

%----------------------------------------------------------------
Here we report systems which are integrable with any positive real
ratio of $r = l_1/l_2 \in \mathbb{R}^+$ but still isolated integer
values of the shift parameter $d$.
Some of them experience bifurcations of phase space with the change
of parameter $r$ (mappings with $'$ index) while the rest of systems
have same set of layers for any value of $r\in(0;\infty)$.
Corresponding level sets of invariant and its bifurcation diagrams
are shown below in Fig.~\ref{fig:PMaps4-C}.
All dynamical properties are listed in Tables~\ref{tab:PMaps4-C1}
and \ref{tab:PMaps4-C2}.
Mapping $\alpha 3.1'$ presents an interesting example:
when $r=1$ outer otherwise nonlinear layer becomes mode-locked
\[
    \nu_2 = \frac{1+r+y}{3+5\,r+4\,y} = \frac{2+y}{4(2+y)} = \frac{1}{4}.
\]

%----------------------------------------------------------------
\begin{figure}[b!]\centering
\includegraphics[width=0.85\columnwidth]{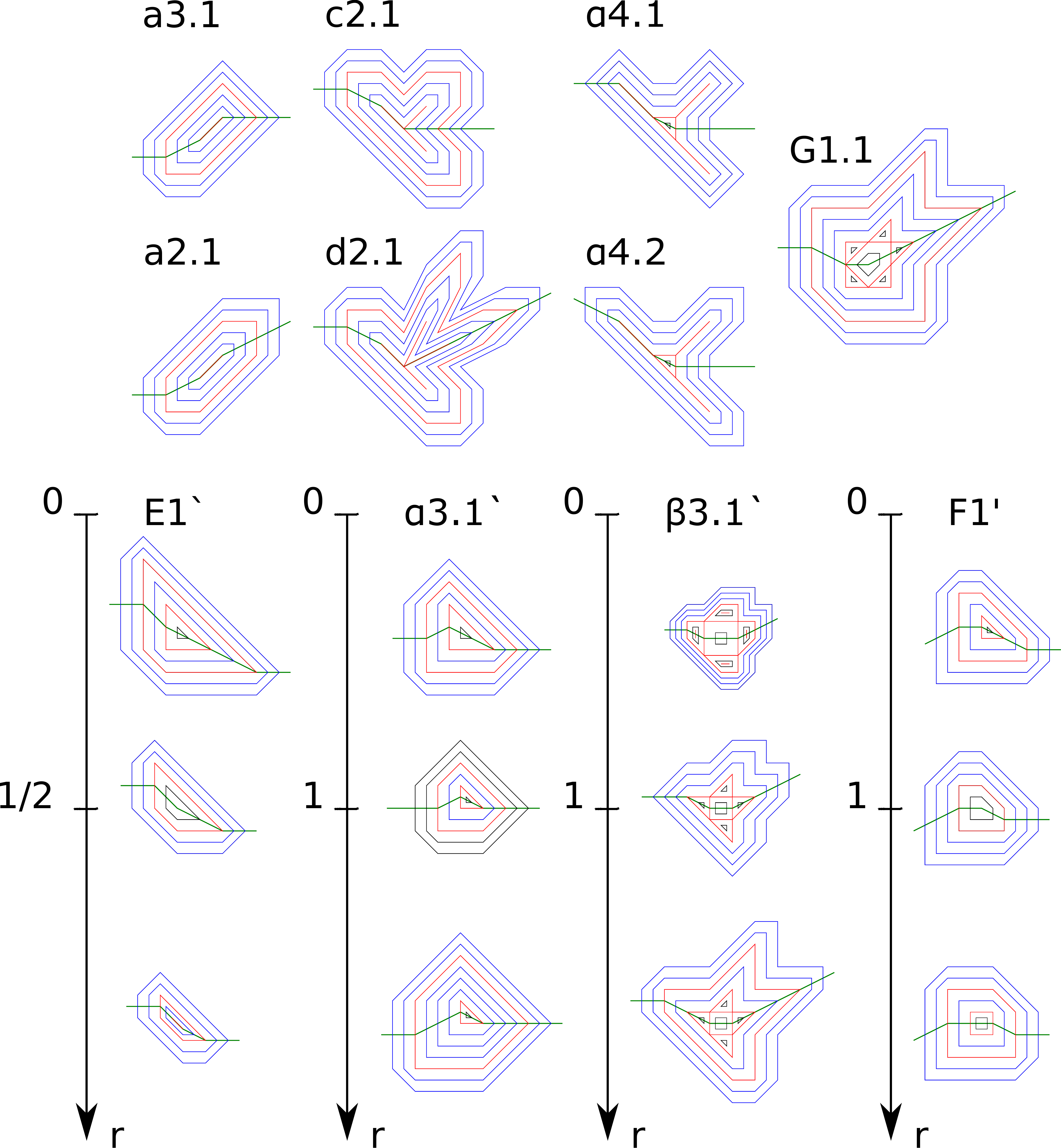}
\caption{\label{fig:PMaps4-C}
    Integer integrable polygon mappings with 4-piece force function
    and isolated values of shift parameter, $d$, and
	continuous ratio of finite pieces, $r\in\mathbb{R}^+$.
	Mappings with $'$ index experience bifurcations of phase space
	with the change of parameter $r$.
	}
\end{figure}
%----------------------------------------------------------------

%===============================================================%
%===============================================================%
\subsection{\label{sec:4P-Tree} Additional cases}

%===============================================================%
\subsubsection{Continuous and semi-continuous {\it d} with unity {\it r}.}

%----------------------------------------------------------------
When $r=1$, we found 2 dynamical systems which are integrable for
any $d \in \mathbb{R}$: $\{0,-1,0,-1\}$ and $\{1,0,1,0\}$.
In addition, map with slopes $\{-2,-1,-2,-1\}$ is integrable for
any $d \geq 4$ (and its twin map $\{-1,-2,-1,-2\}$ is integrable
for any $d \leq 3$).

%===============================================================%
\subsubsection{Non-unity values of {\it r}.}

%----------------------------------------------------------------
For values of $r$ different from 1, the situation is getting more
complicated and more families of mappings appear.
For example, for previously semi-continuous map $\{-2,-1,-2,-1\}$,
we have at least four families with different behaviours of $d$:
\[
\begin{array}{ll}
    d = 4\,L_2 + 1  &\quad (L_2 \geq 1),                    \\
    d = 4\,L_2 + 2  &\quad (L_2 = 1,2,3,\ldots),            \\
    d = 4\,L_2 + 3  &\quad (L_2 = 1,3,4,6,7,9,10,\ldots),   \\
    d = 5\,L_2      &\quad (L_2 \geq 1);
\end{array}
\]
we say at least because our algorithm was limited to the range of
$r=[1,2,...,9,10]$ and we can not guarantee the absence of integrable
mappings out of this range.

%----------------------------------------------------------------
For the continuous mappings  we have
\[
\begin{array}{llll}
    \{0,-1,0,-1\}:  &
    \qquad\qquad\qquad\qquad\qquad\qquad\qquad\qquad\qquad  &
        \{-1,0,-1,0\}:  &                                       \\[0.25cm]
    d = L_2         &\quad (L_2 \geq 0),                    &
        d = 2\,L_2      &\quad (L_2 \geq 0),                    \\
    d = L_2 + 1     &\quad (L_2 \geq 0),                    &
        d = 2\,L_2 + 1  &\quad (L_2 \geq 0),                    \\
    d = 2\,L_2 + 1  &\quad (L_2 \geq 0),                    &
        d = 3\,L_2 + 1  &\quad (L_2 \geq 0),                    \\
    d = 2\,L_2 + 2  &\quad (L_2 = 1,2,3,\ldots),            &
        d = 3\,L_2 + 2  &\quad (L_2 = 1,2,3,\ldots),            \\
    d = 2\,L_2 + 3  &\quad (L_2 = 1,3,4,6,7,9,10,\ldots),   &
        d = 3\,L_2 + 3  &\quad (L_2 = 1,3,4,6,7,9,10,\ldots),   \\
    d = 3\,L_2      &\quad (L_2 \geq 1/2),                  &
        d = 4\,L_2      &\quad (L_2 \geq 1/2),
\end{array}
\]
and 
\[
\begin{array}{llll}
    \{1,0,1,0\}:\qquad\qquad  &
    \qquad\qquad\qquad\qquad\qquad\qquad\qquad\qquad\qquad  &
        \{0,1,0,1\}:  &                                         \\[0.25cm]
    d = 0           &\quad (L_2 \geq 0),                    &
        d = L_2         &\quad (L_2 \geq 0),                    \\
    d = 1           &\quad (L_2 \geq 0),                    &
        d = L_2 + 1     &\quad (L_2 \geq 0),                    \\
    d = L_2 + 1     &\quad (L_2 \geq 0),                    &
        d = 2\,L_2 + 1  &\quad (L_2 \geq 0),                    \\
    d = L_2 + 2     &\quad (L_2 = 1,2,3,\ldots),            &
        d = 2\,L_2 + 2  &\quad (L_2 = 1,2,3,\ldots),            \\
    d = L_2 + 3     &\quad (L_2 = 1,3,4,6,7,9,10,\ldots),   &
        d = 2\,L_2 + 3  &\quad (L_2 = 1,3,4,6,7,9,10,\ldots),   \\
    d = 2\,L_2      &\quad (L_2 = 1/2 \quad\mathrm{and}\quad L2 \geq 1),&
        d = 3\,L_2      &\quad (L_2=1/2,3/2 \quad\mathrm{and}\quad L2 \geq 2).\\
    d = 2           &\quad (L_1:L_2 = 1:4),                 &
                    &
\end{array}
\]

%----------------------------------------------------------------
Finally, we have two more dynamical systems with complicated
dependence of $d$ and $r$:
\[
\begin{array}{llll}
    \{0,-1,0,1\}:\qquad\qquad  &
    \qquad\qquad\qquad\qquad\qquad\qquad\qquad\qquad\qquad  &
        \{1,2,1,0\}:        &                                       \\[0.25cm]
    d = L2 + 2          &\quad (L_2 \geq 0),                    &
        d = 1               &\quad (L_2 = 3,4,5,\ldots),            \\
    d = L2 + 3          &\quad (L_2 \geq 0),                    &
        d =(L_2/2) + 1      &\quad (L_2 = 2,6,10,14,\ldots).        \\
    d = 2\,L_2 + 1      &\quad (L_2 = 1,2,4,5,7,8,10,\ldots),   &
                            &                                       \\
    d = 2\,L_2 + 2      &\quad (L_2 = 1,2,3,\ldots),            &
                            &                                       \\
    d = 2\,L_2 + 3      &\quad (L_2 \geq 0),                    &
                            &                                       \\
    d = (3/2)\,L_2 + 7/2&\quad (L_2 = 1,3,5,7,\ldots),          &
                            &                                       \\
    d = (3/2)\,L_2 + 3  &\quad (L_2 = 2,4,6,8,\ldots),          &
                            &                                       \\
    d = (5/3)\,L_2 +10/3&\quad (L_2 = 1,4,7,10,\ldots),         &
                            &
\end{array}
\]

%----------------------------------------------------------------
Since there are too many mappings in this subsection, we did not
provide any figures or tables with dynamical properties, leaving
them as an exercise for the reader.

\newpage
%===============================================================%
%===============================================================%
%===============================================================%
\section{\label{sec:App}Applications of mappings with polygon invariants}

%----------------------------------------------------------------
In the next three subsections we will discuss the importance and
some practical applications of mappings with polygon invariants
as well as its connection to other dynamical systems.

%===============================================================%
%===============================================================%
\subsection{\label{sec:Panleve}
Connection to Painlev\'e equations and Suris mappings}

%----------------------------------------------------------------
Some integrable mappings with polygon invariants have a direct
connection to discrete Painlev\'e equations (dP).
Painlev\'e equations are nonlinear difference equations with
canonical form
being~\cite{levi1996symmetries,grammaticos2004discrete}:
\begin{equation}\label{eq:PL}
\begin{array}{lll}
\mathrm{dP}_\mathrm{I}:    & \ds \quad X_{n+1}\,X_{n-1} =
    \frac{\lambda^n X_n + 1}{X_n^\sigma},                       &
    \sigma = 0,1,2,                                             \\[0.35cm]
\mathrm{dP}_\mathrm{II}:   & \ds \quad X_{n+1}\,X_{n-1} =
    \frac{\alpha(\lambda^n + X_n)}{X_n^\rho (1+\lambda^n X_n)}, &
    \rho = 0,1,                                                 \\[0.35cm]
\mathrm{dP}_\mathrm{III}:  & \ds \quad X_{n+1}\,X_{n-1} =
    \frac{(X_n+\alpha\, \lambda^n)(X_n+\alpha^{-1} \lambda^n)}
         {(1+\beta\,\lambda^n X_n)(1+\beta^{-1} \lambda^n X_n)}.
\end{array}
\end{equation}
We will be interested in the {\it autonomous} (i.e. mappings
do not have explicit dependence on $n$) limit of
dP$_\mathrm{I-III}$ corresponding to  $\lambda \to 1$.
It is worth mentioning that autonomous limit of Eqs.~(\ref{eq:PL})
could be written in the Suris form by introducing a new set of
variables
$X_n = e^{x_n/\epsilon}$,
$\alpha = e^{A/\epsilon}$,
$\beta=e^{B/\epsilon}$:
\begin{equation}\label{eq:PL_Suris}
\begin{array}{ll}
\mathrm{dP}_\mathrm{I}^\mathrm{S}:    & \ds \quad x_{n+1}+x_{n-1} =
\epsilon\,\log{\left[
    \frac{e^{x_n/\epsilon}+1}{e^{\sigma\,x_n/\epsilon}}
    \right]},                                           \\[0.425cm]
\mathrm{dP}_\mathrm{II}^\mathrm{S}:   & \ds \quad x_{n+1}+x_{n-1} =
     A - \rho\,x_n,                                     \\[0.325cm]
\mathrm{dP}_\mathrm{III}^\mathrm{S}:  & \ds \quad x_{n+1}+x_{n-1} =
\epsilon\,\log{\left[
    \frac{(e^{x_n/\epsilon}+e^{A/\epsilon})(e^{x_n/\epsilon}+e^{-A/\epsilon})}
         {(e^{x_n/\epsilon}+e^{B/\epsilon})(e^{x_n/\epsilon}+e^{-B/\epsilon})}
\right]}.
\end{array}
\end{equation}
While the force function for dP$_\mathrm{II}^\mathrm{S}$ is
simply a linear function, two other cases are specific examples
of integrable exponential Suris mapping.
%for example, the corresponding invariant for the map
%dP$_\mathrm{I}^\mathrm{S}$ in MH form for different values of
%parameter $\sigma$ is
%\[
%\begin{array}{ll}
%\sigma = 0:\qquad    &\ds  \mathcal{K}(p,q) =
%    e^{2p}e^{2q} - e^{2p} - e^{2q} + e^{p}e^{q} - e^{p} - e^{q},\\[0.25cm]
%\sigma = 1:\qquad    &\ds  \mathcal{K}(p,q) =
%    e^{2p}e^{2q} + e^{2p}e^{q} + e^{p}e^{2q},           \\[0.25cm]
%\sigma = 2:\qquad    &\ds  \mathcal{K}(p,q) =
%    e^{2p}e^{q} + e^{p}e^{2q} + e^{2p} + e^{2q}.        \\[0.25cm]
%\end{array}
%\]

%----------------------------------------------------------------
The {\it ultradiscretized} form of Painlev\'e equations (udP)
could be obtained by taking the limit $\epsilon \to 0$ in
Eqs.~(\ref{eq:PL_Suris}).
At this point we should clarify the terminology: by saying
``{\it discrete}'' Painlev\'e equation we refer to the difference
nature of the equation in contrast to nonlinear second-order
ordinary differential Painlev\'e equations.
The addition ``{\it ultra}'' refers to another level of
discretization when the force function is transitioning from a
smooth continuous to a piecewise linear function.
Using identity
$\ds \lim_{\epsilon \to 0} \epsilon
\log{\left[
    \exp{(x/\epsilon)}+\exp{(y/\epsilon)
}\right]} = \max(x,y)$
we arrive to
\begin{equation}\label{eq:udPL}
\begin{array}{ll}
\mathrm{udP}_\mathrm{I}:    & \ds \quad x_{n+1}+x_{n-1} =
    \max{(x_n, 0)} - \sigma\,x_n,                   \\[0.25cm]
\mathrm{udP}_\mathrm{II}:   & \ds \quad x_{n+1}+x_{n-1} =
     A - \rho\,x_n,                                 \\[0.25cm]
\mathrm{udP}_\mathrm{III}:  & \ds \quad x_{n+1}+x_{n-1} =
\max{(x_n, A)}+\max{(x_n, -A)}-\max{(x_n, B)}-\max{(x_n, -B)}.
\end{array}
\end{equation}
The udP equations~(\ref{eq:udPL}) have a symplectic Suris form
$x_{n+1}+x_{n-1} = f(x_n)$ with following ultradiscrete piecewise
linear force functions
\begin{equation}
\label{eq:force_PL}
\begin{array}{l}
\ds f_\mathrm{I}(x)     \,\,    = \frac{1}{2}\left(|x|+x\right) - \sigma\,x =
        \begin{cases}
            (1 - \sigma)\,x,   & \textrm{for } x \geq 0,\\
            - \sigma\,x,       & \textrm{for } x<0,
        \end{cases}                                             \\[0.5cm]
\ds f_\mathrm{II}(x)    \,\,\!  = A - \rho\,x,                         \\[0.1cm]
\ds f_\mathrm{III}(x)           =
    \frac{1}{2} \left( |x+A|+|x-A|-|x+B|-|x-B|\right) =
        \begin{cases}
            0,    & \textrm{for  } x \in [B, \infty), \\
            x-B,  & \textrm{for  } x \in [A,B),       \\
            A-B,  & \textrm{for  } x \in [-A, A),     \\
            -B-x, & \textrm{for  } x \in [-B, -A),    \\
            0,    & \textrm{for  } x \in (-\infty, -B).
        \end{cases}
\end{array}
\end{equation}

\begin{figure}[t!]\centering
\includegraphics[width=\columnwidth]{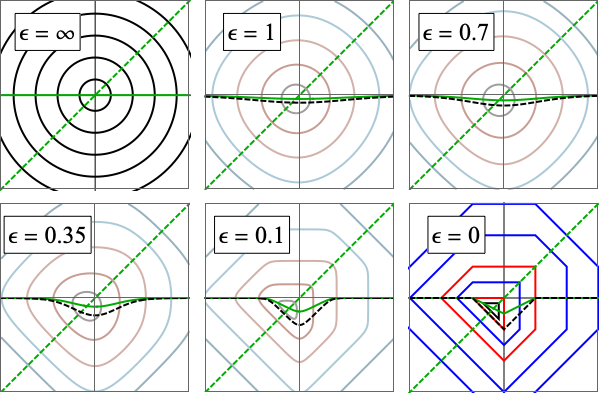}
\caption{\label{fig:PLmaps}
	Constant level sets of invariant on $(x_n,x_{n+1})$ plane for
	discrete Painlev\'e map dP$_\mathrm{III}^\mathrm{S}$ in Suris
	form with $A=0$ and $x$ measured in units of $B$.
	For $\epsilon = \infty$ map degenerate to linear with $f(x)=0$
	and level sets degenerate to concentric circles.
	When $\epsilon = 0$ we observe ultradiscrete limit
	udP$_\mathrm{III}$ corresponding to polygon map $\alpha 3.1'$
	with ratio of finite pieces restricted to unity, $r=1$.
	Black and blue level sets show trajectories in linear
	and nonlinear areas of phase space, red level sets are
	separatrices of motion between different layers.
	Green curve is the second symmetry line $f(x)/2$ and black dashed
	curve is the force function $f(x)$.
	}
\end{figure}

\newpage
%----------------------------------------------------------------
The force function $f_\mathrm{I}$ (corresponding to the first
equation udP$_\mathrm{I}$) is the simplest two-piece linear force
without shift parameter, $d=0$, and with slopes defined as
$(k_1, k_2)=(-\sigma, 1-\sigma)$.
Since $\sigma$ can only take values 0,1 and 2, we
obtain previously considered mappings G, F and E respectively
(see Fig.~\ref{fig:PMaps2-L}).
The second ultradiscretized Painlev\'e equation has a trivial
linear force function $f_\mathrm{II}$ and is integrable for any
values of parameters $A$ and $\sigma$.
In the case of integer slopes this results in 3 mappings with stable
trajectories~\cite{cairns2014piecewise} $\alpha$, $\beta$, $\gamma$
and 2 unstable mappings $a$, $b$ (see Fig.~\ref{fig:PMaps2-L}).
Finally, for the third ultradiscretized Painlev\'e equation
udP$_\mathrm{III}$ (assuming $B>A>0$), we obtain force function
containing five pieces.
Taking the limit $A = \const$ and $B \to \infty$, from
Eq.~(\ref{eq:force_PL}) we have the linear map $\alpha$ with
slope of force function $k_1=-1$.
Considering another limiting case $A \to 0$, one can see that
$f(x)$ could be reduced to one of our 4-piece polygon maps
$\alpha 3.1'$ (slope coefficients
$\{k_1,k_2,k_3,k_4\} = \{0, -1,1,0\}$ and $d=2B$)
with ratio parameter restricted to $r=1$
(note that our map $\alpha 3.1'$ is integrable for arbitrary $r$,
see Fig.~\ref{fig:PMaps4-C}).
Figure~\ref{fig:PLmaps} shows invariant level sets for this last
case for different values of $\epsilon$.
As one can see, approaching the ultradiscrete level $\epsilon\to 0$,
the force function transitions to piecewise linear and level sets
transform to polygons.
When $\epsilon = \infty$ map becomes linear with $\nu = 1/4$;
note that invariant level sets are chosen as circles and not
polygons.
As we previously discussed linear mappings with rational rotation
numbers are degenerate and allow infinitely many invariants of
motion, but in this case the shape is chosen as a limiting case
$\epsilon \to \infty$ for the exponential Suris invariant.

%----------------------------------------------------------------
%----------------------------------------------------------------

%----------------------------------------------------------------
We would like to emphasize that our polygon mappings represent a
much richer family of integrable systems beyond ultradiscretezations
of known Painlev\'e equations udP$_\mathrm{I\text{--}III}$;
ultradiscretized equations of Painlev\'e type are limited to only
5-piece force functions and the set of parameters is very restricted
(e.g. slopes of outer pieces of 5-piece force functions can only be
equal to zero).
%Some of our polygonal maps could be considered as ultradiscretization
%of more general equations in Painlev\'e form given by a ratio of 
%two polynomials
%\begin{equation}
%\label{eq:poly1}
%\ds X_{n+1} X_{n-1} =
%\frac{a_m X^m_n + a_{m-1} X^{m-1}_n + \ldots + a_0}
%     {b_l X^l_n + b_{l-1} X^{l-1}_n + \ldots + b_0},
%\end{equation}
%or equivalently as a limit of Suris form given by a ratio of two
%exponential polynomials
%\begin{equation}
%\label{eq:poly2}
%\ds x_{n+1}+x_{n-1} =
%\epsilon\,\log{\left[
%    \frac{ e^{(m x_n+A_m)/\epsilon}+ e^{((m-1) x_n+A_{m-1})/\epsilon} + \ldots + e^{A_0/\epsilon}}
%         { e^{(l x_n+B_{l})/\epsilon}+ e^{((l-1) x_n+B_{l-1})/\epsilon} + \ldots + e^{B_0/\epsilon}}
%\right]}.
%\end{equation}
%Ultradiscretization of Eq. (\ref{eq:poly2}) will result in a
%piecewise linear force with the number of pieces equal to $l+m+1$
%and integer slope coefficients $k_{i}\in \{-l, \ldots, m\}$.

\newpage
%===============================================================%
%===============================================================%
\subsection{Near-integrable systems}

\begin{figure}[h!]\centering
\includegraphics[width=\columnwidth]{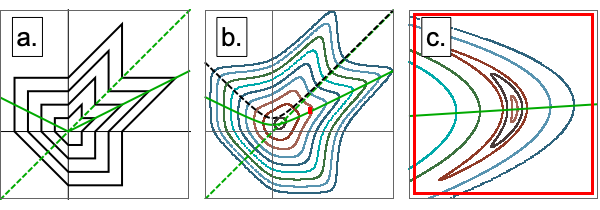}
\caption{\label{fig:Cohen}
    The left plot (a.) illustrates invariant level sets for the
    Brown-Knuth map, force function $f(q) = |q|$.
    The middle plot (b.) displays invariant level sets for the
    Cohen map, $f(q) = \sqrt{q^2 +1}$.
    The right plot (c.) again provides invariant level sets for
    the Cohen map, but on a different scale showing one of the
    island structures.
    The level sets for Cohen map are obtained by tracking.
    The green curve is the second symmetry line $p=f(q)/2$.
    }
\end{figure}

%----------------------------------------------------------------
Since force function is piecewise linear and isn't smooth (not
differentiable at discrete set of points), it might seems that
these polygon mappings are somewhat impractical.
However, in a similar manner to the example in Fig.~\ref{fig:PLmaps}
where the polygon mapping can be smoothed using parameter $\epsilon$,
below we will demonstrate how one can construct not integrable,
but quasi-integrable systems.
One of the possible ways to perform this was first done by
Cohen~\cite{cohen1993,rychlik1998algebraic} in application to
Brown-Knuth map (map H) by introducing a new force function as
\[
    f(q) = \sqrt{q^2+\epsilon^2}\underset{\epsilon \to 0}{\to}|q|.
\]
As one can see it tends to $|q|$ as $\epsilon \to 0$ or at large
amplitudes as $q \to \infty$.
Fig.~\ref{fig:Cohen} shows a comparison between Cohen and Brown-Knuth
maps.
Looking at the tracking results one can see that new $f(q)$
produces a system with surprisingly regular behavior.
However, detailed numerical experiments reveal multiple
island structures at small scales (see Fig. \ref{fig:Cohen} case
(c.)) indicating that analytic invariant doesn't exist.

%----------------------------------------------------------------
%----------------------------------------------------------------

%----------------------------------------------------------------
Thus, in a similar manner, one can first rewrite a general $n$-piece
linear function using one linear and $(n-1)$ modulus functions
in a form $k_i|q-q_i|$ where $q_i$ is a location of $i-th$ vertex,
then substitute all modulus with square roots
$k_i|q-q_i|\to k_i\sqrt{(q-q_i)^2+\epsilon^2}$, resulting in 
\[
\begin{array}{lll}
\ds n=2: & \ds\quad
    \frac{k_1 + k_2}{2}\,q +
    \frac{k_2 - k_1}{2}\,\sqrt{q^2+\epsilon^2} + d
                      &\ds \underset{\epsilon \to 0}{\to}\,\,
d + q\times\left\{\begin{array}{cc}
    k_1, & q <    0, \\
    k_2, & q \geq 0,
\end{array}\right.                                          \\[0.35cm]
\ds n=3: & \ds\quad
    \frac{k_1 + k_3}{2}\,q +
    \frac{k_2 - k_1}{2}\,\sqrt{q^2+\epsilon^2} +
    \frac{k_3 - k_2}{2}\,\sqrt{(q-1)^2+\epsilon^2} -
    \frac{k_3 - k_2}{2} + d
                      &\ds \underset{\epsilon \to 0}{\to}\,\,
d + q\times\left\{\begin{array}{cl}
    k_1, & q <    0,        \\
    k_2, & 0 \leq q < 1,    \\
    k_3, & q \geq 1,
\end{array}\right.                                          \\[0.25cm]
\ldots                & \qquad\ds\ldots.
\end{array}
\]
Fig.~\ref{fig:NearIntegr}    below shows examples of how the
``smoothening'' procedure can be used to produce new nearly
integrable systems.
More importantly, in addition to being near integrable, these
mappings are stable since limiting polygon at larger amplitudes
provides an isolating integral.

%----------------------------------------------------------------
Note that from a practical standpoint, the fact that produced
systems are quasi- and not exactly integrable is not a problem as
long as $\epsilon$ is sufficiently small and chaotic formations
are within needed tolerances. 
Even if a system is originally designed as integrable, small errors
(e.g. rounding errors for simulations, or tolerance errors for
physical experimental setups) will destroy integrability on a
small scale.
Also, the square root function is only one of infinitely many
ways of smoothing force function (e.g. one can use $\arctan$,
$\tanh$ or others) and is used here as an example.

%----------------------------------------------------------------
\begin{figure}[t!]\centering
\includegraphics[width=\columnwidth]{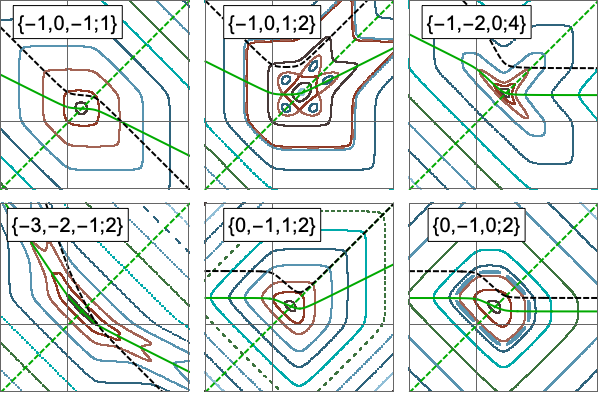}
\caption{\label{fig:NearIntegr}
    Examples of quasi-integrable systems produced by ``smoothening''
    3-piece integrable polygon maps using $\epsilon = 0.05$.
    For each plot the label shows parameters $\{k_1,k_2,k_3;d\}$.
    All level sets are obtained by tracking.
    The green curve is the second symmetry line $f(q)/2$ and the
    black dashed curve is the force function $f(q)$.
    }
\end{figure}
%----------------------------------------------------------------

\newpage
%===============================================================%
%===============================================================%
\subsection{\label{sec:Discrete}Discrete perturbation theory}

%----------------------------------------------------------------
Finally, knowledge of all possible integrable mappings with
polygon invariants will not only help to construct new near
integrable systems, but to understand dynamics and phase space of
more general non-integrable mappings with smooth $f(p)$.
This can be done by discretizing $f(p)$ by a piecewise linear
function;
so we have a perturbation theory where order is determined by
number of pieces and smallness parameter is defined by a smallest
piece of force function.
In order to demonstrate the idea, we provide a comparison of a
phase space for chaotic quadratic
\[
    f(p) = a\,q + q^2
\]
and cubic H\'enon maps
\[
    f(p) = a\,p + q^3
\]
around
specific resonances ($\nu = 1/4$ and stop band $\nu = 1/2$) and
corresponding polygon maps.
As one can see in Fig.~\ref{fig:REMAll} below, even low order
approximations can reveal  important qualitative dynamical features,
including shapes of trajectories in phase space and the topological
structure of the phase portrait.

%----------------------------------------------------------------
\begin{figure}[b!]\centering
\includegraphics[width=\columnwidth]{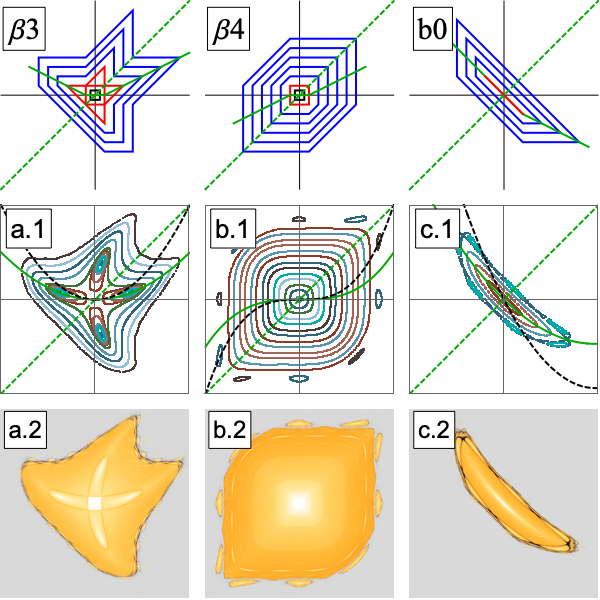}
\caption{\label{fig:REMAll}
    The middle row of plots (a.1 -- c.1) shows tracking for H\'enon
    map with quadratic $f(q) = a\,q + q^2$ (cases a. and c.),
    and cubic force functions $f(q) = a\,q + q^3$ (case b.).
    The value of parameter is (a.) $a=a_{1/4}-0.005$, (b.)
    $a=a_{1/4}+0.005$ and (c.) $a=a_{1/2}+0.05$, where $a_{1/4}=0$
    is the main octupole resonance with $\nu = 1/4$ and $a_{1/2}=-2$
    is the linear-half integer stopband $\nu=1/2$.
    The green curve is the second symmetry line $f(q)/2$ and the black
    dashed curve is the force function $f(q)$.
    All plots correspond to a range $p,q\in [-1,1]$.
    The top row shows corresponding low order approximation via
    mappings with polygon invariants.
    The bottom row of plots (a.2 -- c.2) shows $\log$ of SALI values
    for the same cases, by courtesy of Ivan Morozov (BINP).
	}
\end{figure}
%----------------------------------------------------------------

%----------------------------------------------------------------
An interesting observation can be made looking at the {\it smaller
alignment index} (SALI) color maps.
SALI is an indicator (alternative to traditional maximal Lyapunov
exponent) distinguishing between ordered and chaotic motion
\cite{skokos2003smaller}.
As one can see, there is a significant area around a fixed point
with almost the same color (white).
It means that in this area the motion is quasi-integrable and
almost linear (very small variation of rotation number).
One can note that this area resembles a separatrix of inner linear
layer from corresponding polygon map;
thus, we learn that linear dynamics in the inner layer of integrable
polygon map is not just a consequence of oversimplified
discretization of force function, but a reflection of actual
dynamics --- negligible spread of rotation numbers.
In contrast to smooth perturbation theory when in zeroth order
we have an ellipse $\K (p,q) = p^2 - a\,p\,q + q^2$, here we have
a polygon with degenerate motion.
While in both approaches the rotation number is independent of
amplitude, discrete perturbation theory gives better boundary of
linear area when system is close to some resonant condition, while
smooth perturbation theory will work better far away from any
major resonances or very close to the origin where ellipses are
not distorted.

\newpage
%===============================================================%
%===============================================================%
%===============================================================%
\section{\label{sec:Sum}Summary}

\begin{figure}[bh!]\centering
\includegraphics[width=\columnwidth]{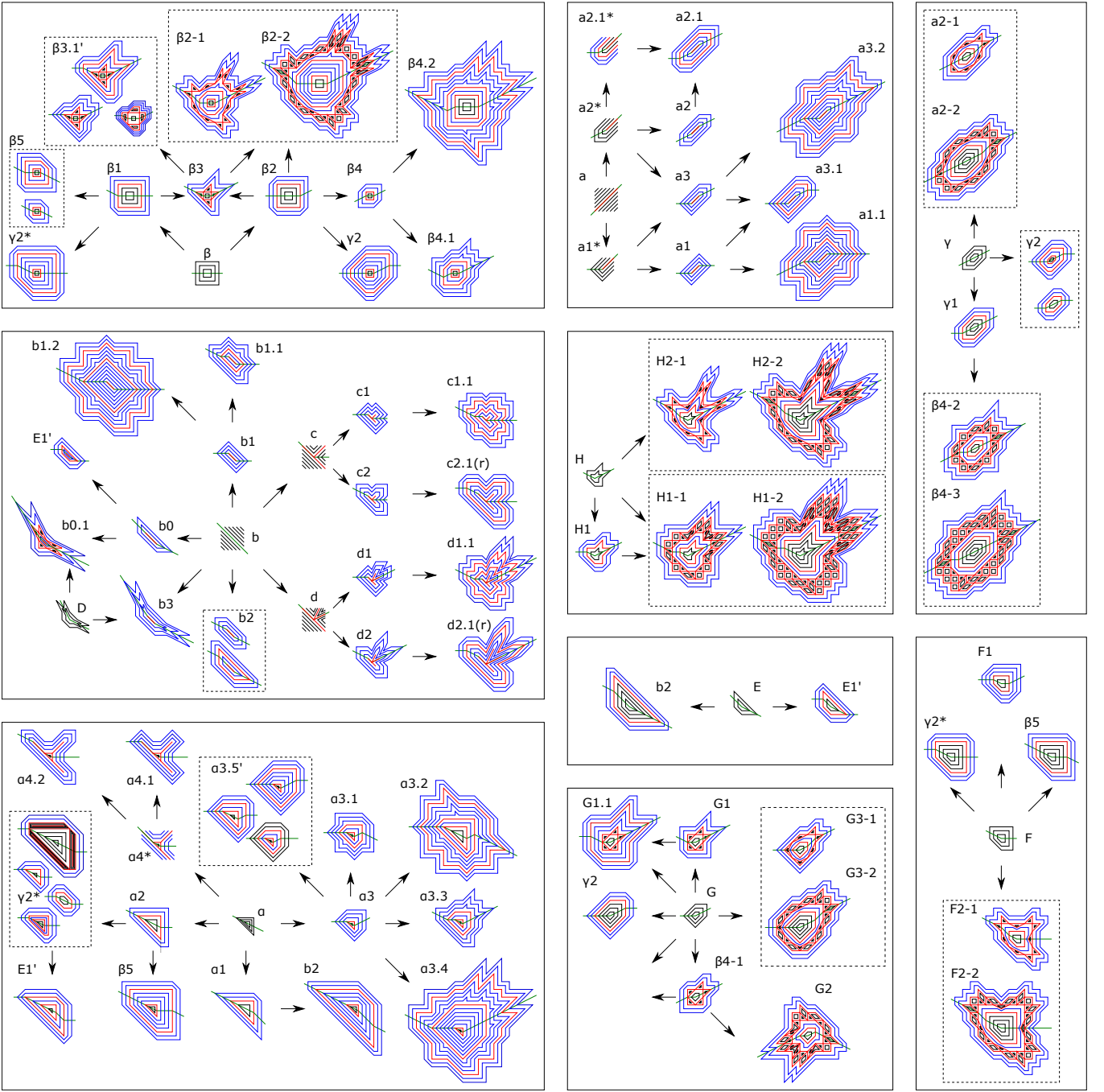}
\caption{\label{fig:PMaps4-All}
    Most (but not all) mappings discovered by an algorithm: 1-
    through 4-piece force functions.
    Maps are organized by dynamics around fixed point;
    systems with the same rotation number around fixed points are
    presented in the same solid box.
    Dashed boxes indicates that map belongs to discrete or
    continuous family.
	}
\end{figure}

%----------------------------------------------------------------
In this article, we presented an algorithm for automated discovery
of new integrable symplectic maps of the plane with polygon
invariants,
and as a result, we reported over 100 new integrable families.
The algorithm successfully rediscovered some famous discrete
Painlev\'e equations as well as some of McMillan-Suris integrable
mappings.
Fig.~\ref{fig:PMaps4-All} presents most, but not all, systems
discovered by the search algorithm.
They are grouped by similarity of dynamics around the fixed point
while arrows are pointed toward systems with more segments in $f$,
thus showing hierarchy.

%----------------------------------------------------------------
%----------------------------------------------------------------

\newpage
%----------------------------------------------------------------
While the discovered systems have a piecewise linear force and
polygon invariants, thus lacking the differentiability, we suggested
a ``smoothening'' procedure which can produce new quasi-integrable
systems without singularities.
Such ``smooth'' polygon maps have potential applications in physics,
engineering, dynamical system theory, and signal processing.
For example, in accelerator physics, such near-integrable systems
could reduce unwanted particle losses in accelerators and, thus,
improve their performance and the achievable beam intensities.
So far, the design of integrable accelerator focusing systems is at
its very beginning, even conceptually.
Fermilab presently operates the first accelerator~\cite{Antipov_2017},
designed to test some of these concepts.

%----------------------------------------------------------------
Finally, we introduced a discrete perturbation theory in order to
relate our polygon mappings to a general symplectic map of a plane.
The theory furthers understanding of shapes of phase space curves
and approximates dynamics for more realistic chaotic systems such
as H\'enon or Chirikov mappings or even integrable ones, such as
McMillan-Suris maps.

%----------------------------------------------------------------
It remains an open question whether one can understand the totality
of all maps with polygon invariants, and not only corresponding to
force functions with integers coefficients.
For example, it is not clear whether there exists some sort of
generating formula allowing to derive integrable maps with $n+1$
piece force function from mappings with $n$-pieces.

%===============================================================%
%===============================================================%
%===============================================================%
\section{Acknowledgments}

%----------------------------------------------------------------
The authors would like to thank Eric Stern (FNAL) and Taylor Nchako
(Northwestern University) for carefully reading this manuscript and
for their helpful comments.
Moreover, we would like to extend our gratitude to Ivan Morozov
(BINP) for multiple discussions and his generous contributions in
preparation of Subsection~\ref{sec:Discrete} and especially
Fig.~\ref{fig:REMAll}.
Finally we would like to express our deep appreciation for Nikolay
Kuropatkin's (FNAL) useful discussion on vertex count of polygons.

%----------------------------------------------------------------
Y. K.~acknowledges funding by the DoE ASCR Accelerated Research in
Quantum Computing program (award No.~DE-SC0020312), DoE QSA, NSF
QLCI (award No.~OMA-2120757), NSF PFCQC program, DoE ASCR Quantum
Testbed Pathfinder program (award No.~DE-SC0019040), U.S. Department
of Energy Award No.~DE-SC0019449, AFOSR, ARO MURI, AFOSR MURI, and
DARPA SAVaNT ADVENT.

%----------------------------------------------------------------
%----------------------------------------------------------------
This manuscript has been authored by Fermi Research Alliance, LLC
under Contract No. DE-AC02-07CH11359 with the U.S. Department of
Energy, Office of Science, Office of High Energy Physics.

%===============================================================%
%===============================================================%
%===============================================================%
%===============================================================%
\appendix
%===============================================================%
%===============================================================%
%===============================================================%
%===============================================================%

\newpage
%===============================================================%
%===============================================================%
%===============================================================%
\section{Calculation of action-angle variables for polygon mappings}
\label{secAPP:Act-Ang}

%===============================================================%
%===============================================================%
\subsection{Action}

%----------------------------------------------------------------
\begin{figure}[t!]\centering
\includegraphics[width=0.75\columnwidth]{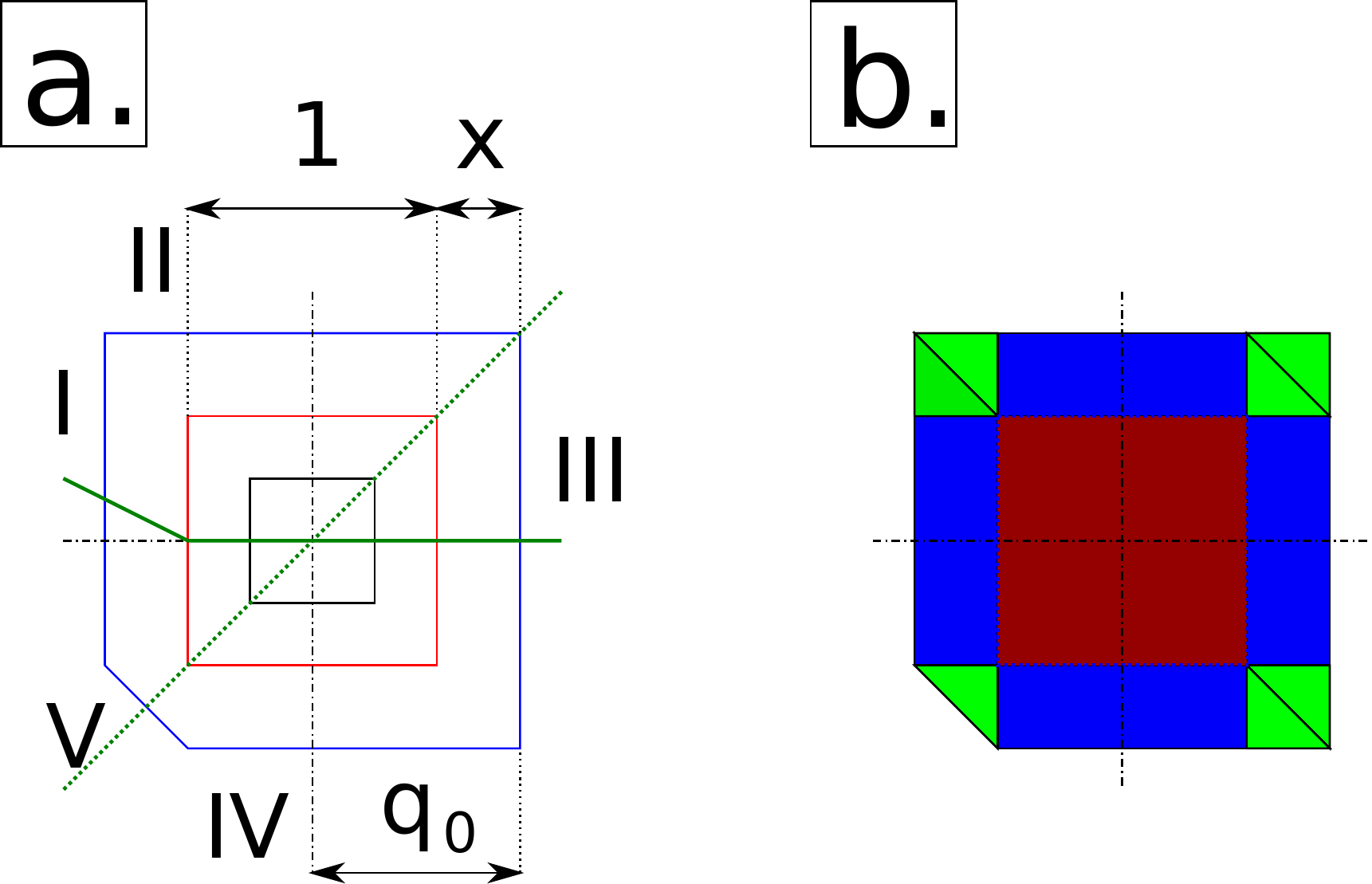}
\caption{\label{fig:JnuCalc}
	Calculation of action-angle variables for map $\beta$1.
    Plot (a.) shows three level sets of invariant: level set from
    linear layer (black), separatrix (red) and level set from
    nonlinear layer (blue).
    Solid and dotted green lines are second ($p=f(q)/2$) and first
    ($p=q$) symmetry lines.
    Segments of nonlinear trajectory are labeled with Roman numerals.
    Plot (b.) shows partition of the area under nonlinear trajectory
    into fundamental rectangles and triangles.
    Area of green triangles is proportional to initial condition
    squared, area of blue rectangles is linearly proportional and
    red square is independent of $q_0$.
	}
\end{figure}
%----------------------------------------------------------------

%----------------------------------------------------------------
Since invariant curves are polygons, the calculation of the area
enclosed under the trajectory is a trivial task.
Each polygon can be split into collection of simple rectangles and
(or) triangles.
Thus, the action integral (as well as partial action integral) has
the form
\begin{equation}
\label{math:Jform}
\begin{array}{l}
\ds J = \frac{1}{2\,\pi}\,\oint p\,\dd q =
    C_1 q_0^2 + C_2 q_0 + C_3,              \\[0.3cm]
\ds J'= \frac{1}{2\,\pi}\,\int_{q_0}^{q'_0} p\,\dd q =
    C'_1 q_0^2 + C'_2 q_0 + C'_3
\end{array}
\end{equation}
where $C^{(')}_{1,2,3}$ are some constants and $(q_0,0)$ is an initial
condition on the trajectory.
The first term represents rectangles (or triangles) with both sides
being proportional to initial conditions, the next term with only one
side of rectangle being proportional to $q_0$ and the other side
determined by map parameters, and the final term represents contribution
from fundamental rectangles independent of initial condition.

%----------------------------------------------------------------
As an example, below we will calculate action and partial action
for the outer layer of nonlinear trajectories of the map $\beta1$,
see Fig.~\ref{fig:JnuCalc}.
Introducing a new variable $x = |q_0 - 1/2|$ which represents a
horizontal distance from separatrix to nonlinear trajectory, we
have
\begin{equation}
\label{math:JBeta1}
\begin{array}{l}
\ds J = \frac{1}{2\,\pi}\,\left(\frac{7}{2}x^2 + 4x + 1\right),     \\[0.3cm]
\ds J'= \frac{1}{2\,\pi}\,\left(x^2 + x + \frac{1}{4}\right).
\end{array}
\end{equation}

%===============================================================%
%===============================================================%
\subsection{Rotation number and angle variable}

%----------------------------------------------------------------
The dynamics of map in action-angle variables is given by
\[
\begin{array}{rl}
    J_n         &\!\! = J_0            ,              \\[0.2cm]
    \theta_n    &\!\! = \theta_0 + 2\,\pi\,\nu\,n,
\end{array}
\]
where $\nu$ is rotation number and $\theta_0$ is an initial phase
which can be defined, let's say, as an initial polar angle on the
trajectory measured around fixed point
\[
    \theta_0 = \arctan \frac{p_0-p_f}{q_0-q_f}.
\]
Below we provide 2 methods of calculation of $\nu$;
both methods are based on Danilov theorem, see
\cite{zolkin2017rotation,nagaitsev2020betatron,mitchell2021extracting}.
In addition, along with the second method we will show how
to verify the integrability of the map.

%===============================================================%
\subsubsection{Method I}
%===============================================================%

%----------------------------------------------------------------
According to Danilov theorem, the rotation number is given as a
derivative of the partial action (Hamiltonian) over the action
variable.
Using chain rule for derivatives and Eq.~(\ref{math:Jform}), we
have
\[
    \nu = \frac{\dd J'}{\dd J}
        = \frac{\dd J'/\dd q_0}{\dd J/\dd q_0}
        = \frac{C'_2 + 2\,C'_1 q_0}{C_2 + 2
    \,C_1 q_0}.
\]
As one can see, for all mappings, the dependence of rotation number
on linear amplitude is always a rational function of a degree 1.
In addition, one can conclude that for all rational (irrational)
initial conditions $x\in\mathbb{Q}$ ($\in\mathbb{R}\setminus\mathbb{Q}$)
the rotation number is rational (irrational) as well, so the
trajectory is periodic (quasi-periodic).

%----------------------------------------------------------------
Again using mapping $\beta1$ as an example we have
\[
    \nu = \frac{4|q_0|}{1+14|q_0|} = \frac{1+2x}{4+7x},
    \qquad |q_0|>1/2.
\]
As we expect, for $x=0$ we have $\nu = 1/4$ matching rotation
number on the separatrix and inner linear layer (limiting case
for map $\beta$), and, for large amplitudes
$\ds\lim_{x\to\infty} \nu = 2/7$ (limiting case map F).

%===============================================================%
\subsubsection{Method II}
%===============================================================%

%----------------------------------------------------------------
Here we present another method which does not involve direct
calculation of action and partial action.
Instead, one can directly use Danilov formula
\[
\nu = \frac{\ds \oint       \left(\frac{\pd \K}{\pd p}\right)^{-1}\,\dd q}
           {\ds \int_q^{q'} \left(\frac{\pd \K}{\pd p}\right)^{-1}\,\dd q}
\]
where integrals are taken along the constant level set of invariant
$\K (p,q) = \const$.
Since level set is polygon, each integral can be seen as a sum of
integrals along its sides.
If polygon has vertical sides, corresponding integrals should be
replaced with (see~\cite{zolkin2017rotation,nagaitsev2020betatron})
\[
\int \left( \frac{\pd \K}{\pd p}\right)^{-1}\,\dd q
\qquad\rightarrow\qquad
\int \left(-\frac{\pd \K}{\pd q}\right)^{-1}\,\dd p.
\]

%----------------------------------------------------------------
Let's use the already familiar mapping $\beta$1 as an example.
First, we label sides of invariant polygon corresponding to nonlinear
layer using Roman numerals (see Fig.~\ref{fig:JnuCalc}).
Then, we write down the equation for each side of polygon using
$q_0: (0,q_0)\in\K(p,q)=\const$ (see Table~\ref{tab:Beta1Segments}).
In general one can use any point $(p_0,q_0)$ belonging to the
invariant curve, and, $p_0 = 0$ is chosen due to a convenience
without any loss of generality.
Now we can use initial condition $q_0$ as an invariant of motion
$\K(p,q) \equiv q_0$ and find it from the equation of the polygon segment
(third column in table).
Then, the integral in denominator can be calculated as
\[
\oint \left( \frac{\pd \K}{\pd p}\right)^{-1}\,\dd q =
\underbrace{\int_{-1/2}  ^{1/2+x}    1 \,\dd p}_{\mathrm{I}  } +
\underbrace{\int_{-1/2-x}^{1/2+x}    1 \,\dd q}_{\mathrm{II} } +
\underbrace{\int_{1/2+x} ^{-1/2-x} (-1)\,\dd p}_{\mathrm{III}} +
\underbrace{\int_{1/2+x} ^{-1/2}   (-1)\,\dd q}_{\mathrm{IV} } +
\underbrace{\int_{-1/2}  ^{-1/2-x} (-1)\,\dd q}_{\mathrm{V}  } =
4 + 7x.
\]
Using initial condition $(q_0,0)$, the integral in numerator
is
\[
\int_{(q_0,0)}^{(0,p_1)} \left( \frac{\pd \K}{\pd p}\right)^{-1}\,\dd q =
\underbrace{\int_{0}     ^{1/2+x} 1\,\dd p}_{\mathrm{I} } +
\underbrace{\int_{-1/2-x}^{0}     1\,\dd q}_{\mathrm{II}} =
1 + 2x.
\]
One can see that the resulting rotation number is the same as in Method I:
\[
    \nu = \frac{1 + 2x}{4 + 7x}.
\]

%----------------------------------------------------------------
Finally, using explicit expressions for each side of the polygon
we can verify an integrability of the map.
For that we will check that both symmetries are satisfied.
The first symmetry is that for the entire polygon we have:
\[
    \K(q,p) = \K(p,q).
\]
It is easy to see that vertical side I (III) is the symmetric pair
to side IV (II), while segment V is its own pair since its invariant
under $q \leftrightarrow p$.
The second symmetry is guaranteed by McMillan condition, i.e.
vertically opposite sides are equidistant from the second symmetry
line or alternatively the sum of opposite sides is equal to the
force function:
\[
\begin{array}{llll}
\mathrm{II+IV}&\!\!\!:\quad q_0 + (-q_0)           &\!\!= 0
    &\!\!= f(q>-1/2),           \\[0.25cm]
\mathrm{II+V} &\!\!\!:\quad q_0 + (- q - q_0 -1/2) &\!\!= -q -1/2
    &\!\!= f(q \leq -1/2).
\end{array}
\]
Additionally, vertical sides I and III are split in halves by
symmetry line.
Therefore we verify an integrability since both symmetries hold
for any $q_0$.
In a similar manner, all mappings presented in this article were
confirmed to be integrable.

%----------------------------------------------------------------
%----------------------------------------------------------------
\begin{table}[t!]
\centering
\begin{tabular}{p{3cm}p{4cm}p{4cm}p{3.1cm}p{3.1cm}}
\hline\hline
Side \#
    & Equation                  
    & $\K(q,p)\equiv q_0$ 
    & $(\pd \K / \pd p)^{-1}$
    & $-(\pd \K / \pd q)^{-1}$                      \\\hline
    &                   &           &       &       \\[-0.25cm]
I:  & $q=-q_0$          & $-q$      & ---   & 1     \\[0.25cm]
II: & $p= q_0$          & $ p$      & 1     & ---   \\[0.25cm]
III:& $q= q_0$          & $ q$      & ---   &-1     \\[0.25cm]
IV: & $p=-q_0$          & $-p$      &-1     & ---   \\[0.25cm]
V:  & $p=-q-q_0-1/2$    & $-p-q-1/2$&-1     & 1     \\[0.1cm]
\hline\hline
\end{tabular}
\caption{
        Equation of a side, expression for the invariant along
        with its partial derivatives for the polygon in outer
        nonlinear layer of the map $\beta 1$.
        }
\label{tab:Beta1Segments}
\end{table}   
%----------------------------------------------------------------
%----------------------------------------------------------------

\newpage
%===============================================================%
%===============================================================%
%===============================================================%
\section{Pseudo-code for the search algorithm}
\label{secAPP:Alg}

%----------------------------------------------------------------
%----------------------------------------------------------------
% Algorithm 1: Piecewise linear continuous integer force function
%----------------------------------------------------------------
%----------------------------------------------------------------
\begin{algorithm}[h!]
\DontPrintSemicolon
\SetAlgoLined
\SetKwInOut{Input}{Input}\SetKwInOut{Output}{Output}
\Input{$V^\mathrm{cutoff}\in\mathbb{Z}$ --- cutoff parameter for polygon vertex count,          \\
       $q_\mathrm{ini}^\mathrm{max}\in\mathbb{Z}\,\,
       (\sum_{i=2}^{n} l_i < q_\mathrm{ini}^\mathrm{max} < r_\mathrm{max})$ ---
       max initial condition.}
\Output{$integrable\in\mathbb{B}$ --- integrability of map.}
\BlankLine
\tcc{Map is integrable unless unstable or chaotic orbit is detected}
$integrable \leftarrow \mathrm{True}$                                                           \\
\tcc{Scan over all initial conditions from fixed point $q^*$ to $q_\mathrm{max}$}
{\bf for} $q=q^*,\ldots,q_\mathrm{ini}^\mathrm{max}$                                            \\
\{                                                                                              \\
    $\quad$         \tcc{Compute number of vertexes for given orbit}
    $\quad$         $V \leftarrow vertex(q)$                                                    \\
    $\quad$         \tcc{If motion is stable}
    $\quad$         {\bf if} $V \neq -1$ {\bf then}                                             \\
    $\qquad$        \tcc{If the number of vertices in the orbit polygon is larger than
                         a cutoff value, $V^\mathrm{cutoff}$}
    $\qquad$        {\bf if} $V > V^\mathrm{cutoff}$ {\bf then}                                \\
    $\quad\qquad$        \tcc{Exclude map if any of orbits is chaotic}
    $\quad\qquad$   $\,\!integrable \leftarrow \mathrm{False}$                                  \\
    $\quad$        \tcc{Exclude map if any of orbits is unstable}
    $\quad$        {\bf else} $integrable \leftarrow \mathrm{False}$                            \\
\{                                                                                              \\
\BlankLine
\Return{$integrable$}
%----------------------------------------------------------------
\caption{Map analyzer.
        Algorithm verifies if a map is integrable with invariant level sets
        being polygons.
        Algorithm needs function $vertex(q)$ and force function.}
\end{algorithm}
%----------------------------------------------------------------
%----------------------------------------------------------------

%----------------------------------------------------------------
%----------------------------------------------------------------
% Algorithm 1: Piecewise linear continuous integer force function
%----------------------------------------------------------------
%----------------------------------------------------------------
\begin{algorithm}[h!]
\DontPrintSemicolon
\SetAlgoLined
\SetKwInOut{Input}{Input}\SetKwInOut{Output}{Output}
\Input{$n \equiv \mathrm{dim}\,\mathbf{k} \geq 2$ --- number of segments,       \\
        $q\in\mathbb{R}$ --- coordinate,        \\
        $\mathbf{k}=(k_1,k_2,\ldots,k_n)\in\mathbb{Z}^n$ ---
        vector of slopes of segments,           \\
        $\mathbf{l}\,\,\,\!=(l_2,l_3,\ldots,l_{n-1})\in\mathbb{R}^{n-2}$ ---
        vector of finite segments' length.}
\Output{$f(q,\mathbf{k},\mathbf{l})\in\mathbb{R}$ ---
        value of force function.}
\BlankLine
\tcc{Compute force function value depending on segment}
{\bf switch} $q$                                                                                \\
\{                                                                                              \\
    $\quad$     \tcc{1-st segment}
    $\quad$     {\bf case $q \leq 0$}                                                           \\
    $\qquad$        $f \leftarrow k_1 q$                                                        \\
    $\quad$     \tcc{2-nd segment}
    $\quad$     {\bf case $0                    < q \leq l_2$}                                  \\
    $\qquad$        $f \leftarrow k_2 q$                                                        \\
    $\quad$     \tcc{3-rd segment}
    $\quad$     {\bf case $l_2                  < q \leq l_2 + l_3$}                            \\
    $\qquad$        $f \leftarrow k_3(q-l_2) + k_2 l_2$                                         \\
%    $\quad$     \tcc{4-th segment}
%    $\quad$     {\bf case $l_2 + l_3            < q \leq l_2 + l_3 + l_4$}                      \\
%    $\qquad$        $f \leftarrow k_4(q-[l_2+l_3]) + k_2 l_2 + k_3 l_3$                         \\
    $\quad$     \ldots                                                                          \\
    $\quad$     \tcc{$(n-1)$-th segment}
    $\quad$     {\bf case $\sum_{i=2}^{n-2} l_i < q \leq \sum_{i=2}^{n-1} l_i$}                 \\
    $\qquad$        $f \leftarrow k_{n-1}(q-\sum_{i=2}^{n-2} l_i) + \sum_{i=2}^{n-2} k_i l_i$   \\
    $\quad$     \tcc{$n$-th segment}
    $\quad$     {\bf case $                       q > q \sum_{i=2}^{n-1} l_i$}                  \\
    $\qquad$        $f \leftarrow k_n(q-\sum_{i=2}^{n-1} l_i) + \sum_{i=2}^{n-1} k_i l_i$       \\
\}                                                                                              \\
\BlankLine
\Return{$f$}
%----------------------------------------------------------------
\caption{Continuous piecewise linear force function
$f(q,\mathbf{k},\mathbf{l})$ with first vertex at the origin.
\[
f(l_{s-1} < q \leq l_s,\mathbf{k},\mathbf{l}) =
k_s(q-\sum_{i=2}^{s-1}l_i) + \sum_{i=2}^{s-1}k_i l_i,
\qquad\qquad
l_1 = 0,\,l_{0,n} = \mp \infty.
\]
}
\end{algorithm}
%----------------------------------------------------------------
%----------------------------------------------------------------

%----------------------------------------------------------------
%----------------------------------------------------------------
% Algorithm 2: Orbit quantifier
%----------------------------------------------------------------
%----------------------------------------------------------------
\begin{algorithm}[h!]
\DontPrintSemicolon
\SetAlgoLined
\SetKwInOut{Input}{Input}\SetKwInOut{Output}{Output}
\Input{ $\!\!q_\mathrm{ini}\in\mathbb{R}$ --- initial coordinate on a second
        symmetry line,                                                      \\
        $\mathbf{k}=(k_1,k_2,\ldots,k_n)\in\mathbb{Z}^n$ ---
        vector of slopes of segments,                                       \\
        $\mathbf{l}\,\,\,\!=(l_2,l_3,\ldots,l_{n-1})\in\mathbb{R}^{n-2}$ ---
        vector of finite segments' length,                                  \\
        $d\in\mathbb{R}$ --- force function shift parameter,                \\
        $N\in\mathbb{Z}$ --- number of iterations,                          \\
        $r_\mathrm{max} \in \mathbb{R}\,\,(\gg \sum_{i=2}^{n-1} l_i)$ ---
        max distance allowed in simulation,                                 \\
        $\eta \in \mathbb{R}\setminus\mathbb{Q}\,\,(\ll 1)$ --- small
        irrational parameter to shift initial conditions,                   \\
        $\epsilon \in \mathbb{R}\,\,(\ll 1)$ --- collinearity cutoff parameter.
        }
\Output{$V$ --- number of vertex of orbit ($V=-1$ if orbit is unstable).}

\BlankLine

$V \leftarrow 0$                                                            \\
\tcc{Initial condition on 2-nd symmetry line with small irrational shift}
$q \leftarrow q_\mathrm{ini} + \eta$                                        \\
$p \leftarrow [f(q,\mathbf{k},\mathbf{l})+d]/2$                             \\
\tcc{Orbit, $\vec{\zeta} = (\zeta_1,\zeta_2,\ldots,\zeta_N)$}
{\bf for} $i=1,\ldots,N$                                                    \\
\{                                                                          \\
    $\quad$         $\zeta_i \leftarrow (q,p)$                              \;
    $\quad$         $q' \leftarrow p$                                       \;
    $\quad$         $p' \leftarrow-q + f(p) + d $                           \;
    $\quad$         $q \leftarrow q'$                                       \;
    $\quad$         $p \leftarrow p'$                                       \;
    $\quad$         \tcc{Stop if radius of point in the phase space is $>r_\mathrm{max}$}
    $\quad$         {\bf if} $q^2+p^2 > r_\mathrm{max}^2$ {\bf then}        \\
    $\quad$         \{                                                      \\
    $\qquad$            $V \leftarrow -1$                                   \\
    $\qquad$            {\bf break}                                         \\
    $\quad$         \}                                                      \\
    $\quad$         {\bf else}                                              \\
    $\quad$         \{                                                      \\    
    $\qquad$            \tcc{Stop if orbit is periodic}
    $\qquad$            {\bf if}
                        $q' = q_\mathrm{ini} \land p' = [f(q_\mathrm{ini},\mathbf{k},\mathbf{l})+d]/2$
                        {\bf then}                                          \\
    $\quad\qquad$       {\bf break}                                         \\
    $\quad$         \}                                                      \\
\}                                                                          \\

\BlankLine

\tcc{Compute number of polygon vertexes, $V$, if the orbit is stable ($V \neq -1$)}
{\bf if} $V \neq -1$ {\bf then}  \\
\{          \\
    $\quad$        Sort $\vec{\zeta} = (\zeta_1,\ldots,\zeta_N)$:
                    $\forall\,0 < k < N \quad
                    \left[\arctan\frac{p_k}{q_k} >
                    \arctan\frac{p_{k+1}}{q_{k+1}}\right]$\;
    $\quad$        Add $\zeta_1$ and $\zeta_2$ to the tail of $\vec{\zeta}$:
                    $\vec{\zeta} \leftarrow
                    (\zeta_1,\zeta_2,\ldots,\zeta_{N-1},\zeta_N,\zeta_1,\zeta_2)$\;
    $\quad$        {\bf for} $j=1,\ldots,N$         \\
    $\quad$        \{          \\
    $\qquad$        \tcc{Compute dot product between displacement vectors}
   $\qquad$      $\ds\alpha_j = \frac{(p_{j+2} - p_{j+1} ) (p_{j+1} - p_{j}) +
   (q_{j+2} - q_{j+1}) (q_{j+1} - q_j)}{\sqrt{(p_{j+2} - p_{j+1} )^2 + (q_{j+2} - q_{j+1} )^2 }\sqrt{(p_{j+1} - p_{j} )^2 + (q_{j+1} - q_{j} )^2}}$\;
     $\qquad$        \tcc{New polygon vertex detection}   
     $\qquad$       {\bf if} $||\alpha_j| - 1| > \epsilon$ {\bf then}   \\
     $\quad\qquad$  V++     \\
    $\quad$        \}          \\
\}          \\

\Return{$V$}
%----------------------------------------------------------------
\caption{Function $vertex(q)$. Algorithm counts number of vertexes
$V$ for a given orbit with initial condition on 2-nd symmetry line
$(q,p) = (q_\mathrm{ini},f(q_\mathrm{ini})/2)$ or returns value of
$(-1)$ if trajectory is unstable.}
\end{algorithm}
%----------------------------------------------------------------
%----------------------------------------------------------------

\newpage
%===============================================================%
%===============================================================%
%===============================================================%
\section{Action-angle variables for 4-piece mappings}
\label{secAPP:Nu}

%----------------------------------------------------------------
In Tables~\ref{tab:PMaps4-I1}--\ref{tab:PMaps4-C2} below we provide
rotation number $\nu$, action $J$ and partial action $J'$ for integer
integrable mappings with 4-piece force function.

%----------------------------------------------------------------
%----------------------------------------------------------------
\begin{table}[h]
\centering
\begin{tabular}{p{2cm}p{5cm}p{5cm}p{5cm}}
\hline\hline
Map         & $\nu_1,\,J_1,\,J'_1$              & $\nu_2,\,J_2,\,J'_2$                  & $\nu_3,\,J_3,\,J'_3$                  \\\hline
            &                                   &                                       &                                       \\[-0.25cm]
$\alpha$3.1 & $\frac{1+x}{3+5x},\,x\in[0;1]$    & $\frac{2+2y}{8+7y},\,y\in[0;\infty]$  &                                       \\[0.2cm]
            & $\frac{5}{2}x^2+3x+\frac{1}{2}$   & $\frac{7}{2}y^2+8y+J_1(1)$            &                                       \\[0.2cm]
            & $\frac{1}{2}x^2+x+\frac{1}{6}$    & $y^2+2y+J'_1(1)$                      &                                       \\[0.2cm]
$\alpha$3.2 & $\frac{3+x}{9+5x},\,x\in[0;1]$    & $\frac{4+3y}{14+10y},\,y\in[0;2]$     & $\frac{10+2z}{34+7z},\,z\in[0;\infty]$\\[0.2cm]
            & $\frac{5}{2}x^2+9x+\frac{9}{2}$   & $5y^2+14y+J_1(1)$                     & $\frac{7}{2}z^2+34z+J_2(2)$           \\[0.2cm]
            & $\frac{1}{2}x^2+3x+\frac{3}{2}$   & $\frac{3}{2}y^2+4y+J'_1(1)$           & $z^2+10z+J'_2(2)$                     \\[0.2cm]
$\alpha$3.3 & $\frac{1+x}{3+5x},\,x\in[0;1]$    & $\frac{2+2y}{8+9y},\,y\in[0;\infty]$  &                                       \\[0.2cm]
            & $\frac{5}{2}x^2+3x+\frac{1}{2}$   & $\frac{9}{2}y^2+8y+J_1(1)$            &                                       \\[0.2cm]
            & $\frac{1}{2}x^2+x+\frac{1}{6}$    & $y^2+2y+J'_1(1)$                      &                                       \\[0.2cm]
$\alpha$3.4 & $\frac{1+x}{3+5x},\,x\in[0;3]$    & $\frac{4+3y}{18+14y},\,y\in[0;2]$     & $\frac{10+2z}{46+9z},\,z\in[0;\infty]$\\[0.2cm]
            & $\frac{5}{2}x^2+3x+\frac{1}{2}$   & $7y^2+18y+J_1(3)$                     & $\frac{9}{2}z^2+46z+J_2(2)$           \\[0.2cm]
            & $\frac{1}{2}x^2+x+\frac{1}{6}$    & $\frac{3}{2}y^2+4y+J'_1(3)$           & $z^2+10z+J'_2(2)$                     \\[0.2cm]
$\beta$4.1  & $\frac{1+x}{4+6x},\,x\in[0;1]$    & $\frac{2+2y}{10+9y},\,y\in[0;\infty]$ &                                       \\[0.2cm]
            & $3x^2+4x+1$                       & $\frac{9}{2}y^2+10y+J_1(1)$           &                                       \\[0.2cm]
            & $\frac{1}{2}x^2+x+\frac{1}{4}$    & $y^2+2y+J'_1(1)$                      &                                       \\[0.2cm]
$\beta$4.2  & $\frac{3+x}{12+6x},\,x\in[0;1]$   & $\frac{4+3y}{18+13y},\,y\in[0;2]$     & $\frac{10+2z}{44+9z},\,z\in[0;\infty]$\\[0.2cm]
            & $3x^2+12x+9$                      & $\frac{13}{2}y^2+18y+J_1(1)$          & $\frac{9}{2}z^2+44z+J_2(2)$           \\[0.2cm]
            & $\frac{1}{2}x^2+3x+\frac{9}{4}$   & $\frac{3}{2}y^2+4y+J'_1(1)$           & $z^2+10z+J'_2(2)$                     \\[0.2cm]

$\delta$2   & $\frac{1+x}{3+2x},\,x\in[0;1/2]$  & $\frac{3+16y}{8+42y},\,y\in[0;1/2]$   & $\frac{11+3z}{29+8z},\,z\in[0;\infty]$\\[0.2cm]
            & $x^2+3x+3$                        & $21y^2+8y+J_1(1/2)$                   & $4z^2+29z+J_2(1/2)$                   \\[0.2cm]
            & $\frac{1}{2}x^2+x+1$              & $8y^2+3y+J'_1(1/2)$                   & $\frac{3}{2}z^2+11z+J'_2(1/2)$                     \\[0.2cm]

E2          & $\frac{3+5x}{8+14x},\,x\in[0;1]$  & $\frac{4}{11},\,y\in[0;1]$            & $\frac{8+3z}{22+8z},\,z\in[0;\infty]$ \\[0.2cm]
            & $7x^2+8x+4$                       & ---                                   & $4z^2+22z+11+J_1(1)$                  \\[0.2cm]
            & $\frac{5}{2}x^2+3x+\frac{3}{2}$   & ---                                   & $\frac{3}{2}z^2+8z+4+J'_1(1)$         \\[0.2cm]
G2          & $\frac{1+2x}{5+8x},\,x\in[0;1]$   & $\frac{3}{13},\,y\in[0;1]$            & $\frac{6+2z}{26+9z},\,z\in[0;\infty]$ \\[0.2cm]
            & $4x^2+5x+\frac{5}{2}$             & ---                                   & $\frac{9}{2}z^2+26z+\frac{39}{2}+J_1(1)$\\[0.2cm]
            & $x^2+x+\frac{1}{2}$               & ---                                   & $z^2+6z+\frac{9}{2}+J'_1(1)$          \\[0.1cm]
\hline\hline
\end{tabular}
\caption{
        Rotation number, $\nu_i$, action, $J_i$, and partial
        action, $J_i'$, for the $i$-th layer of the integer
        polygon mappings with 4-piece force and integrable for 
        isolated integer values of shift parameter, $d$, and
        ratio of finite pieces, $r$. Part 1.
        }
\label{tab:PMaps4-I1}
\end{table}   
%----------------------------------------------------------------
%----------------------------------------------------------------

%----------------------------------------------------------------
%----------------------------------------------------------------
\begin{table}[t!]
\centering
\begin{tabular}{p{2cm}p{5cm}p{5cm}p{5cm}}
\hline\hline
Map     & $\nu_1,\,J_1,\,J'_1$          & $\nu_2,\,J_2,\,J'_2$              & $\nu_3,\,J_3,\,J'_3$                      \\\hline
        &                               &                                   &                                           \\[-0.25cm]
a1.1    & $\frac{x}{2+4x},\,x\in[0;2]$  & $\frac{2+2y}{10+9y},\,y\in[0;2]$  & $\frac{6+z}{28+5z},\,z\in[0;\infty]$      \\[0.2cm]
        & $2x^2+2x$                     & $\frac{9}{2}y^2+10y+J_1(2)$       & $\frac{5}{2}z^2+28z+J_2(2)$               \\[0.2cm]
        & $\frac{1}{2}x^2$              & $y^2+2y+J'_1(2)$                  & $\frac{1}{2}z^2+6z+J'_2(2)$               \\[0.2cm]
a3.2    & $\frac{x}{2+5x},\,x\in[0;2]$  & $\frac{2+2y}{12+11y},\,y\in[0;2]$ & $\frac{6+z}{34+6z},\,z\in[0;\infty]$      \\[0.2cm]
        & $\frac{5}{2}x^2+2x$           & $\frac{11}{2}y^2+12y+J_1(2)$      & $3z^2+34z+J_2(2)$                         \\[0.2cm]
        & $\frac{1}{2}x^2$              & $y^2+2y+J'_1(2)$                  & $\frac{1}{2}z^2+6z+J'_2(2)$               \\[0.2cm]
b1.1    & $\frac{1+x}{2+4x},\,x\in[0;1]$& $\frac{2+2y}{6+7y},\,y\in[0;\infty]$ &                                        \\[0.2cm]
        & $2x^2+2x$                     & $\frac{7}{2}y^2+6y+J_1(1)$        &                                           \\[0.2cm]
        & $\frac{1}{2}x^2+x$            & $y^2+2y+J'_1(1)$                  &                                           \\[0.2cm]
b1.2    & $\frac{1+x}{2+4x},\,x\in[0;3]$& $\frac{4+3y}{14+11y},\,y\in[0;2]$ & $\frac{10+2z}{36+7z},\,z\in[0;\infty]$    \\[0.2cm]
        & $2x^2+2x$                     & $\frac{11}{2}y^2+14y+J_1(3)$      & $\frac{7}{2}z^2+36z+J_2(2)$               \\[0.2cm]
        & $\frac{1}{2}x^2+x$            & $\frac{3}{2}y^2+4y+J'_1(3)$       & $z^2+10z+J'_2(2)$                         \\[0.2cm]
c1.1    & $\frac{2+x}{6+4x},\,x\in[0;2]$& $\frac{4+3y}{14+11y},\,y\in[0;2]$ & $\frac{10+2z}{36+7z},\,z\in[0;\infty]$    \\[0.2cm]
        & $2x^2+6x$                     & $\frac{11}{2}y^2+14y+J_1(2)$      & $\frac{7}{2}z^2+36z+J_2(2)$               \\[0.2cm]
        & $\frac{1}{2}x^2+2x$           & $\frac{3}{2}y^2+4y+J'_1(2)$       & $z^2+10z+J'_2(2)$                         \\[0.2cm]
d1.1    & $\frac{2+x}{8+5x},\,x\in[0;2]$& $\frac{4+3y}{18+14y},\,y\in[0;2]$ & $\frac{10+2z}{46+9z},\,z\in[0;\infty]$    \\[0.2cm]
        & $\frac{5}{2}x^2+8x$           & $7y^2+18y+J_1(2)$                 & $\frac{9}{2}z^2+46z+J_2(2)$               \\[0.2cm]
        & $\frac{1}{2}x^2+2x$           & $\frac{3}{2}y^2+4y+J'_1(2)$       & $z^2+10z+J'_2(2)$                         \\[0.1cm]
\hline\hline
\end{tabular}
\caption{
        Rotation number, $\nu_i$, action, $J_i$, and partial
        action, $J_i'$, for the $i$-th layer of the integer
        polygon mappings with 4-piece force and integrable for 
        isolated integer values of shift parameter, $d$, and
        ratio of finite pieces, $r$. Part 2.
        }
\label{tab:PMaps4-I2}
\end{table}
%----------------------------------------------------------------
%----------------------------------------------------------------

%----------------------------------------------------------------
%----------------------------------------------------------------
\begin{table}[h!]
\centering
\begin{tabular}{p{2.4cm}p{2.45cm}p{4.75cm}p{2.45cm}p{4.75cm}}
\hline\hline
Map             & $\nu_0$       & $\nu_1$                             & $\nu_\mathrm{isl}$, $\#_C$  &
    $\nu_2$                                                                                         \\\hline
                &               &                                     &                             &
                                                                                                    \\[-0.25cm]
F2-$n$          & $\frac{2}{7}$ & $\frac{2n+2x}{7n+6x},\,x\in[0;1/2]$ & $\frac{1+2n}{3+7n}$, $n$    &
    $\frac{2(1+2n)+2y}{2(3+7n)+7y},\,y\in[0;\infty]$                                                \\[0.2cm]
G3-$n$          & $\frac{1}{5}$ & $\frac{n}{5n+2x},\,x\in[0;1/2]$     & $\frac{n}{1+5n}$, $n$       &
    $\frac{2n+y}{2(1+5n)+5y},\,y\in[0;\infty]$                                                      \\[0.2cm]
H1-$n$          & $\frac{2}{9}$ & $\frac{2n+x}{9n+5x},\,x\in[0;1]$    & $\frac{1+2n}{5+9n}$, $n+1$  &
    $\frac{2(1+2n)+2y}{2(5+9n)+9y},\,y\in[0;\infty]$                                                \\[0.2cm]
H2-$n$          & $\frac{2}{9}$ & $\frac{2n+2x}{9n+8x},\,x\in[0;1/2]$ & $\frac{1+2n}{4+9n}$, $n$    &
    $\frac{2(1+2n)+2y}{2(4+9n)+9y},\,y\in[0;\infty]$                                                \\[0.1cm]
\hline\hline
\end{tabular}
\caption{Rotation number, $\nu_i$, for the $i$-th layer of the
        integer polygon mappings with 4-piece force and integrable
        for isolated integer value of shift parameter $d$ and
        discrete ratio of finite pieces, $r\in\mathbb{N}^+$.
        $\nu_\mathrm{isl}$ and $\#_C$ are the rotation number in mode-locked area
        and number of chains in group of linear islands, respectively.
        Part 1.}
\label{tab:PMaps4-D1}
\end{table}
%----------------------------------------------------------------
%----------------------------------------------------------------

%----------------------------------------------------------------
%----------------------------------------------------------------
\begin{table}[h!]
\centering
\begin{tabular}{p{2.4cm}p{1.5cm}p{3.2cm}p{1.5cm}p{3.2cm}p{1.5cm}p{3.2cm}}
\hline\hline
Map                     & $\nu_0$       & $\nu_1$                           & $\nu_\mathrm{sep}$    &
    $\nu_2$                          & $\nu_\mathrm{isl}$, $\#_C$   & $\nu_3$                                           \\\hline
                        &               &                                   &                       &
                                     &                              &                                                   \\[-0.25cm]
$\beta$2-$n$            & $\frac{1}{4}$ & $\frac{n+x}{4n+5x},\,x\in[0;n]$   & $\frac{2}{9}$         &
    $\frac{2n+y}{9n+4y},\,y\in[0;1]$ & $\frac{1+2n}{4+9n}$, $n$     & $\frac{2(1+2n)+2z}{2(4+9n)+9z},\,z\in[0;\infty]$  \\[0.1cm]
\hline\hline
\end{tabular}
\caption{Rotation number, $\nu_i$, for the $i$-th layer of the
        integer polygon mappings with 4-piece force and integrable
        for isolated integer value of shift parameter $d$ and
        discrete ratio of finite pieces, $r\in2\mathbb{N}^+$ and
        $n=r/2$.
        $\nu_\mathrm{isl}$ and $\#_C$ are the rotation number in mode-locked area
        and number of chains in group of linear islands, respectively.
        Part 2.}
\label{tab:PMaps4-D2}
\end{table}
%----------------------------------------------------------------
%----------------------------------------------------------------

%----------------------------------------------------------------
%----------------------------------------------------------------
\begin{table}[ht!]
\centering
\begin{tabular}{p{2cm}p{3cm}p{6cm}p{6cm}}
\hline\hline
Map         &                   & $\nu_1,\,J_1,\,J'_1$              & $\nu_2,\,J_2,\,J'_2$                          \\\hline
            &                   &                                   &                                               \\[-0.25cm]
$\alpha$3.1'&                   & $\frac{1+x}{3+5x},\,x\in[0;r]$    & $\frac{1+r+y}{3+5r+4y},\,y\in[0;\infty]$      \\[0.2cm]
            &                   & $\frac{5}{2}x^2+3x+\frac{1}{2}$   & $2y^2+(3+5r)y+J_1(r)$                         \\[0.2cm]
            &                   & $\frac{1}{2}x^2+x+\frac{1}{6}$    & $\frac{1}{2}y^2+(1+r)y+J'(r)$                 \\[0.2cm]
$\beta$3.1' & $(r\leq 1)$       & ---                               & $\frac{1+r+y}{4+4r+5y},\,y\in[0;\infty]$      \\[0.2cm]
            &                   & ---                               & $\frac{5}{2}y^2+(4+4r)y+1+4r-2r^2$            \\[0.2cm]
            &                   & ---                               & $\frac{1}{2}y^2+(1+r)y+\frac{1}{4}+r-\frac{r^2}{2}$   \\[0.2cm]
            & $(r>1)$           & $\frac{2+2x}{8+9x},\,x\in[0;r-1]$ & $\frac{2r+y}{9r-1+5y},\,y\in[0;\infty]$       \\[0.2cm]
            &                   & $\frac{9}{2}x^2+8x+3$             & $\frac{5}{2}y^2+(9r-1)y+J_1(r-1)$             \\[0.2cm]
            &                   & $x^2+2x+\frac{3}{4}$              & $\frac{1}{2}y^2+2r\,y+J'_1(r-1)$              \\[0.2cm]
E1'         & $(2l_1\leq l_2)$  & $\frac{l_2-2l_1+3x}{3l_2-6l_1+8x},\,x\in[0;l_1]$
                                & $\frac{l_1+l_2+y}{2l_1+3l_2+4y},\,y\in[0;\infty]$                                 \\[0.2cm]
            &                   & $4x^2+3(l_2-2l_1)x+\frac{(l_2-2l_1)^2}{2}$
                                & $2y^2+(2l_1+3l_2)y+J_1(l_1)$                                                      \\[0.2cm]
            &                   & $\frac{3}{2}x^2+(l_2-2l_1)x+\frac{(l_2-2l_1)^2}{6}$
                                & $\frac{1}{2}y^2+(l_1+l_2)y+J'_1(l_1)$                                             \\[0.2cm]
            & $(2l_1>l_2)$      & $\frac{l_1-\frac{l_2}{2}+3x}{2l_1-l_2+8x},\,x\in[0;l_2/2]$
                                & $\frac{l_1+l_2+y}{2l_1+3l_2+4y},\,y\in[0;\infty]$                                 \\[0.2cm]
            &                   & $4x^2+2(l_1-\frac{l_2}{2})x$      & $2y^2+(2l_1+3l_2)y+J_1(l_2/2)$                \\[0.2cm]
            &                   & $\frac{3}{2}x^2+(l_1-\frac{l_2}{2})x$ & $\frac{1}{2}y^2+(l_1+l_2)y+J'_1(l_2/2)$   \\[0.2cm]
F.1'        & $(r\leq 1)$       & $\frac{1-r+2x}{3(1-r)+7x},\,x\in[0;r]$& $\frac{1+r+y}{3+4r+5y},\,y\in[0;\infty]$  \\[0.2cm]
            &                   & $\frac{7}{2}x^2+3(1-r)x+\frac{1}{2}(1-r)^2$& $\frac{5}{2}y^2+(3+4r)y+J_1(r)$      \\[0.2cm]
            &                   & $x^2+(1-r)x+\frac{1}{6}(1-r)^2$   & $\frac{1}{2}y^2+(1+r)y+J'_1(r)$               \\[0.2cm]
            & $(r>1)$           & $\frac{r-1+2x}{4(r-1)+7x},\,x\in[0;1]$& $\frac{1+r+y}{3+4r+5y},\,y\in[0;\infty]$  \\[0.2cm]
            &                   & $\frac{7}{2}x^2+4(r-1)x+(r-1)^2$  & $\frac{5}{2}y^2+(3+4r)y+J_1(1)$               \\[0.2cm]
            &                   & $x^2+(r-1)x+\frac{1}{4}(r-1)^2$   & $\frac{1}{2}y^2+(1+r)\,y+J'_1(1)$             \\[0.1cm]
\hline\hline
\end{tabular}
\caption{Rotation number, $\nu_i$, action, $J_i$, and partial
        action, $J_i'$, for the $i$-th layer of the integer
        polygon mappings with 4-piece force and integrable for 
        isolated value of shift parameter $d$ and
        continuous ratio of finite pieces, $r\in\mathbb{R}^+$. Part 1.}
\label{tab:PMaps4-C1}
\end{table}
%----------------------------------------------------------------
%----------------------------------------------------------------

%----------------------------------------------------------------
%----------------------------------------------------------------
\begin{table}[h!]
\centering
\begin{tabular}{p{5cm}p{6cm}p{6cm}}
\hline\hline
Map         & $\nu_1,\,J_1,\,J'_1$              & $\nu_2,\,J_2,\,J'_2$                      \\\hline
            &                                   &                                           \\[-0.25cm]
a2.1        & $\frac{x}{2+6x},\,x\in[0;r]$      & $\frac{r+y}{2+6r+5y},\,y\in[0;\infty]$    \\[0.2cm]
            & $3x^2+2x$                         & $\frac{5}{2}y^2+(2+6r)y+J_1(r)$           \\[0.2cm]
            & $\frac{1}{2}x^2$                  & $\frac{1}{2}y^2+r\,y+J_1'(r)$             \\[0.2cm]
a3.1        & $\frac{x}{2+5x},\,x\in[0;r]$      & $\frac{r+y}{2+5r+4y},\,y\in[0;\infty]$    \\[0.2cm]
            & $\frac{5}{2}x^2+2x$               & $2y^2+(2+5r)y+J_1(r)$                     \\[0.2cm]
            & $\frac{1}{2}x^2$                  & $\frac{1}{2}y^2+r\,y+J_1'(r)$             \\[0.2cm]
c2.1        & $\frac{2+2x}{6+7x},\,x\in[0;r]$   & $\frac{2+2r+y}{6+7r+4y},\,y\in[0;\infty]$ \\[0.2cm]
            & $\frac{7}{2}x^2+6x$               & $2y^2+(6+7r)y+J_1(r)$                     \\[0.2cm]
            & $x^2+2x$                          & $\frac{1}{2}y^2+(2+2r)y+J_1'(r)$          \\[0.2cm]
d2.1        & $\frac{2+2x}{8+9x},\,x\in[0;r]$   & $\frac{2+2r+y}{8+9r+5y},\,y\in[0;\infty]$ \\[0.2cm]
            & $\frac{9}{2}x^2+8x$               & $\frac{5}{2}y^2+(8+9r)y+J_1(r)$           \\[0.2cm]
            & $x^2+2x$                          & $\frac{1}{2}y^2+(2+2r)y+J_1'(r)$          \\[0.2cm]
G1.1        & $\frac{2+2x}{10+9x},\,x\in[0;r]$  & $\frac{2+2r+y}{10+9r+5y},\,y\in[0;\infty]$\\[0.2cm]
            & $\frac{9}{2}x^2+10x+5$            & $\frac{5}{2}y^2+(10+9r)y+J_1(r)$          \\[0.2cm]
            & $x^2+2x+1$                        & $\frac{1}{2}y^2+(2+2r)y+J'_1(r)$          \\[0.2cm]
$\alpha$4.1 & $\frac{1+2r+ x}{3+6r+4x},\,x\in[0;\infty]$    &                               \\[0.2cm]
            & $2x^2+(3+6r)x+\frac{1}{2}$                    &                               \\[0.2cm]
            & $\frac{1}{2}x^2+(1+2r)x+\frac{1}{6}$          &                               \\[0.2cm]
$\alpha$4.2 & $\frac{1+2r+2x}{3+6r+7x},\,x\in[0;\infty]$    &                               \\[0.2cm]
            & $\frac{7}{2}x^2+(3+6r)x+\frac{1}{2}$          &                               \\[0.2cm]
            & $x^2+(1+2r)x+\frac{1}{6}$                     &                               \\[0.1cm]
\hline\hline
\end{tabular}
\caption{Rotation number, $\nu_i$, action, $J_i$, and partial
        action, $J_i'$, for the $i$-th layer of the integer
        polygon mappings with 4-piece force and integrable for 
        isolated value of shift parameter $d$ and
        continuous ratio of finite pieces, $r\in\mathbb{R}^+$. Part 2.}
\label{tab:PMaps4-C2}
\end{table}
%----------------------------------------------------------------
%----------------------------------------------------------------

$\,$

\newpage

$\,$

\newpage

$\,$

\newpage

$\,$

\newpage
%===============================================================%
%===============================================================%
%===============================================================%
\section{\label{secAPP:Zoo} Properties of invariant polygons for Zoo maps}

%----------------------------------------------------------------
%----------------------------------------------------------------
\begin{table}[ht!]
\centering
\begin{tabular}{p{1cm}p{4cm}p{4cm}p{4cm}p{4cm}}
\hline\hline
Zone        & Gingerbreadman        & Rabbit                & Octopus               & Crab                  \\\hline
            &                       &                       &                       &                       \\[-0.25cm]
            &
\includegraphics[width=0.22\columnwidth]{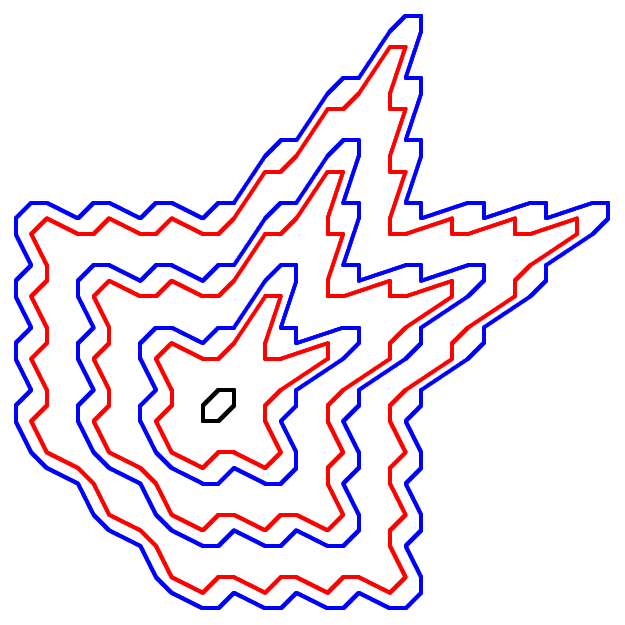}  &
\includegraphics[width=0.22\columnwidth]{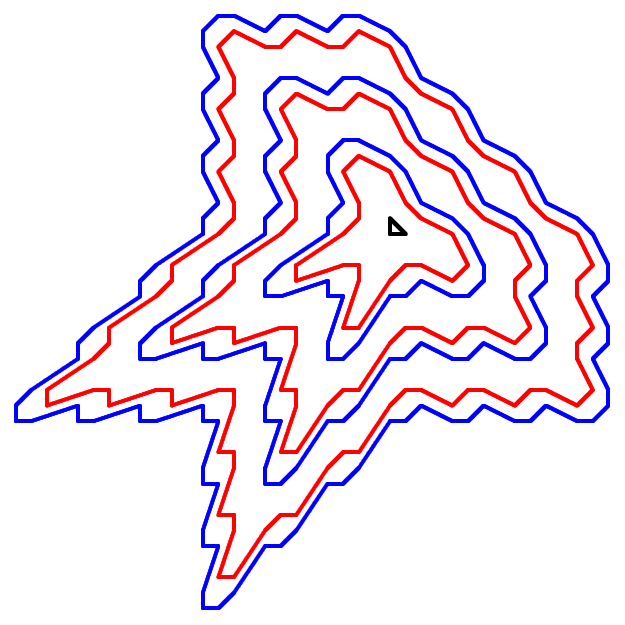}          &
\includegraphics[width=0.22\columnwidth]{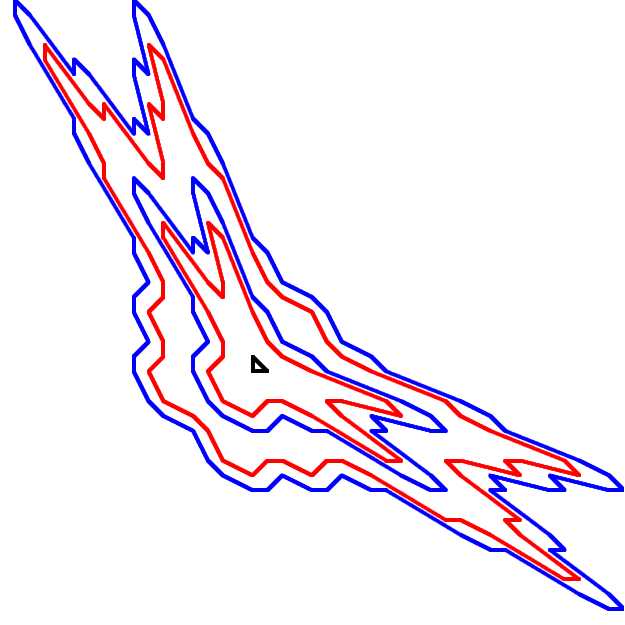}         &
\includegraphics[width=0.22\columnwidth]{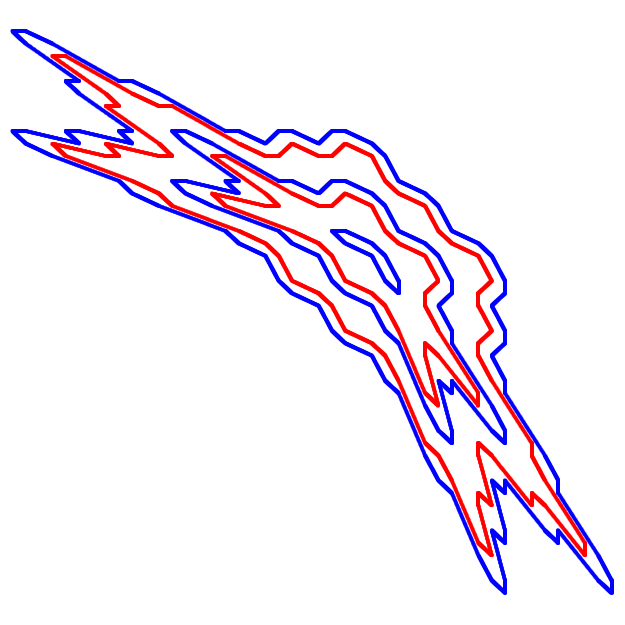}           \\\hline
            &                       &                       &                       &                       \\[-0.25cm]
A$_0$       & $\frac{1}{6}$         & $\frac{1}{3}$         & $\frac{1}{3}$         & $\frac{4}{9}$         \\[0.2cm]
a$_0$       & ---                   & ---                   & ---                   & $\frac{1}{2}$         \\[0.2cm]
a$_0^+$     & ---                   & ---                   & ---                   & $\frac{5}{11}$        \\[-0.3cm]
            &                       &                       &                       &                       \\\hline
            &                       &                       &                       &                       \\[-0.25cm]
A$_k$       & $\frac{6k+1}{27k+6}$  & $\frac{6k+1}{27k+3}$  & $\frac{15k+1}{36k+3}$ & $\frac{15k+4}{36k+9}$ \\[0.2cm]
a$_k$       & $\frac{4k}{18k+1}$    & $\frac{4k}{18k-1}$    & $\frac{10k-1}{24k-2}$ & $\frac{10k+1}{24k+2}$ \\[0.2cm]
a$_k^+$     & $\frac{10k+1}{45k+7}$ & $\frac{10k+1}{45k+2}$ & $\frac{25k}{60k+1}$   & $\frac{25k+5}{60k+11}$\\[0.2cm]
a$_k^-$     & $\frac{10k-1}{45k-2}$ & $\frac{10k-1}{45k-7}$ & $\frac{25k-5}{60k-11}$& $\frac{25k}{60k-1}$   \\[-0.3cm]
            &                       &                       &                       &                       \\\hline
            &                       &                       &                       &                       \\[-0.25cm]
B$_k$       & $\frac{6k-1}{27k-3}$  & $\frac{6k-1}{27k-6}$  & $\frac{15k-4}{36k-9}$ & $\frac{15k-1}{36k-3}$ \\[0.2cm]
b$_k$       & $\frac{2k-1}{9k-4}$   & $\frac{2k-1}{9k-5}$   & $\frac{5k-3}{12k-7}$  & $\frac{5k-2}{12k-5}$  \\[0.1cm]
\hline\hline
\end{tabular}
\caption{Rotation number, $\nu$, for invariant polygons of Zoo maps.
        First row shows images of few invariant polygons responsible
        for zone structure.
        Each map has two sets of zones: ones bounded from outside by
        polygons A$_k$ (shown in blue) and ones bounded by B$_k$
        (shown in red) with $k>1$.
        First three mappings also have an inner linear zone A$_0$
        (shown in black) and Crab map has chaotic inner zone bounded
        by A$_0$ (shown in blue).
        Inside each zone B$_k$ there is a set of linear polygon
        islands (hexagons) b$_k$.
        Inside each zone A$_k$ there are 3 sets of linear polygon
        islands: set of big hexagons a$_k$ and inner and outer (with
        respect to a$_k$) sets a$_k^-$ and a$_k^+$.
        Inner zone A$_0$ of the crab map contains only unstable fixed
        point, so it is formed around 2 sets of islands, a$_0$
        (based on a stable 2-cycle) and smaller islands a$_0^+$
        around a$_0$.
        At large amplitudes, $k \rightarrow \infty$, all frequencies
        tend to $2/9$ (Gingerbreadman and Rabbit) or $5/12$ (Octopus
        and Crab).
        Periods of orbits A,B$_k$ and number of islands a,b$_k$
        (or a$_k^\pm$) is equal to denominator of corresponding $\nu$.
        }
\label{tab:Zoo}
\end{table}
%----------------------------------------------------------------
%----------------------------------------------------------------

\newpage
%\bibliography{papers}

%

\end{document}